\documentclass{article}

\usepackage[final]{template}
\usepackage{graphicx}
\usepackage{multirow}

\usepackage[utf8]{inputenc} % allow utf-8 input
\usepackage[T1]{fontenc}    % use 8-bit T1 fonts
\usepackage{hyperref}       % hyperlinks
\usepackage{url}            % simple URL typesetting
\usepackage{booktabs}       % professional-quality tables
\usepackage{amsfonts,amsmath,dsfont}       % blackboard math symbols
\usepackage{nicefrac}       % compact symbols for 1/2, etc.
\usepackage{microtype}      % microtypography
\usepackage{xcolor}         % colors
\usepackage{verbatim}
\usepackage{algorithm}
\usepackage{algorithmic}

\newtheorem{proposition}{Proposition}
\newtheorem{theorem}{Theorem}

\newtheorem{lemma}{Lemma}

\newtheorem{remark}{Remark}

\newtheorem{assumption}{Assumption}
\newtheorem{corollary}{Corollary}

\title{\vspace{-0.1in} 
On Robust Clustering of Temporal Point Process 
\vspace{-0.1in}}

\author{%
Yuecheng Zhang \\
    School of Mathematical Science, Fudan University\\
    ~~ \\
    \textbf{Guanhua Fang ~~ Wen Yu} \\
    Department of Statistics and Data Science, Fudan University
}

\begin{document}

\bibliographystyle{plainnat}
\maketitle

\bigskip
\begin{abstract}
Clustering of event stream data is of great importance in many application scenarios, including but not limited to, e-commerce, electronic health, online testing, mobile music service, etc.
Existing clustering algorithms fail to take outlier data into consideration and are implemented without theoretical guarantees.
In this paper, we propose a robust temporal point processes clustering framework which works under mild assumptions and meanwhile addresses several important issues in the event stream clustering problem. 
Specifically, we introduce a computationally efficient model-free distance function to quantify the dissimilarity between different event streams so that the outliers can be detected and the good initial clusters could be obtained. 
We further consider an expectation-maximization-type algorithm incorporated with a Catoni's influence function for robust estimation and fine-tuning of clusters. 
%Our proposed method works under mild assumptions.
We also establish the theoretical results including algorithmic convergence, estimation error bound, outlier detection, etc.
Simulation results corroborate our theoretical findings and
real data applications show the effectiveness of our proposed methodology.
\end{abstract}

\noindent%
\vfill

\newpage

\section{Introduction}

In recent applications, many real-world data can be characterized by time-stamped event sequences/streams.
For example, in e-commerce \citep{xu2014path}, the actions taken by a customer in purchasing and viewing the items on the website can form an event sequence. 
In electronic health \citep{enguehard2020neural}, the messages sent by a patient through an AI medical assistant can be viewed as the sequence of events.
In online testing \citep{xu2018latent}, the students take steps to complete the complex problem-solving questions on the computer and their response history can be treated as an event stream.
In mobile music service \citep{carneiro2011towards}, the users can search and play different song tracks and their listening history will be recorded and hence be treated as an event sequence.
Such event data is complicated and entails a lot of individual-level information, which is particularly useful for personalized treatment and recommendation \citep{hosseini2017recurrent, wang2021modeling, cao2021deep}.

To explore the underlying patterns and structures of event stream data, one of the primary tasks is user/individual clustering \citep{yan2019recent}. That is, given a collection of event sequences, we aim to identify groups displaying similar user/individual behaviors.
In recent years, there are quite a few literature investigating on this topic. The existing methods on event stream clustering can be mainly summarized into two categories, namely distance-based clustering and model-based clustering. The methods in the former category measured the similarity among distinct event sequences based on extracted features or pre-specified metrics. For example, \cite{berndt1994using} introduced a dynamic time warping approach to detect the similar patterns. \cite{pei2013clustering} used the discrete Frechet distance to construct the similarity matrix.
The methods in the second one adopted a temporal point process (TPP) framework, where the event sequences are assumed to follow a mixture of point process models. Most popular algorithms fall into this category. 
\cite{xu2017dirichlet} proposed a Dirichlet mixture of Hawkes processes, which is the first attempt in TPP clustering.
\cite{yin2021row} considered a mixture of multi-level log-Gaussian Cox processes and developed an efficient semi-parametric estimation algorithm.
\cite{zhang2022learning} introduced a mixture of neural temporal point processes framework, which first incorporates the TPP clustering with neural network techniques.

Despite the recent progress in TPP clustering mentioned above, there are still some fundamental practical issues remaining. In real world applications, there could exist quite many noisy data. That is, a collection of observed event sequences can not be assumed to exactly follow a mixture of temporal point processes. Instead, a small proportion of event sequences should be treated as outliers. Ignoring this could lead to biased or unreliable classification results.
Consequently, it comes with another issue that how to properly determine whether an observed event sequence is an outlier or not. Unlike the case in panel data where we could use Eculidean distance, Manhattan distance, or other well-specified metric to quantitatively detect the outliers, there is no consensus on the metric to be used for event stream data.
Last but not least, in the current literature, there is no theoretical study on the performance of TPP clustering or the convergence property of proposed algorithms even in the setting without outlier event streams. 
With the existence of outliers, developing the new TPP clustering methodology and the related theoretical guarantees are hence non-trivial tasks.  

%However, research on event stream clustering is limited.
%TPP framework is one of the most widely used framework.

%Moreover, there are two concerns. 1. definition of distance without model specification. 2. outliers.

In this work, we make an attempt to address the above issues. In particular, we propose a robust TPP clustering framework that is less sensitive to the outliers and provides reasonable classification results with theoretical guarantees. Our method works under very mild assumptions that (i) the ``inlier" event stream follows a mixture of non-homogeneous Poisson (NHP) processes while the ``outlier" event stream can be any arbitrary sequence and (ii) we do not assume the specific temporal point process formula for modeling the ``inlier" event stream. 
The clustering algorithm consists of two components, initialization and robust estimation. 
In the first component, we construct a distance function induced by the cubic spline \citep{de1972calculating} to quantify the dissimilarity between different event sequences and use the new distance for outlier screening to get a subset which presumably contains the ``inlier" event streams only. We then apply the $K$-means++ \citep{arthur2007k} method to such subset to determine the initial center of each group and compute the initial probability of how likely each sample belongs to each group based on the distance from the center.  
In the second component, in order to fine-tune the clusters, we adopt an expectation-maximization (EM \citep{dempster1977maximum})-type estimation procedure to iteratively maximize a pseudo likelihood function over a working model space. 
(Since we neither specify the formula of ``inlier" event sequences nor assume the distribution of ``outlier" event streams, then it is impossible to write down the exact likelihood function. Therefore we use a pseudo likelihood as the alternative objective. The working model space considered here is the span of linear combinations of cubic spline functions.)
Moreover, in the M-step, the estimation equation is incorporated with a Catoni-type \citep{catoni2012challenging} influence function which is known to be robust and enjoys many computational and theoretical advantages. The gradient decent is used for updating the parameters.

The technical contributions of this work are summarized as follows.
(a) We introduce a new model-free metric to quantitatively characterize the distance between distinct event sequences. The proposed metric is computationally efficient compared with the existing one (e.g. discrete Frechet distance). Moreover, it can be generalized to a shift-invariant version.  
(b) We propose a robust estimation procedure which utilizes the a Catoni's influence function. We explicitly give out the gradient formula to update the working model parameters. In terms of computational complexity, it only requires an additional step to compute the adjusted weight (which re-weights the possibility of being in a particular group and reduces the impact of outliers) for each sample.
(c) A complete theoretical analysis is provided. Under mild conditions, we show the effectiveness of the proposed algorithm. For the initialization component, it can return a set of high-quality centers. For the robust estimation component, it enjoys a linear convergence rate. 
With the help of Catoni's influence function, the method is robust and has a relatively high break-down point. 
When the model is correctly-specified and the tuning parameter is carefully chosen, the error bound of the estimated parameter is nearly optimal and the algorithm can detect all outliers with high probability.
To the best of our knowledge, this is the first theoretical work in studying the convergence of TPP clustering.

The rest of paper is organized as follows. 
A preliminary of event stream data, temporal point process model, Catoni's influence function, and related existing work are provided in Section \ref{sec:2}. 
The main methodology of robust clustering is described in Section \ref{sec:3}. 
We provide the corresponding theoretical analyses in Section \ref{sec:4}.
In section \ref{sec:5}, simulation studies are carried out to show the effectiveness of the new method.
Two real data applications are given in Section \ref{sec:6} to show the superior performance of our proposed algorithm.
Finally, a concluding remark is given in Section \ref{sec:7}.

%\vspace{-2mm}

\section{Preliminary}\label{sec:2}

%\vspace{-2mm}

\subsection{Data Format}

We consider the following event stream data, $\Big\{ (t_{n1}, ..., t_{ni}, ..., t_{n M_n}); n = 1, ..., N \Big\}$, where $t_{ni}$ is the $i$-th event time stamp of the $n$-th individual, $M_n$ is the number of events observed for individual $n$, and $N$ is the total number of individuals. For the notional simplicity, we may use $S_n$ to denote observation sequence of individual $n$, i.e., $S_n = (t_{n1}, ..., t_{ni}, ..., t_{n M_n})$. To help readers to gain more intuitions, we provide two real data examples in Table \ref{tab:example1} and Table \ref{tab:example2}, which show the event stream sequence of an randomly selected user from the internet protocol television (IPTV) data and music listening (Last.FM 1K) data, respectively.

\begin{minipage}{\textwidth}
\begin{minipage}[t]{0.46\textwidth}
\makeatletter\def\@captype{table}
\scalebox{0.9}{
\begin{tabular}{ccc}
        \toprule
          & id & time \\
        \midrule
        1 & 55357201 & 2012/01/01 18:33:15 \\
        2 & 55357201 & 2012/01/01 18:34:55 \\
        % 3 & 55357201 & 2012/01/02 20:44:07 \\
        % 4 & 55357201 & 2012/01/02 21:36:42 \\
        % 5 & 55357201 & 2012/01/02 21:47:19 \\
        % 6 & 55357201 & 2012/01/02 22:01:48 \\
        $\cdots$   & $\cdots$   & $\cdots$   \\
        % 4144 & 55357201 & 2012/11/28 01:59:02 \\
        4145 & 55357201 & 2012/11/28 02:01:42 \\
        4146 & 55357201 & 2012/11/28 02:04:01 \\
        \bottomrule
    \end{tabular}\label{tb:iptv}
    }
\caption{{\small IPTV dataset. "id": user identifier. "time": the time stamp when the user started to watch a TV program.}}
\label{tab:example1}
\end{minipage}
\hspace{2mm}
\begin{minipage}[t]{0.46\textwidth}
\makeatletter\def\@captype{table}
\scalebox{0.9}{
    \begin{tabular}{ccc}
        \toprule
          & user\_id & time \\
        \midrule
        1 & user000685 & 2005/12/10 06:23:10 \\
        2 & user000685 & 2005/12/10 06:26:35 \\
        % 3 & user000685 & 2005/12/10 06:28:16 \\
        % 4 & user000685 & 2005/12/10 06:33:03 \\
        % 5 & user000685 & 2005/12/10 06:39:43 \\
        % 6 & user000685 & 2005/12/10 06:44:29 \\
        $\cdots$   & $\cdots$   & $\cdots$   \\
        % 84440 & user000685 & 2009/05/20 05:24:57 \\
        84441 & user000685 & 2009/05/22 06:44:01 \\
        84442 & user000685 & 2009/05/23 11:12:10 \\
        \bottomrule
    \end{tabular}
    }\label{tb:lastfm}
\caption{{\small Last.FM 1K Dataset. "user\_id": user identifier. "time": the time stamp when the user played a song track.}}
\label{tab:example2}
\end{minipage}
\end{minipage}

\vspace{3mm}

To mathematically characterize the event stream data, it is appropriate to adopt the framework of TPP \citep{daley2003introduction}, also known as the counting process. For any increasing event time sequence $0 < t_1 < t_2 < ... < t_M$, we let $N(t) := \sharp\{i: t_i \leq t \}$ be the number of events observed up to time $t$. Then we can define the conditional intensity function,
$
\lambda^{\ast}(t) := \lim_{dt \rightarrow 0}
\mathbb E[N[t, t+dt) | \mathcal H_t]/dt
$,
where $N[t, t+dt) := N(t+dt) - N(t)$ and 
$\mathcal H_t := \sigma(\{N(s); s < t\})$ is the history filtration before time $t$.
Intensity $\lambda^{\ast}(t)$ describes the dynamic of the event process and is of great importance and interest for statistical modelling.

% A temporal point process (TPP) (\citet{daley2003introduction}) defines a probability distribution over variable-length event sequences in an interval $[0, T]$. A TPP realization $X$ consists of strictly increasing arrival times $\left(t_1, \ldots, t_N\right)$, where $N$, the number of events, is itself a random variable. A TPP is characterized by its conditional intensity function $\lambda^*(t):=\lambda\left(t \mid \mathcal{H}_t\right)$ that is equal to the rate of arrival of new events given the history $\mathcal{H}_t=\left\{t_j: t_j<t\right\}$. Equivalently, a TPP can be specified with the integrated intensity function $\Lambda^*(t)=\int_0^t \lambda^*(u) d u$.

\subsection{Robustness}

% Figure: real data illustrate the existence of outliers. 

In event stream analysis, one could always observe that a few individuals may behave very differently from the majority of the users \citep{gupta2013outlier, sani2019repairing}. Therefore, we need to take into account the potential existence of outliers and develop robust methods to alleviate estimation bias.
In the literature of robust $M$-estimation, there exist different types of methods to estimate population mean, including but not limited to, median of mean \citep{bubeck2013bandits}, geometric median \citep{hsu2016loss}, Huber's estimator \citep{huber1992robust}, trimmed mean \citep{lugosi2021robust}, robust empirical mean \citep{prasad2020robust}, and Catoni's estimator \citep{catoni2012challenging}.
As discussed in the seminal work \citep{catoni2012challenging}, Catoni's estimator is shown to have sub-Gaussian non-asymptotic error bound with optimal multiplicative constant.
Furthermore, as shown in the recent work \citep{bhatt2022minimax}, Catoni's estimator has the highest break-down point compared with other computational friendly methods, i.e., trimmed mean and robust empirical mean.
Moreover, according to the numerical results in \cite{fang2023empirical}, Catoni's estimator could achieve the best empirical performance among all methods mentioned above.
As a result, we will focus on Catoni's estimator in the remaining sections.

To be mathematically formal, given a set of observations $\{X_i\}_{i=1}^n$, a Catoni's estimator is defined to be the solution to the following non-linear equation,
$
\sum_{i=1}^n \phi(\alpha(X_i - \mu)) = 0$, with respect to $\mu$,
where the influence function $\phi$ is non-decreasing and satisfies
\begin{eqnarray}\label{eq:require:cat}
  -\log \left(1-x+x^2 / 2\right) \leq \phi(x) \leq \log \left(1+x+x^2 / 2\right),
\end{eqnarray}
and $\alpha$ is a tuning parameter. 
Throughout the paper, we choose the following
specific formula,

% \begin{lemma}[Theorem 3.1 (\cite{bhatt2022minimax})]
% Let $\delta \in(0,1)$ such that $\delta \geq 2 e^{-n / 4}$. Let $\left\{X_i\right\}_{i=1}^n$ be i.i.d random variables with mean $\mu$ and $\mathbb{E}\left|X_1-\mu\right|^2 \leq v$. Let the corruption parameter $\eta \in\left[0, \frac{1}{4 A}\right)$ for $A>0$. 
% Robust Catoni's M-estimator $\widehat{\mu}$ with parameter
% $$
% \alpha=\frac{1}{v^{1 / 2}}\left(\Omega A \eta+\frac{2 \log (2 / \delta)}{n}\right)^{1 / 2}
% $$
% satisfies, with probability at least $1-\delta$,
% $$
% |\widehat{\mu}-\mu|<v^{1 / 2} \frac{\frac{(\Omega+4)}{2 \Omega^{1 / 2}} A^{1 / 2} \eta^{1 / 2}+\left(\frac{2 \log (2 / \delta)}{n}\right)^{1 / 2}}{1-(\Omega+4) A \eta / 2-2 \log (2 / \delta) / n} .
% $$
% \end{lemma}

% Clearly, $\phi(x) = x$ corresponds to the empirical mean. We constructed the following robust function $\phi: \mathbb{R}\rightarrow\mathbb{R}$ such that 

{\small 
\begin{equation}\label{eq:rob}
\phi(x)=\left\{
\begin{aligned}
&\log(1+x+0.5\cdot x^2) & & x\leq 2, \\
& 0.032/9\cdot (x-9.5)^3+1.5+\log(5) & & 2<x\leq 9.5, \\
& 1.5+\log(5) & & x> 9.5 ,
\end{aligned}
\right.
\end{equation}
}
for $x\in \mathbb{R^+}$ and $\phi(0) = 0$. When $x < 0$, define $\phi(x):=-\phi(-x)$.
It is not hard to see that the constructed $\phi(x)$ has the continuous second derivative, which facilitates the theoretical analyses. 

\begin{remark}
    The constant (e.g. 9.5) in \eqref{eq:rob} could be modified. Here the only principle in choosing $\phi$ is that it satisfies \eqref{eq:require:cat} and is sufficiently smooth, that is, the second derivative is continuous. 
\end{remark}

%\vspace{-2mm}

\subsection{Clustering}

%\vspace{-2mm}

In many real applications, we could observe strong clustering effects, that is, individuals can be classified into groups according to whether their behaviors are similar or not.
For the classical panel data, the clustering problem has been investigated thoroughly.  
$K$-means \citep{lloyd1982least}, an iterative refinement technique by clustering the samples into the nearest class centers according to a certain well-defined metric (e.g. Euclidean distance), is arguably the most widely used method. 
% $K$-nearest neighbors (KNN, \cite{fix1989discriminatory}), a non-parametric supervised learning method aiming to classify a new sample based on the labels of $K$ closest training samples, is also popular in machine learning field.
% Hierarchical clustering \citep{johnson1967hierarchical}, a systematic method of cluster analysis, can build a hierarchy of clusters.  
% Spectral clustering \citep{von2007tutorial} makes use of the spectrum /eigenvalues of the similarity matrix constructed from the raw data to perform dimensionality reduction and to cluster in low-dimensional space. 
Other methods including $K$-nearest neighbors (KNN, \cite{fix1989discriminatory}), hierarchical clustering \citep{johnson1967hierarchical}, and spectral clustering \citep{von2007tutorial} are also popular in the literature.
Model-based method \citep{reynolds2009gaussian} is another important line of clustering algorithms in the statistical literature. By introducing augmented latent variable which indicates the class label, the expectation-maximization (EM, \cite{dempster1977maximum}) algorithm is widely adopted in many areas including social science \citep{little1989analysis}, psychonometrics \citep{rubin1982algorithms}, quantitative genetics \citep{zhan2011stochastic}, etc.

For analyzing event stream data, there is no unanimous method yet.  
Existing methods can be divided into two categories, distance-based clustering \citep{berndt1994using, bradley1998refining, peng2008distance} and model-based clustering \citep{luo2015multi, xu2017dirichlet, yin2021row}.
The former one quantifies the similarities
between event streams based on some extracted features and then applies classical 
clustering algorithms such as $K$-means, spectral clustering, etc.
The latter one assumes that event streams are generated from some underlying parametric
mixture models of point processes so that the likelihood function can be derived and EM algorithm could be applied.
% For example,  \cite{xu2017dirichlet} proposed a Dirichlet mixture of Hawkes processes (DMHP). \cite{yin2021row} introduced a mixture model of Multi-level Marked Point Processes (MM-MPP).

However, none of above mentioned methods is robust to outliers or provides any theoretical guarantee to ensure the correct clustering results. In this work, we try to propose a new algorithm enjoying the merits of both metric-based and model-based methods.
We use a metric-based component for screening outliers and obtaining good initializations of group centers.
We use a model-based component for fine-tuning the model parameters and final clustering results.

\subsection{How to define a suitable distance}

Note that our primary goal is to cluster different individuals based on their observed event time sequences.
It is urgent to introduce a suitable metric distance to quantify the dissimilarity between distinct event sequences.
Unlike the classical situations that each individual / subject has the same number of covariates / features, the length of time sequences in our setting could vary among different people. 
Therefore Euclidean distance cannot be applied, at least directly, to the event stream data. How to define a reasonable metric becomes a non-trivial task.

Most existing distances for TPPs are based on the random time change theorem \citep{brown2002time}. 
Such metrics suffer severe non-identifiability issues. Two very different event streams can be very close under such metrics.
More failure modes can be found in \cite{pillow2009time}.
Detailed explanations can be found in the supplementary.

In the literature, there also exists an intensity-free metric called discrete Frechet distance \citep{eiter1994computing, pei2013clustering}. It can be used to measure the difference between any two polygonal curves in the metric space. However, in terms of computation, it requires dynamic programming technique, which leads to quadratic computational complexity. That is, the computational time is proportional to the square of number of observed event numbers. Hence, it is not a desired method when the data size becomes larger.
Therefore, we need to seek a different type of distance which will be described in later sections.
% Therefore, we want to design a distance between different samples directly based on the intensity function $\hat{\lambda}(\cdot)$.

\section{Robust Clustering Algorithm}\label{sec:3}

\subsection{Distance Induced via Cubic Spline}

For any two event streams, $S_A = (t_{A,1}, ..., t_{A,N_A})$ and $S_B = (t_{B,1}, ..., t_{B, N_B})$, we consider quantifying the distance between them by adopting the cubic splines.
We suppose that event streams are observed within time interval $[0,T]$ or they are periodic with the same period $T$.
Then we define the following distance, 
\begin{equation}\label{eq:dis}
    d(S_A,S_B):=\int_{0}^T \left|\hat{\lambda}_{S_A}\left(t\right)/\sqrt{M_A}-\hat{\lambda}_{S_B}\left(t\right)/\sqrt{M_B}\right|dt ,
\end{equation}
where $M_A$ and $M_B$ is the number of events of sequence $S_A$ and $S_B$ and $\hat{\lambda}_S(\cdot)$ is the estimated intensity function by fitting cubic splines to event stream $S$.
Moreover, if we want to make the distance to be shift invariant, we can adopt the following generalized definition,
\begin{equation}\label{eq:dis:shift}
    \tilde d(S_A,S_B):= \min_{s \in [0,T]} \int_{0}^T \left|\hat{\lambda}_{S_A}\left(t + s\right)/\sqrt{M_A}-\hat{\lambda}_{S_B}\left(t\right)/\sqrt{M_B}\right|dt,
\end{equation}
where $\hat \lambda_S(t + s) = \hat \lambda_S(t + s - T)$ when $t + s > T$.
\eqref{eq:dis:shift} becomes useful when event sequences are collected from users of different countries which are in different time zones.

In order to compute $\hat \lambda_S(\cdot)$ for a fixed event stream $S$,
we need to construct basis functions in the form of cubic splines. Note that the event streams are assumed to be periodic.
Therefore, we also enforce the basis to be periodic as well, that is, its value, the first- and second-order derivatives are all continuous at the boundaries.
The detailed construction procedure of basis is given in the supplementary.
% Cubic splines can be used to model cyclic data as well. When we have cyclic data, it means the function we are approximating repeats itself periodically. In this case, we want the spline to be continuous not only in its first and second derivatives but also across the boundaries of the cyclic data.
% To achieve this, we typically use a cubic spline with natural boundary conditions. The natural boundary conditions set the second derivatives at the endpoints to be zero, which means the spline becomes linear beyond the endpoints. This ensures that the spline naturally repeats itself because the second derivative (curvature) is zero at the endpoints, making it smooth and continuous across the cyclic.
We then estimate $\hat \lambda_S(t)$ by $\sum_{h=1}^H b_{h,S} \kappa_h(t)$, where $H$ is the number of basises, $\kappa_h(t)$ is the $h$-th basis, and
$\{b_{h,S}\}$'s satisfy
\begin{eqnarray}\label{def:bh}
(b_{1,S}, ..., b_{H,S}) = \arg\max_{(b_1,...,b_H)} \left\{\sum_{i=1}^{N_S} \log \lambda(t_i) - \int_0^T \lambda(t)dt \right\} 
\end{eqnarray}
with $\lambda(t) = \sum_{h=1}^H b_{h} \kappa_h(t)$.
Note that \eqref{def:bh} is essentially a convex optimization problem which can be efficiently solved.
Computation of \eqref{eq:dis} or \eqref{eq:dis:shift} scales linearly with the lengths of event sequences.
Therefore, the proposed metric is more computationally friendly than the discrete Frechet distance.

Note that we divide the estimated intensity by the square root of the number of events in \eqref{eq:dis}. This is due to the following observation.
\begin{proposition}\label{pro:poi}
Suppose $S=(t_1,t_2,\cdots)$~ follows a homogeneous Poisson process with intensity $\lambda$ and $f(\cdot)$ is a bounded function in $[0,T]$. The variance of $\sum_i f(t_i)/\sqrt{N(T)}$ is approximately $(\int_0^T f^2(t)dt/T)\cdot(1/4+O(1/\lambda))$.
\end{proposition}

According to Proposition \ref{pro:poi}, we rescale the intensity function to make the distance function be insensitive to the magnitude of intensity. Thus we can classify different individuals based on their intrinsic patterns instead of the absolute value of event number.
% the internal variances between different classes as close as possible to eliminate the influence of different sample variances between different classes.

To end this subsection, we show that $d(S_A, S_B)$ ($\tilde d(S_A, S_B)$) given in \eqref{eq:dis} (\eqref{eq:dis:shift}) is a proper distance function. 
Here $d(S_A, S_B)$ is called as a distance function if it satisfies three properties: (i) the distance between an event sequence and itself is always zero, (ii) the distance between distinct event sequences is always positive and symmetric, and (iii) the distance satisfies the triangle inequality.

\begin{theorem}\label{thm:dis}
The function defined in \eqref{eq:dis} or \eqref{eq:dis:shift} is a distance function.
\end{theorem}

Theorem \ref{thm:dis} is proved in the supplementary. Without validating these, directly applying existing clustering algorithms may fail without theoretical guarantees.

\subsection{Clustering with Robust Estimation}

In this section, we propose a clustering algorithm based on a mixture model \citep{fraley2002model, mclachlan2019finite}.
In particular, we assume the observed event sequences 
$\mathbf{S}=\left\{S_n\right\}_{n=1}^N$ are generated from mixture 
%temporal point processes
non-homogeneous Poisson processes
with $K$ classes and possible outlier sequences.
All of them has the same period $T$.
If an event sequence belongs to class $k \in [K]$, then its corresponding population-level intensity, or rate, is $\lambda_k^{\ast}(t)$.
At the moment, we do not put any structural assumption on $\lambda_k^{\ast}(t)$'s.
Instead, we consider the following working model, that is, $\lambda_k^{\ast}(t)$ can be approximated by 
% Given a set of event sequences $\boldsymbol{S}=\left\{\boldsymbol{S}_n\right\}_{n=1}^N$, where $\boldsymbol{S}_n=\left\{t_i\right\}_{i=1}^{M_n}$ contains a series of events time stamps $t_i \in\left[0, T_n\right]$, and $M_n$ is the number of events of $\boldsymbol{S}_n$.
% For the event sequence belonging to the $k$-th cluster its intensity function at time $t$ is assumed to have the following representation,
\begin{eqnarray}\label{eq:model}
\lambda_k(t) :=\sum_{h=1}^H b_{k,h} \kappa_h(t),
\end{eqnarray}
where $\kappa_h(t)$ is the $h$-th basis function defined in the last section.
We write $\boldsymbol{B}_k :=\left[b_{k,h}\right] \in \mathbb{R}_{0+}^{H}$ as the coefficient parameter in non-homogeneous Poisson process of class $k$, $\mathbf{B} := \{\boldsymbol{B}_k\}_{k=1}^K$ as the whole parameter for simplicity.

% Mixture models (\cite{mclachlan2019finite}) have gained growing popularity in the last decades as a tool for model-based clustering (\cite{fraley2002model}).  

According to the classical mixture models \citep{xu2017dirichlet, zhang2022learning} with no outliers, we let $Z_n$ denote the latent label for the $n$-th event stream. In other words, $Z_n = k$ represents that the $n$-th event sequence belongs to $k$-th class.
If there is \textbf{no} outlier, we can write down the probability of an event stream $S$ as
$p(S; \mathbf{B}) = \sum_k \pi_k \cdot \operatorname{NHP}\left(S |  \boldsymbol{B}_k\right)$ with 
$$\operatorname{NHP}\left(S \mid \boldsymbol{B}_k\right) := p(S|Z = k) = \prod_i \lambda_k\left(t_i\right) \exp \left(- \int_0^{L(S) \cdot T} \lambda_k(t) d t\right),$$
% \begin{eqnarray}
%    p(S; \mathbf{B}) &=& \sum_k \pi_k \cdot \operatorname{NHP}\left(S |  \boldsymbol{B}_k\right), \label{eq:prob} \\ ~\text{with}~ \operatorname{NHP}\left(S \mid \boldsymbol{B}_k\right) &:=& p(S|Z = k) = \prod_i \lambda_k\left(t_i\right) \exp \left(- \int_0^{L(S) \cdot T} \lambda_k(t) d t\right), \nonumber
% \end{eqnarray}
where $\pi_k$ 's are class probabilities, $\operatorname{NHP}\left(S \mid \boldsymbol{B}_k\right)$ is the conditional probability of the event sequence $S$ if it belongs to class $k$, and $L(S)$ is the number of periods in event sequence $S$. 
% $$
% \begin{aligned}
%  \boldsymbol{s} \mid k, \boldsymbol{B}\sim \operatorname{HP}\left(\boldsymbol{s}\mid\boldsymbol{B}_k\right).
% \end{aligned}
% $$
We write $\mathbf Z = \{Z_n\}_{n=1}^N$. Then the (pseudo) joint likelihood of $\mathbf S$ and $\mathbf Z$ is 
% $$
% \begin{aligned}
% & p(\boldsymbol{S}, \boldsymbol{Z}, \boldsymbol{\pi}\mid \boldsymbol{B})=p(\boldsymbol{S} \mid \boldsymbol{Z}, \boldsymbol{B}) p(\boldsymbol{Z}) , \\
% & p(\boldsymbol{S} \mid \boldsymbol{Z}, \boldsymbol{B})=\prod_{n, k} \operatorname{HP}\left(\boldsymbol{S}_n \mid \boldsymbol{B}_k\right)^{z_{n k}}, \\
% & p(\boldsymbol{Z} \mid \boldsymbol{\pi})=\prod_{n, k}\left(\pi^k\right)^{z_{n k}}.
% \end{aligned}
% $$

% Given $\mathbf{Z}$, we use $K$ NHPPs to model the conditional probability of $\mathbf{S}$. The full likelihood can be written as
$$
\begin{aligned}
 p(\mathbf S, \mathbf Z; \mathbf B) = \prod_{n=1}^N \prod_{k=1}^K\left[\pi_k \operatorname{NHP}\left(S_n \mid \boldsymbol{B}_k\right)\right]^{\mathbf 1\{Z_{n} = k\}}
\end{aligned}
$$
and the (pseudo) marginal likelihood of $\mathbf S$ is 
\begin{eqnarray}\label{eq:pesudo}
    p(\mathbf S; \mathbf B) = 
    \prod_{n=1}^N \left\{ \sum_{k=1}^K \pi_k \operatorname{NHP}\left(S_n \mid \boldsymbol{B}_k\right) \right\}.
\end{eqnarray}
Then the goal becomes to compute the maximizer,
$\mathbf{B}_{o p t} :=\arg \max_{\mathbf{B}} p(\mathbf{S} ; \mathbf{B})$.
% \begin{eqnarray}\label{eq:em:obj}
%     \mathbf{B}_{o p t} :=\arg \max_{\mathbf{B}} p(\mathbf{S} ; \mathbf{B}).
% \end{eqnarray}

\begin{remark}
Here we call \eqref{eq:pesudo} as the pseudo likelihood since it is not the exact likelihood function.
This is because we treat all $N$ observed event sequences as inliers even if it is not. 
In other words, we try to estimate the group centers and model parameters under the mis-specified setting.
\end{remark}

In order to solve $\mathbf{B}_{o p t}$, the standard and most popular computational approach is the expectation-maximization (EM) algorithm \citep{dempster1977maximum} in the literature.
However, due to the existence of outliers, we cannot directly apply the EM algorithm. We make the modification to it by using Catoni influence function to reweight each observed event sequence. 
At time step $t$, E-step and M-step are given as follows.

\noindent \textbf{E-step}. 
We first compute the posterior $p(\mathbf Z | \mathbf S; \mathbf B^{(t-1)})$, where $\mathbf B^{(t-1)}$ is the parameter estimate in the previous step. It is not hard to find that 
\begin{eqnarray}\label{eq:posterior:e-step}
    p(\mathbf Z | \mathbf S; \mathbf B^{(t-1)}) &=& \prod_{n=1}^N p(Z_n | S_n; \mathbf B^{(t-1)})  = \prod_{n=1}^N \prod_{k=1}^K (r_{nk}^{(t)})^{\mathbf 1\{Z_n = k\}}
\end{eqnarray}
with 
\begin{eqnarray}\label{eq:def:rnk}
    r_{nk}^{(t)} = \frac{\rho_{nk}^{(t)}}{\sum_{k'} \rho_{nk'}^{(t)}},
\end{eqnarray}
where $\rho_{nk}^{(t)} := \pi_{k}^{(t-1)} \cdot \operatorname{NHP}(S_n|\boldsymbol{B}_k^{(t-1)})$. For simplicity, we write $p(\mathbf Z | \mathbf S; \mathbf B^{(t-1)})$ as $q^{(t)}(\mathbf Z)$.
Thus the $Q$-function, the expectation of the complete log-likelihood over $q^{(t)}(\mathbf Z)$, is 
\begin{align}\label{eq:Qfunc}
\mathcal{Q}(\mathbf{B} | \mathbf B^{(t-1)}) & =\mathbb{E}_{q^{(t)}(\mathbf Z)}[\log p(\mathbf{S} \mid \mathbf{Z},\mathbf{B})]+C =\sum_{n=1}^N \sum_{k=1}^K r_{n k}^{(t)} \log \operatorname{NHP} \left(S_n \mid \boldsymbol{B}_k\right)+C.
\end{align}

\noindent \textbf{M-step}.
The classical routine is to find the estimate $\mathbf B^{(t)} = \arg\max_{\mathbf B} \mathcal Q(\mathbf B|\mathbf B^{(t-1)})$. In our setting, we have the following observation that $\mathbf B^{(t)} \equiv (\boldsymbol{B}_k^{(t)})_{k=1}^K$ with 
\[\boldsymbol{B}_k^{(t)} := \arg\max_{\boldsymbol{B}_k} \sum_{n=1}^N r_{nk}^{(t)} \log \operatorname{NHP}(S_n | \boldsymbol{B}_k),\]
which can be equivalently written as 
$
    \boldsymbol{B}_k^{(t)} := \arg\max_{\boldsymbol{B}_k} \mu_{avg}^{(t)}(\boldsymbol{B}_k)
$ with $\mu_{avg}^{(t)}(\boldsymbol{B}_k)$ being the solution to 
\begin{eqnarray}\label{eq:mu:def}
\sum_{n=1}^N  r_{n k}^{(t)} \left(\log \operatorname{NHP}\left(S_n \mid \boldsymbol{B}_k\right)- \mu \right)=0
\end{eqnarray}
with respect to $\mu$.

Given the existence of outliers, we instead consider the following robust estimator 
\begin{eqnarray}\label{eq:def:robust:est}
    \boldsymbol{B}_k^{(t)} := \arg\max_{\boldsymbol{B}_k} \hat{\mu}_{\phi}^{(t)}(\boldsymbol{B}_k),
\end{eqnarray}
where $\hat{\mu}_{\phi}^{(t)}(\boldsymbol{B}_k)$ is the solution to
\begin{eqnarray}\label{eq:mu:robust}
\sum_{n=1}^N  r_{n k}^{(t)}  \cdot L(S_n) \cdot {\color{black} \phi_{\rho}} \left(\log \operatorname{NHP}\left(S_n \mid \boldsymbol{B}_k\right) / L(S_n)- \mu \right)=0
\end{eqnarray}
with respect to $\mu$, where $ \phi_{\rho}(x) := \rho^{-1} \cdot \phi(\rho \cdot x)$ with $\phi(x)$ defined in \eqref{eq:rob} and $\rho$ being a tuning parameter. (The following results will not be affected, if we also allow $\rho$ depends on class index $k$.)
Especially, when $\phi(x)$ is an identity function, \eqref{eq:mu:robust} reduces to \eqref{eq:mu:def} up to a multiplicative constant (free of $\boldsymbol{B}_k$).
To solve \eqref{eq:def:robust:est}, we consider to use gradient descent-type method. In particular, we can compute the gradient with explicit formula which is given in the following proposition.

\begin{proposition}\label{pro:mstep}
    The gradient $\varrho^{(t)}_k$ of $\hat{\mu}_{\phi}^{(t)}(\boldsymbol{B}_k)$ with respect to parameter $\boldsymbol{B}_k$ at $\boldsymbol{B}_k^{(t-1)}$ (i.e. $\varrho^{(t)}_k := \frac{\partial \hat{\mu}_{\phi}^{(t)}(\boldsymbol{B}_k)}{\partial \boldsymbol{B}_k}|_{\boldsymbol{B}_k^{(t-1)}}$) is 
    \begin{eqnarray}
        \label{eq:gradient}
    \sum_{n=1}^N\frac{r_{n k}^{(t)} w_{n k}^{(t)}}{\sum_{n=1}^N  r_{n k}^{(t)} w_{n k}^{(t)}L(S_n)}\cdot\frac{\partial\log \operatorname{NHP}\left(S_n \mid \boldsymbol{B}_k\right)}{\partial\boldsymbol{B}_k}\mid_{\boldsymbol{B}_k^{(t-1)}},
    \end{eqnarray}
    where %$w_{n k}^{(t)}= \phi'_{\rho}\left(\log \operatorname{NHP} \left(S_n \mid \boldsymbol{B}^{(t-1)}_k\right)-\hat{\mu}_{\phi}(\boldsymbol{B}^{(t-1)}_k)\right)$
    $w_{n k}^{(t)}= \phi'_{\rho}\left(\log \operatorname{NHP} \left(S_n \mid \boldsymbol{B}^{(t-1)}_k\right) / L(S_n) -\hat{\mu}_{\phi}(\boldsymbol{B}^{(t-1)}_k)\right)$.
\end{proposition}

According to Proposition \ref{pro:mstep}, we actually adjust $\boldsymbol{B}_k$'s gradient via influence function $\phi_{\rho}$. Here $w_{n k}^{(t)}$ can be viewed as the adjusted weight of $n$-th event stream. By the construction of $\phi_{\rho}$, it can be checked that $w_{n k}^{(t)} \in [0, 1]$. When $w_{n k}^{(t)}$ is close to one, it indicates the strong confidence that event stream $n$ is more likely to belong to class $k$. On the other hand, if $w_{nk}^{(t)}$ is close to zero, it implies the corresponding event stream could be an outlier or is at least far away from class $k$. 
If an event sequence $n$ is truly an outlier, then its weights $w_{nk}$'s are uniformly small for all $k \in [K]$. Then it has negligible influence to the gradient according to \eqref{eq:gradient}, which in turn implies the robustness of our proposed method.
To sum up, the parameter update is 
\begin{eqnarray}\label{eq:para:update}
\boldsymbol{B}_k^{(t)} = \boldsymbol{B}_k^{(t-1)} - \text{lr} \cdot \varrho^{(t)}_k ~~ \text{for}~ k \in [K],
\end{eqnarray}
where $\text{lr}$ is the learning rate/step size.
When $\|\boldsymbol{B}_k^{(t)} - \boldsymbol{B}_k^{(t-1)}\| \leq \epsilon$ ($\epsilon$ is a small tolerance parameter), we stop the E- and M-steps.  
Lastly, for class probabilities, we can update $\{\pi_k\}_{k=1}^K$ by $\pi_k^{(t)}=\sum_{n=1}^N r_{n k}^{(t)}/N$.

In the case of time shift, we need to assign a shift parameter to each event sequence. We let $\text{shift}_n$ be the time zone of $n$-th event stream.  
In addition to update $\boldsymbol{B}$ at time step $t$, we also update 
\begin{align}\label{eq:update:shift}
    \operatorname{shift}_n^{(t)}=\underset{\operatorname{shift}_n\in\{\frac{T}{H_{\text{shift}}},\frac{2\cdot T}{H_{\text{shift}}},\cdots,T\}}{\operatorname{argmin}}\int_0^T \left|\hat{\lambda}_{S_n}(u+\operatorname{shift}_n)-\hat{\lambda}_{Z_n^{(t)}}(u)\right|du,
\end{align}
where $H_{\text{shift}}$ represents the number of possible time shifts (e.g. $H_{\text{shift}}$ can be seen as the 24 time zones), $\hat{\lambda}_{S_n}(\cdot)$ is obtained from \eqref{def:bh} and $\hat{\lambda}_{Z_n^{(t)}}(\cdot)$ is the estimated intensity function of class $Z_n^{(t)}$ with $Z_n^{(t)} = \arg\max_{k} r_{nk}^{(t)}$.
Again, when $u+\operatorname{shift}_n>T$, we define $\hat{\lambda}_{S_n}(u+\operatorname{shift}_n):=\hat{\lambda}_{S_n}(u+\operatorname{shift}_n-T)$.

The algorithm of robust clustering is summarized in Algorithm \ref{alg1}.

\begin{algorithm}[h]
	%\textsl{}\setstretch{1.8}
	\renewcommand{\algorithmicrequire}{\textbf{Input: Sequences $S = \{S_n \}_{n=1}^N$}}
	\renewcommand{\algorithmicensure}{\textbf{Output:}}
	\caption{Robust clustering}
	\label{alg1}
	\begin{algorithmic}[1]
		\STATE \textbf{Input} {Sequences $S = \{s_n \}_{n=1}^N$}, tolerance parameter $\epsilon$.
        \STATE ------ \textbf{Initialize clusters} ------
        \STATE Run Algorithm \ref{alg:kmean+} to get $r_{nk}^{ini}$ (and $\{\text{shift}_n^{ini}\}$ if necessary).  
        \STATE Compute initial $\mathbf{B}^{(0)}$ by maximizing $\mathcal L(\mathbf{B})$ specified in \eqref{eq:Qfunc} with $r_{nk}^{(0)}$ replaced by $r_{nk}^{ini}$.
        \STATE Compute initial $\pi_k^{(0)} = \sum_n r_{nk}^{ini} / \sum_{n,k} r_{nk}^{ini}$ and set $t = 0$.
        \STATE ------ \textbf{Fine-tune clusters} ------
		\REPEAT
		\STATE Compute $r_{n k}^{(t)}$ according to Eq.\eqref{eq:def:rnk}.
		\STATE Compute $\pi_k^{(t)}=\sum_{n=1}^N r_{n k}^{(t)}/N$.
        \STATE Compute $w_{n k}^{(t)}= \phi'_{\rho}\left(\log \operatorname{NHP} \left(s_n \mid \boldsymbol{B}^{(t-1)}_k\right) / L(S_n) -\hat{\mu}_{\phi}(\boldsymbol{B}^{(t-1)}_k)\right)$.
		\STATE Update $\boldsymbol{B}_{k}^{(t)}$ according to Eq.\ref{eq:para:update}.
        \STATE Update the shift parameter according to Eq. \eqref{eq:update:shift}, if necessary.
        \STATE Increase $t$ by one.
		\UNTIL $\|\boldsymbol{B}_{k}^{(t)}-\boldsymbol{B}_{k}^{(t-1)}\|\leq\epsilon, ~\forall k\in [K]$.  
		\ENSURE $\hat{\mathbf{B}}$, $\{\hat r_{n k}\}$.
	\end{algorithmic}
\end{algorithm}

\subsection{Initialization}

A major weakness of EM-type algorithm is that it can only return local optimal solutions. With bad initialization, the algorithm may give the erroneous classification results which could be very different from the true underlying clusters. As we find in the numerical study, this issue becomes even worse under the temporal point process settings. 

\cite{arthur2007k} introduced the $K$-means++ algorithm, an extended $K$-means method, to alleviate local convergence issues. $K$-means++ has since gained popularity for its ability to produce high-quality initial centers, leading to faster convergence and better clustering performance.
Following the main ideas of $K$-means++, we propose a robust $K$-means++ initialization algorithm. 
It mainly consists of two steps, (i) outlier screening
and (ii) inlier weighting.

\textbf{Outlier screening}.
We first introduce several tuning parameters $M$, $ N'$, $\beta$, and $\alpha$. $M$ is an integer which is much smaller than $N$, $N'$ is the pre-determined number of inliers, $\beta$ is the  
screening speed ($\beta \in (0,\frac{N'}{N})$), and $\alpha \in (0,1)$ is the quality parameter.
Outlier screening iteratively repeats the following procedures until it finds $N'$ inliers. 

At round 0, we set $\mathcal S_{in}$ to be the empty set. For $n$-th event sequence, we calculate its corresponding distance set $\mathcal{D}_n^{(0)}$,
where 
$
    \mathcal{D}_n^{(0)} := \{d(S_n, S_{n,m}^{(0)})\}_{m=1}^M
$
with $S_{n,m}^{(0)}$ being a uniformly randomly selected sample from the whole dataset $\mathbf S$ and metric function $d$ being defined according to \eqref{eq:dis} (or \eqref{eq:dis:shift} when shift is considered).
We then compute the lower $\alpha$-quantile $q_{n,\alpha}^{(0)}$ of $\mathcal{D}_n^{(0)}$.
We rank $\{q_{n,\alpha}^{(0)}\}_{n=1}^{N}$ from the smallest to the largest and add the first $\lfloor \beta \cdot N \rfloor$ samples into $\mathcal S_{in}$.

At round $t \geq 1$, for event sequence $n$ not in $\mathcal S_{in}$, we similarly calculate its corresponding distance set $\mathcal D_n^{(t)}$,
where 
\begin{eqnarray}\label{eq:d:t}
    \mathcal{D}_n^{(t)} := \{d(S_n, S_{n,m}^{(t)})\}_{m=1}^M ~~ \text{with} ~~ S_{n,m}^{(t)}~ \text{being a uniformly randomly selected sample from } \mathcal S_{in}.  
\end{eqnarray}
We similarly compute its lower  $\alpha$-quantile $q_{n,\alpha}^{(t)}$ of $\mathcal{D}_n^{(t)}$. We then rank $\{q_{n,\alpha}^{(t)}\}_{n \notin \mathcal S_{in}}$ from the smallest to the largest and add the first $\min\{\lfloor \beta \cdot |\mathbf S \backslash \mathcal S_{in}| \rfloor, N' - |\mathcal S_{in}|\}$ samples into $\mathcal S_{in}$.
We repeat this procedure until $\mathcal S_{in}$ reaches $N'$.
(Here we let $M \ll N$ since the computation of distance sets could be time consuming.)

In summary, the above procedure recursively detects inliers. If an event sequence is closer to the center of inliers, then it is more likely to be detected in very early rounds. If an event sequence is far from other samples, then it is hard to be included in set $\mathcal S_{in}$. 

\textbf{Inlier weighting}. 
After obtaining $\mathcal S_{in}$, a set tentatively consisting of inliers only, we then perform $K$-means++ algorithm \citep{arthur2007k, georgogiannis2016robust, deshpande2020robust} onto it. The detailed steps are given as follows.

$(a)$ Select the first center $c_1$: Choose one event stream uniformly at random from $\mathcal S_{in}$. 

$(b)$ Select subsequent centers $c_k$'s: For the next center, randomly select the event stream with the probability proportional to the square of the distance from it to the nearest existing center. That is, $p(S_n) = \frac{D(S_n)^2}{\sum_{S \in \mathcal S_{in}} D(S)^2}$, where $D(S)=\min_{k' \in [k-1]} d(S,c_{k'})$. 
%This probability distribution ensures that the selected centroids are well-separated.

$(c)$ Repeat step $(b)$ until $K$ centers are chosen.

% K-means++ helps K-means converge faster and often with better results compared to random initialization. And the centroids selected by K-means++ are more spread out, reducing the likelihood of converging to suboptimal local minima. Unlike random initialization, K-means++ yields consistent results across multiple runs.

% Considering the Outlier, in Steps $(b)$ we may choose outlier with a high probability. So we should detect the outliers at the first step.

% \cite{georgogiannis2016robust} and \cite{deshpande2020robust} have designed some algorithms to improve the robustness of k-means algorithm and k-means++ algorithm. In order to improve the robustness, we must first identify and remove outliers. In other words, given a set $\mathcal{S} \subseteq \mathbb{R}^d$ of $n$ points, an integer $k>0$, and an outlier parameter $0<\alpha<1$, the objective of the robust $k$-means++ is to find a set $C \subseteq \mathbb{R}^d$ of $k$ centers that minimizes $\Upsilon_{\mathcal{C}}(\mathcal{S}_{in}):=\sum_{x \in \mathcal{S}_{in}} \min _{c \in \mathcal{C}}d(x,c)^2$, where $\mathcal{S}_{in}$ is the inlier set that we select before.

% $\forall n \in\{1,2,\cdots,N\}$, we randomly select a subset of samples of size $M(<N)$ and denote it as $\mathcal{S}_n:=\{s^n_m \}_{m=1}^M$. Refer to equation \eqref{eq:dis} and \eqref{eq:shi}, we calculate the distance between $s_n$ and every samples in $\mathcal{S}_n$, and we write the set of distances as 
% \begin{eqnarray}
%     \mathcal{D}_n:=\{d(s_n,s^n_m)\}_{m=1}^M.
% \end{eqnarray}

We denote $K$ selected centers by $\mathcal C^{ini} = \{c_k\}_{k=1}^K$.
To make the subsequent classification more robust, we also design the initial weight for sequence $S_n$ in $\mathcal S_{in}$ of being in class $k$ as 
\begin{eqnarray}\label{eq:wei}
    r_{n k}=\frac{\psi_{\alpha_k}(d(S_n,c_k))}{\sum_{n\in \mathcal{S}_{in}} \psi_{\alpha_k}(d(S_n,c_k))},
\end{eqnarray}
where $\alpha_k$ is the the median of the set $\{d(S_n,c_k)\}_{n\in \mathcal{S}_{in}}$ and $\psi_{\alpha}(x) = \psi(x/\alpha)$ with $\psi(x):= \phi'(x) \equiv x/(1+x+0.5\cdot x^2)$. 
The reason of doing this inlier weighting is to reduce the weights of a few outliers that may still remain in $\mathcal{S}_{in}$.
For $n \notin \mathcal S_{in}$, we let $r_{nk} \equiv 0$ for any $k \in [K]$.

In the case of data shift, we also return the initial shift parameter. For event stream $S_n$, we set
\begin{equation}\label{eq:shi:ini}
\operatorname{shift}_{n}=\underset{\operatorname{shift}\in\{\frac{T}{H_{\text{shift}}},\frac{2\cdot T}{H_{\text{shift}}},\cdots,T\}}{\operatorname{argmin}}\int_0^T \left|\hat{\lambda}_{S_n}(t+\operatorname{shift}) -\hat{\lambda}_{c_{k_n}}(t) \right|dt,
\end{equation}
where $c_{k_n}=\operatorname{argmin}_{c_k\in\mathcal{C}}d(S_n,c_{k})$.

The algorithm of initialization is summarized in Algorithm \ref{alg:kmean+}.

\begin{algorithm}[h]
	%\textsl{}\setstretch{1.8}
	\renewcommand{\algorithmicrequire}{\textbf{Input: Data $\mathbf S = \{S_n \}_{n=1}^N$ and tuning parameters $\alpha$, $\beta$, $M$}}
	\renewcommand{\algorithmicensure}{\textbf{Output:}}
	\caption{Robust Initialization}\label{alg:kmean+}
	\begin{algorithmic}[1]
		\STATE \textbf{Input:} {Data $\mathbf{S} = \{s_n \}_{n=1}^N$ and tuning parameters $\alpha$, $\beta$, $N'(<N)$}
            \STATE \textbf{Outlier Screening:} set $\mathcal S_{in} = \emptyset$.
            \REPEAT 
		\STATE For event stream $n$ not in $\mathcal{S}_{in}$, compute $\mathcal D_n^{(t)}$ and $q_{n,\alpha}^{(t)}$ according to \eqref{eq:d:t}. Rank the quantiles $q_{n,\alpha}$'s in the increasing order and add the first $\min\{\lfloor \beta \cdot |\mathbf S \backslash \mathcal S_{in}| \rfloor, N' - |\mathcal S_{in}|\}$ samples into $\mathcal S_{in}$.
            \UNTIL $\left|\mathcal{S}_{in}\right|\geq N'$.
		  \STATE \textbf{Inlier weighting:} follow steps (a)-(c) to get $K$ centers $\{c_1, ..., c_K\}$.
            \STATE Compute the weight matrix $\{r_{nk}\}$'s according to \eqref{eq:wei}.
            \STATE Compute the initial shift parameter $\text{shift}_n$ of $S_n$ according to \eqref{eq:shi:ini}, if necessary. 
		\ENSURE  Weight matrix $\{r_{n k}\}$, shift parameters $\{\operatorname{shift}_n\}$, inlier set $\mathcal{S}_{in}$; centers $\mathcal{C}^{ini}$.
	\end{algorithmic}  
\end{algorithm}

\section{Theoretical results}\label{sec:4}

Previously, we have not put any requirement on the observed event sequences yet. In this section, we theoretically show that our proposed algorithm works under mild conditions.
To start with, we introduce several technical assumptions.

\begin{assumption}\label{asm:1}
    Suppose the dataset has the following decomposition, $\mathbf{S} = \mathcal{S}_{inlier}\cup\mathcal{S}_{outlier} = \mathcal{S}_{1} \cup ... \cup \mathcal{S}_{K} \cup\mathcal{S}_{outlier}$. Here $\mathcal{S}_{outlier}$ is the set of outlier event sequences, $\mathcal{S}_{k}$ is the set of inlier event streams that belong to class $k$, and $\mathcal{S}_{inlier}$ is the union of all interior samples. $\mathcal S_1, ..., \mathcal S_K, \mathcal S_{outlier}$ are non-overlapping. Assume $ \max_{S_{n_1},S_{n_2} \in \mathcal{S}_{k}} d(S_{n_1},S_{n_2}) < \min\{\min_{S_{n_1} \in \mathcal{S}_{k}, S_{n_2} \in \mathcal S_{outlier}} d(S_{n_1},S_{n_2}), \min_{S_{n_1}, S_{n_2} \in \mathcal S_{outlier}} d(S_{n_1},S_{n_2})\}$ for any $k \in [K]$.
\end{assumption}

Here Assumption \ref{asm:1} requires that, for any $k \in [K]$, the upper bound of the distance between two different sequences in $\mathcal{S}_{k}$ is smaller than the distance between any two sequences in $\mathcal{S}_{inlier}$ and $\mathcal{S}_{outlier}$, and it is also smaller than the distance between any two outliers. 
With the help of Assumption \ref{asm:1}, it guarantees that outliers can be identified. 
In fact, this assumption can be relaxed. The requirement that $\min_{S_{n_1}, S_{n_2} \in \mathcal S_{outlier}} d(S_{n_1},S_{n_2})$ is larger than the maximum distance between inliers is not necessary. We can allow the distance between a small number of outliers to be close, which will not affect our results.

\begin{assumption}\label{asm:2}
    There is a lower bound $\pi_{low} > 0$ for the proportion of each inlier cluster, that is, $\pi_k \geq \pi_{low}$ for $k \in [K]$.
\end{assumption}

Assumption \ref{asm:2} ensures ``inlier" identifiability, i.e., every inlier cluster is not drained and inliers will not be treated as outliers. On the other hand, if some outliers, whose number is much less than $\pi_{low}\cdot N$, are close together, they will not be recognized as a new cluster. 

\begin{assumption}\label{asm:3}
    The space of model parameters $\boldsymbol{B}_{k}$'s defined in \eqref{eq:model} is bounded. That is, there exists $\Omega_B>0$ such that $\left\|\boldsymbol{B}_{k}\right\|_1<\Omega_B$ for all $k=1,2,\dots,K$.
\end{assumption}

Assumption \ref{asm:3} is a standard technical condition \citep{lehmann2006theory, casella2021statistical} that parameters are in the compact and bounded space.

\begin{assumption}\label{asm:4}
    There exist $\tau$ and $\Omega$ such that $0 < \tau \leq \lambda_k^{\ast}(t) \leq \Omega$ for all $t \in [0,T]$ and $k=1,2,\dots,K$.
\end{assumption}

Assumption \ref{asm:4} is also a classical technical requirement \citep{cai2022latent, fang2023group} to ensure that the intensity function is bounded away from zero and from above.

We next define the true working model parameter,
\begin{eqnarray}\label{eq:true:work:para}
    \boldsymbol{B}_k^{\ast} = \arg\max_{[b_{k,h}]} \bigg\{ \int_0^T (\log \lambda_k(t)) \cdot \lambda_k^{\ast}(t) dt - \int_0^T \lambda_k(t) dt \bigg\} ~~ \forall k \in [K]
\end{eqnarray}
with $\lambda_k(t)$ being defined in \eqref{eq:model}. We write $\lambda_{\boldsymbol{B}_k^{\ast}}(t) = \sum_{h=1}^{H} b_{k,h}^{\ast} \kappa_h(t)$. Then $\lambda_{\boldsymbol{B}_k^{\ast}}(t)$ is the intensity function closest to $\lambda_k^{\ast}(t)$ within the working model space.

\begin{assumption}\label{asm:5}
    For any two different classes $k$ and $k'$, there exists a constant $C_{gap} > 0$ such that, if event stream $S$ belongs to Class $k$, then it holds $\mathbb{E}[\log \operatorname{NHP}(S | \boldsymbol{B}_{k'}^*)]<\mathbb{E} [\log \operatorname{NHP} (S|\boldsymbol{B}_k^*)] - C_{gap} \cdot L$, $\forall ~ k' \neq k$.
\end{assumption}

Assumption \ref{asm:5} ensures ``class" identifiability that $\boldsymbol{B}_k^{\ast} \neq \boldsymbol{B}_{k'}^{\ast}$ when $k \neq k'$. In other words, event streams from different classes can be distinguished by our working model, the non-homogeneous Poisson process.
Here we assume that all the event streams have the same number of periods $L$ for simplicity. When the number periods are different, Assumption \ref{asm:5} still holds if $L$ is replaced by $\min_n L(S_n)$.

Next we show that our initialization algorithm can return a set of high-quality centers. To see this, we need to introduce the following quantities.
Define $\Upsilon_{\mathcal{C}}(\mathcal{S}_{in}):=\sum_{S\in\mathcal{S}_{in}} \min_{c\in \mathcal{C}}d\left(S,c\right)^2$. We also define $\mathcal{C}_{\mathrm{OPT}}$ is the set that minimizes $\Upsilon_{\mathcal C}(\mathcal{S}_{in})$ over all possible $\mathcal C$. Therefore, $\Upsilon_{\mathcal{C}_{\mathrm{OPT}}}(\mathcal{S}_{in}) = \min_{\mathcal C} \Upsilon_{\mathcal{C}}(\mathcal{S}_{in})$. $\Upsilon_{\mathcal{C}}(\mathcal{S}_{in})$ evaluates the quality of $\mathcal C$, i.e., smaller $\Upsilon_{\mathcal{C}}(\mathcal{S}_{in})$ is, better $\mathcal C$ is.

\begin{theorem}\label{thm:2}
Apply Algorithm \ref{alg:kmean+} and get $\mathcal C^{ini}$. It holds that $E[\Upsilon_{\mathcal C^{ini}} (\mathcal{S}_{in})|\mathcal{S}_{in}] \leq 16 (\ln K+2) \Upsilon_{\mathcal{C}_{\mathrm{OPT}}}(\mathcal{S}_{in})$, where $K$ is the number of clusters.
\end{theorem}

The above theorem indicates that, given the screening set $\mathcal S_{in}$, the set $\mathcal C^{ini}$ is nearly optimal up to a multiplicative constant in the average sense.
Furthermore, when $L$ becomes large, Theorem \ref{thm:2} implies that the algorithm can well identify centers from $K$ different classes. See the following theorem.

\begin{theorem}\label{thm:initia}
    Let $\mathcal{C}_{lack}$ be any set such that it consists of $K$ event streams, but at least two of them are from the same true underlying class. When $L\rightarrow\infty$, we have $\Upsilon_{\mathcal{C}_{lack}}(\mathcal{S}_{in})>16 (\ln K +2) \Upsilon_{\mathcal{C}_\mathrm{OPT}}(\mathcal{S}_{in})$ with high probability under Assumptions \ref{asm:1}, \ref{asm:2} and \ref{asm:5}.
\end{theorem}

Then we illustrate that the gradient descent step in Algorithm \ref{alg1} leads to the local convergence property with high probability when $L$ is large enough. For $k \in [K]$, we define function $\mu(\boldsymbol{B}_k\mid \boldsymbol{B}_k^*)$ which satisfies 
$$
\begin{aligned}
    \mathbb{E}_S \left[w_k\left(S ; \boldsymbol{B}_k^*\right)\phi_\rho\left(\log \operatorname{NHP}(S\mid \boldsymbol{B}_k ))/L-\mu(\boldsymbol{B}_k\mid \boldsymbol{B}_k^*)\right)\right]=0,
\end{aligned}
$$
where $w_k(S;\boldsymbol{B}):={\pi_k \operatorname{NHP}(S\mid \boldsymbol{B}_k)}/{\sum_j\pi_j \operatorname{NHP}(S\mid \boldsymbol{B}_k)}$.

\begin{theorem}\label{thm:local}
Suppose Assumption \ref{asm:3}, \ref{asm:4}, \ref{asm:5}, and $\eta := |\mathcal S_{outlier}| / N < \left(4 \cdot \sup_x|\phi(x)|\right)^{-1}$ hold. There exists a constant $a>0$ such that $C_{gap} - 2a-3 \bar m_c\log\left((\tau+a/T)/\tau\right)>0$; if $\left\|\boldsymbol{B}_k^t-\boldsymbol{B}_k^*\right\|<a$ for $k \in [K]$ and  learning rate $\text{lr} = 2/(\lambda_{\max }+\lambda_{\min })$, then update \eqref{eq:para:update} satisfies
\begin{eqnarray}
    \left\|\boldsymbol{B}_k^{(t+1)}-\boldsymbol{B}_k^*\right\| \leq \frac{\lambda_{\max }-\lambda_{\min }+2 \gamma}{\lambda_{\max }+\lambda_{\min }}\left\|\boldsymbol{B}_k^{(t)}-\boldsymbol{B}_k^*\right\|+\epsilon^{unif},
\end{eqnarray}
where $\lambda_{\max}$ and $\lambda_{min}$ are the largest and smallest eigenvalue of $-\Delta \mu(\boldsymbol{B}_k\mid \boldsymbol{B}_k^*)$ ~(the second derivative matrix of $- \mu(\boldsymbol{B}_k\mid \boldsymbol{B}_k^*)$), 
$\bar m_c := \sup_k \int_0^T \lambda_k^{\ast}(t)dt$,
$\gamma$ is a parameter satisfying $\gamma \leq\frac{\lambda_{\min }}{4}$ for sufficiently large $L$, and $\epsilon^{unif} = O_p(L\exp(-GL)/\sqrt{N}+(\rho+1)(1/\sqrt{NL}+\rho/L + \log N/(\rho N) + \eta/\rho))$.
\end{theorem}

Theorem \ref{thm:local} implies that $\|\boldsymbol{B}_k^{(t)}-\boldsymbol{B}_k^*\|$ decreases geometrically until it has the same order of $\epsilon^{unif}$.
Moreover, the consequence of Theorem \ref{thm:initia} and Theorem \ref{thm:local} is that $\boldsymbol{B}^{(0)}$ obtained in Algorithm \ref{alg1} will eventually satisfy $\left\|\boldsymbol{B}_k^{(0)}-\boldsymbol{B}_k^*\right\| < a$ as $L \rightarrow \infty$. Hence our robust clustering algorithm enjoys linear convergence speed. 
Note that we require the proportion of outlier samples is no greater than $ 100 \cdot \frac{1}{4 \cdot \sup_x|\phi(x)|}$ \%, which indicates that our proposed method can have a higher break-down point when we use the influence function with a smaller upper bound. (According to the definition of Catoni's-type influence function, the highest possible break-down point is no larger than 36\% \cite{bhatt2022minimax}. )

\begin{corollary}\label{cor:error:bound}
    Under the same conditions specified in Theorem \ref{thm:local}, we choose $\rho = \sqrt{L\cdot(\log N/N + \eta)}$. Then it holds 
    $
    \|\hat{\boldsymbol{B}}_k - \boldsymbol{B}_k^{\ast}\| = O_p\left(\sqrt{\log N/(NL) + \eta/L} + \epsilon\right)$,
    where $\epsilon$ is the tolerance parameter in Algorithm \ref{alg1}. 
\end{corollary}
As we can see, the estimation error consists of two parts, $\sqrt{\log N / (NL)}$ and $\sqrt{\eta / L}$. The former one corresponds to the stochastic variability caused by the inlier event streams and the latter one is the price we need to pay when there exist $100 \cdot \eta$ percent outlier event streams.
Note that in the robust statistical literature \citep{lugosi2021robust,bhatt2022minimax}, the minimax $M$-estimator enjoys the rate of $1/\sqrt{\text{sample size}} + \sqrt{\text{proportion of outliers}}$. Hence our proposed estimator is (nearly) statistically optimal. 

In addition to the convergence of working model parameter, we also show that Algorithm \ref{alg1} can identify almost all outliers under certain additional assumptions. 
We say an outlier event stream $S$ is indistinguishable by the working NHP model if 
$\int_0^T (\lambda_o(t)-\lambda_k^{\ast}(t))\log\lambda_{\boldsymbol{B}_k^{\ast}}(t) dt=0$ for some $k \in [K]$, where $S$ is generated according to intensity $\lambda_o(t)$.
We then define $\mathcal{S}_{indis} := \{S \in \mathcal S_{outlier} | S ~ \text{is indistinguishable}\}$ to be the set of indistinguishable event streams.
On the other hand, the outliers detected by our proposed method can be constructed as $\hat{\mathcal{S}}_{outlier} := \{S_n | \phi'_{\rho}\left(\log \operatorname{NHP} \left(S_n \mid \hat{\boldsymbol{B}}_k\right) / L(S_n) -\hat{\mu}_{\phi}(\hat{\boldsymbol{B}}_k)\right)<\epsilon_{bound}; \forall k \in [K]\}$, where we can set $\epsilon_{bound}=0.1$.
In other words, an event stream is treated as the outlier if its adjusted weight for any class is less than the cutoff 0.1.

\begin{theorem}\label{thm:robust}
    Under Assumptions \ref{asm:1} - \ref{asm:5}, 
    % and assume that the working model space contains the true underlying $\lambda_k^{\ast}(t)$'s. 
    it holds
$\mathbb{P}\left(\hat{\mathcal{S}}_{outlier}=\mathcal{S}_{outlier}\backslash \mathcal S_{indis} \right)\rightarrow 1 ~~ \text{as}~~ L \rightarrow \infty$,
    if we choose $\rho = L^{\beta}$ (with $0 < \beta < \frac{1}{2}$).
\end{theorem}

Note that set $\mathcal S_{indis}$ is of measure zero if $\lambda_o(t)$ is uniformly randomly selected from a continuous function space. Therefore, generically speaking, all outliers can be identified out as suggested by Theorem \ref{thm:robust}. 

% \begin{remark}
%     In Theorem \ref{thm:robust}, we can relax the assumption in the following way. The distance between $\lambda_k^{\ast}(t)$ and the working model space is not larger than $L^{- \beta_0}$ with $\beta_0 > \beta$.
% \end{remark}

%Lastly, when the true underlying intensities $\lambda_k^{\ast}(t)$'s can be exactly represented by the linear combinations of $\kappa_h(t)$, we then have the following error bound of estimator $\hat{\boldsymbol{B}}_k$.

%As both $L$ and $N$ become large, $\hat{\boldsymbol{B}}_k$ hence converges to the true model parameter.

% In summary, according to Theorem \ref{thm:robust} and lemma \ref{lem:log-fun}, we can conclude that under appropriate robust function constraints, as the number of cycles tends to infinity, we will reduce the weight of samples of other classes of this parameters through the robust function.

% \begin{remark}
%     The value log-likelihood function of $s_o$ is close to $\hat{\mu}_{\phi}(\boldsymbol{B})$ if and only if $\int_0^T (\lambda_o(t)-\lambda_{\boldsymbol{B}}(t))\log\lambda_{\boldsymbol{B}}(t) dt=0$, where $\lambda_o(t)$ is the intensity function of $s_o$ and $\lambda_{\boldsymbol{B}}(t)$ is the intensity function with parameter $\boldsymbol{B}$.
% \end{remark}

%\vspace{-5mm}

\section{Simulation Study}\label{sec:5}

%\vspace{-4mm}

To demonstrate the feasibility and the efficiency of our robust clustering method, we compare it with the other two baseline methods. One method is a standard EM algorithm with random initialization of $\boldsymbol{B}^{(0)}$, $\pi_k^{(0)}$'s and identity influence function, and the other one is almost the same to the proposed algorithm but with random initialization. 
%The random initialization here is to randomly divide the initial samples into $K$ classes, where $K$ is the number of clusters specified in advance. 

The simulation settings are described as follows. 
We first consider to generate inlier event sequences according to the following intensity functions with a total of $K = 4$ classes,
{\small
\begin{align*}
\lambda_1^{\ast}(t)&=5/3\exp(-{(t+4.8)^2}/{10})+5/3\exp(-{(t-2.4)^2}/{50}),\\
\lambda_2^{\ast}(t)&=5/3\exp(-{(t-6)^2}/{4})+15/4\exp(-{(t-21.6)^2}/{4}),\\
\lambda_3^{\ast}(t)&=15/4\exp(-{(t-4.8)^2}/{1.5})+35/12\exp(-{(t-12)^2}/{1})+15/4\exp(-{(t-19.2)^2}/{1.5}),\\
\lambda_4^{\ast}(t)&=10/3\exp(-{(t-21.6)^2}/{40})+5/3\exp(-{(t-26.4)^2}/{10}),
\end{align*}}
where $t\in [0,T]$ with $T = 24$ (corresponding to 24 hours). 
At the same time, we consider the three types of outlier event sequences according to the following intensity functions:
{\small
\begin{align*}
\lambda_{out1}(t)&=125/6\cdot(U+0.1)\text{ , where }U\sim U(0,1),\\
\lambda_{out2}(t)&=125/18\cdot(U+0.1)+ 125/3\cdot\exp(-{(t-24\cdot B_1)^2}/{0.5})\text{ , where }U\sim U(0,1)\text{ and } B_1 \sim U(0,1), \\
\lambda_{out3}(t)&=25/2\cdot\exp(-{(t-24\cdot B_1)^2}/{0.02})+25/3\cdot\exp(-{(t-24\cdot B_2)^2}/{0.02})\\
&+25/6\cdot\exp(-{(t-24\cdot B_3)^2}/{0.02}) \text{ , where }B_i\sim U(0,1) \quad \forall i\in\{1,2,3\}. \\
\end{align*}}

% {\color{red}{(I use the capital letters to stand for random variables.)}}

%\vspace{-8mm}

Based on the formula, we can find that 
outlier event sequence of the first type follows a homogeneous Poisson process,
the outlier intensity function of the second type has a unimodel shape, and 
the third one has three modes.
% the second type of outlier adds the displacement parameter $b_1$ to become a unimodal distribution. The third type of outlier is composed of three unimodal distributions superimposed. In order to prevent the distributions between outliers from converging, we adjust the window parameter to $0.02$. 
Based on the intensity value, we can observe that  
the number of events in the first two type outliers are generally larger than those of inliers, while the number of events in the third type outliers are slightly smaller than those of inliers. 
% Each outlier here contains at least one of the intensity random parameter $u$ and the position random parameter $b$, because we require that the outliers are not distributed in the same type.

For each setting, we generate 60 event sequences for each inlier class and 60 event sequences according to one of the three outlier intensities. In total, there are $N = 60 \times 4 + 60 = 300$ samples.
We let the number of periods $L \in \{1, 2, 4\}$.
In addition, we also consider to shift the $n$-th sample by $\text{shift}_n$ which is an integer uniformly sampled between 0 and 23. 
We apply our proposed method and two baselines by setting number of classes equal to 4, 5, or 6.
All the above settings are repeated for $100$ times.
In the experiment, we set tuning parameter $\rho$ for class $k$ to be $0.6\cdot\sqrt{\int_0^T\log^2 \lambda_k^{(0)}(t) \cdot \lambda_k^{(0)}(t)dt}$, $\epsilon = 0.1$, $\alpha=0.2$, $\beta=0.3$, $M=50$, and $N'=0.75 \cdot N$.
% For each different outlier, we mixed $240$ real samples and $60$ outlier samples together, and applied three methods to cluster them respectively. Regardless of outlier, the true number of categories here are $4$. We chose three cases: $K=4$, $5$, and $6$ for the number of clustering classes. In addition, we also conducted experiments under different numbers of cycles, including $L=1$, $2$, and $4$. Finally, for each set of experiments above, we also took into account shift as well, that is, randomly shifting each sample by $u$, where $u$ is a random integer between $0$ and $23$. 

We use the clustering purity \citep{schutze2008introduction} to evaluate the performances of three methods.
To be specific, the purity index is defined as 
\begin{eqnarray}\label{eq:purity}
\operatorname{purity}(\hat{\mathcal S}, \mathcal S^{\ast})=\frac{1}{N} \sum_k \max_{k'} \left|\hat{\mathcal S}_k \cap \mathcal S_{k'}^{\ast} \right|,
\end{eqnarray}
where $\hat{\mathcal S} = \{\hat{\mathcal S}_1, ..., \hat{\mathcal S}_{\hat K}\}$ and $\mathcal S^{\ast} = \{\mathcal S_1^{\ast}, ..., \mathcal S_{K^{\ast}}^{\ast}\}$ are two partitions of of the data set according to the estimated labels and true underlying labels.
It is easy to see that the range of purity value is between $0$ and $1$. The higher the purity value is, the better clustering result is. Moreover, the purity is non-decreasing as $\hat K$ increases. In other words, for a fixed algorithm, the purity will get larger if we wish to cluster the data into more classes.

% Its calculation can be thought of as follows: For each cluster, count the number of data points from the most common class in said cluster. Now take the sum over all clusters and divide by the total number of data points. Formally, given some set of clusters $\Omega=\left\{\omega_1, \omega_2, \ldots, \omega_K\right\}$ and some set of clusters $\mathbb{C}=\left\{c_1, c_2, \ldots, c_J\right\}$, both partitioning $N$ data points, purity can be defined as:

% $$
% \operatorname{purity}(\Omega, \mathbb{C})=\frac{1}{N} \sum_k \max _j\left|\omega_k \cap c_j\right|.
% $$

The results are summarized in Table \ref{tab:sim1} to Table \ref{tab:sim3}. As seen from the three tables, the proposed method uniformly outperforms the other two baselines by a big margin under all settings.
As $K$ varies from 4 to 6, the purity returned by the two baseline methods is always smaller than that of the proposed method. This suggests our method is truly robust even with mis-specified number of classes.  
As number of periods $L$ increases, the purity increases and converges to 1, which confirms our theoretical results.
When time shift is considered, the two baselines can only give very low purity values while the result given by our proposed method is still quite descent.
According to the construction of outliers, our method seems to be more effective when the outliers tend to consist of more events (i.e., outlier type 1 and type 2 have larger intensity values).

\begin{table}[]
    \centering
    \scalebox{0.8}{
    \begin{tabular}{c|c|ccc|ccc}
    \toprule  %添加表格头部粗线
    \multicolumn{1}{c|}{\multirow{2}*{Time}}&\multicolumn{1}{c|}{\multirow{2}*{Algorithm}}& \multicolumn{3}{|c|}{No shift}& \multicolumn{3}{|c}{shift}\\
    \cline{3-8}
    \multicolumn{1}{c|}{} &\multicolumn{1}{c|}{} & $K=4$ & $K=5$ & $K=6$ & $K=4$ & $K=5$ & $K=6$ \\
    \midrule  %添加表格中横线
    \multicolumn{1}{c|}{\multirow{3}*{$L=1$}}& Standard & 0.5056 & 0.6111 & 0.7480 & 0.3868 & 0.4506 & 0.5046\\
    \multicolumn{1}{c|}{} &Robust & 0.5225 & 0.6438 & 0.7590 & 0.4198 & 0.4896 & 0.5172\\
    \multicolumn{1}{c|}{} &Robust \& Initialization & \textbf{0.9026} & \textbf{0.9758} & \textbf{0.9797} & \textbf{0.6420} & \textbf{0.6678} & \textbf{0.6910}\\
    \midrule 
    \multicolumn{1}{c|}{\multirow{3}*{$L=2$}}& Standard & 0.4648 & 0.5495 & 0.6688 & 0.3857 & 0.4740 & 0.5351\\
    \multicolumn{1}{c|}{} &Robust & 0.4849 & 0.6090 & 0.7205 & 0.4046 & 0.5023 & 0.5739\\
    \multicolumn{1}{c|}{} &Robust \& Initialization & \textbf{0.9240} & \textbf{0.9916} & \textbf{0.9988} & \textbf{0.7313} & \textbf{0.7728} & \textbf{0.7910}\\
    \midrule  %添加表格中横线
    \multicolumn{1}{c|}{\multirow{3}*{$L=4$}}& Standard & 0.3950 & 0.4725 & 0.5650 & 0.3703 & 0.4581 & 0.5368\\
    \multicolumn{1}{c|}{} &Robust & 0.4051 & 0.5153 & 0.6550 & 0.3958 & 0.4900 & 0.5921\\
    \multicolumn{1}{c|}{} &Robust\& Initialization & \textbf{0.9150} & \textbf{0.9925} & \textbf{1 } & \textbf{0.7610 }& \textbf{0.8130} & \textbf{0.8147}\\
    \bottomrule %添加表格底部粗线
    \end{tabular}
    }
    \caption{{\small Purity indices returned by three algorithms under the setting of outlier type 1.}}
    \label{tab:sim1}
\end{table}

\begin{table}[]
    \centering
    \scalebox{0.8}{
    \begin{tabular}{c|c|ccc|ccc}
    
    \toprule  %添加表格头部粗线
    \multicolumn{1}{c|}{\multirow{2}*{Time}}&\multicolumn{1}{c|}{\multirow{2}*{Algorithm}}& \multicolumn{3}{|c|}{No shift}& \multicolumn{3}{|c}{shift}\\
    \cline{3-8}
    \multicolumn{1}{c|}{} &\multicolumn{1}{c|}{} & $K=4$ & $K=5$ & $K=6$ & $K=4$ & $K=5$ & $K=6$ \\
    \midrule  %添加表格中横线
    \multicolumn{1}{c|}{\multirow{3}*{$L=1$}}& Standard & 0.3996 & 0.5283 & 0.6302 & 0.3132 & 0.3481 & 0.3856\\
    \multicolumn{1}{c|}{} &Robust & 0.5835 & 0.6859 & 0.8017 & 0.3901 & 0.4442 & 0.4636\\
    \multicolumn{1}{c|}{} &Robust \& Initialization &\textbf{ 0.9520} & \textbf{0.9796} & \textbf{0.9791} & \textbf{0.6544} & \textbf{0.6838} & \textbf{0.6939}\\
    \midrule  %添加表格中横线
    \multicolumn{1}{c|}{\multirow{3}*{$L=2$}}& Standard & 0.4239 & 0.5246 & 0.6019 & 0.3057 & 0.3573 & 0.4180\\
    \multicolumn{1}{c|}{} &Robust & 0.5445 & 0.6440 & 0.7115 & 0.3784 & 0.4548 & 0.5095\\
    \multicolumn{1}{c|}{} &Robust \& Initialization   & \textbf{0.9264} & \textbf{0.9838} & \textbf{0.9988} & \textbf{0.7431} & \textbf{0.7681} & \textbf{0.8141}\\
    \midrule  %添加表格中横线
    \multicolumn{1}{c|}{\multirow{3}*{$L=4$}}& Standard & 0.4025 & 0.4950 & 0.5798 & 0.3197 & 0.3784 & 0.4169\\
    \multicolumn{1}{c|}{} &Robust & 0.4975 & 0.5725 & 0.6625 & 0.4235 & 0.4963 & 0.5374\\
    \multicolumn{1}{c|}{} &Robust \& Initialization   & \textbf{0.9225} & \textbf{0.9850} & \textbf{1}      & \textbf{0.7969} & \textbf{0.8026} & \textbf{0.8233}\\
    \bottomrule %添加表格底部粗线
    \end{tabular}
    }
    \caption{{\small Purity indices returned by three algorithms under the setting of outlier type 2. }}
    \label{tab:sim2}
\end{table}

\begin{table}[]
    \centering
        \scalebox{0.8}{
    \begin{tabular}{c|c|ccc|ccc}
    \toprule  %添加表格头部粗线
    \multicolumn{1}{c|}{\multirow{2}*{Time}}&\multicolumn{1}{c|}{\multirow{2}*{Algorithm}}& \multicolumn{3}{|c|}{No shift}& \multicolumn{3}{|c}{shift}\\
    \cline{3-8}
    \multicolumn{1}{c|}{} &\multicolumn{1}{c|}{} & $K=4$ & $K=5$ & $K=6$ & $K=4$ & $K=5$ & $K=6$ \\
    \midrule  %添加表格中横线
    \multicolumn{1}{c|}{\multirow{3}*{$L=1$}}& Standard & 0.8975 & 0.9733 & 0.9764 & 0.4520 & 0.4978 & 0.5152\\
    \multicolumn{1}{c|}{} &Robust & 0.8623 & 0.9613 & 0.9774 & 0.4560 & 0.5006 & 0.5198\\
    \multicolumn{1}{c|}{} &Robust \& Initialization & \textbf{0.9161} &\textbf{ 0.9753} & \textbf{0.9783} & \textbf{0.6420} & \textbf{0.6418} & \textbf{0.6853}\\
    \midrule  %添加表格中横线
    \multicolumn{1}{c|}{\multirow{3}*{$L=2$}}& Standard & 0.9069 & 0.9882 & 0.9907 & 0.4810 & 0.5169 & 0.5467\\
    \multicolumn{1}{c|}{} &Robust & 0.8811 & 0.9656 & 0.9887 & 0.4874 & 0.5240 & 0.5568\\
    \multicolumn{1}{c|}{} &Robust \& Initialization   & \textbf{0.9592} & \textbf{0.9928} & \textbf{0.9984} & \textbf{0.6366} & \textbf{0.7167} & \textbf{0.7695}\\
    \midrule  %添加表格中横线
    \multicolumn{1}{c|}{\multirow{3}*{$L=4$}}& Standard & 0.8873 & 0.9624 & 0.9875 & 0.5042 & 0.5348 & 0.5611\\
    \multicolumn{1}{c|}{} &Robust & 0.8750 & 0.9525 & 0.9900 & 0.5151 & 0.5450 & 0.5818\\
    \multicolumn{1}{c|}{} &Robust \& Initialization   & \textbf{0.9574} & \textbf{0.9900} & \textbf{1}      & \textbf{0.6735} & \textbf{0.7356} & \textbf{0.8083}\\
    \bottomrule %添加表格底部粗线
    \end{tabular}
    }
    \caption{{\small Purity indices returned by three algorithms under the setting of outlier type 3.}}
    \label{tab:sim3}
\end{table}

To end this section, we explain the reason why we do not include another baseline, the EM algorithm with proposed initialization but without robust influence function, in our simulation.
Such baseline method may have obvious defects.
Consider a case that the inlier event streams are from homogeneous Poisson process of four classes, whose intensities are 1, 2, 3, and 4, respectively. There are 30 event sequences for each class and one outlier event sequence which follows a Poisson process with intensity 100.  In this case, even if we start from the true values, it still leads to bad classification result if $\phi_{\rho}$ is not used. To see this, after the first iteration, the outlier will be classified into class 4 and the intensity parameter of this class will be updated to approximately $(30 \times 4+100)/31 \approx 7.10$. After the second iteration, event streams from class 3 and 4 will be mixed together and the intensity parameter of four classes will be approximately $1$, $2$, $3.5$, and $100$, respectively. Then the algorithm converges in the next iteration. Therefore, outlier is classified into a single class and purity index is no larger than 0.75. This indicates the usefulness of $\phi_{\rho}$.

\section{Real Data Application}\label{sec:6}

\noindent \textbf{IPTV dataset}~
The IPTV log-data set \citep{6717182} used in our study are collected
from a large-scale Internet Protocol television (IPTV) provider, China Telecom, in Shanghai,
China. As a privacy protection, anonymous data is used in this
study. The log-data records viewing behaviors of users, which
is composed of anonymous user logs, time stamps (which are
at the precision of one second) of the beginnings and the
endings of viewing sessions.
%, and names of TV programs. 
The log-data is family-based and each family has only one user
ID. For the family with more than one Television, all viewing
behaviors are also recorded under the same user account. 
% We sampled a subset of IPTV data over several months, giving
% rise to a data set with nearly 0.3 million subscribers who
% watched over 3,000 TV programs during 90 days. For training an inference
% model, we
The data collector randomly selected $302$ users from the data set and
collected their household structures and their watching history from 2012 January 1st to 2012 November 30th through phone surveys with the help of China Telecom. 
 On average, each household has $10-15$ events per day.

We do some preprocessing on the IPTV data. By exploratory analysis, we can see a strong evidence that households' watching behavior is periodic with period equal to 24 hours (i.e. $T = 24$).
For each household, we construct an event sequence with number of periods $L = 7$ based on the raw data as follows.
Let period $l \in \{1,2,...7\}$ corresponds to Monday, Tuesday, ..., Sunday. 
Note that our working model is the non-homogeneous Poisson process which enjoys the independent increment property.
Thus superposition of sub event sequences in different periods will not affect the estimation results. 
We then superpose data from 5 randomly selected days (Mondays, ..., Sundays) into each period.
Those households with insufficient data are excluded. In the end, we construct $N = 297$ clean event sequences with $T = 24$ and $L = 7$. The choices of tuning parameters are specified the same as those in the simulation studies.

% For each sample, we first divide it into seven parts by day of the week, from Monday to Sunday. For each part, we treat it as a cycle. But obviously the number of days contained in different parts is different. We need to select part of the data in days for experiment. Since the number of event per day is relatively small, each period combines $5$ days of data. Since the length of each sample is different, we randomly selected $5$ days as the experimental subjects. Those with a length of less than $5$ days were discarded. In the end, we retained 297 samples, each sample is a data containing 7 periods from Monday to Sunday.

Since we do not know the true underlying class labels for each household, the purity index cannot be computed.
Instead, we use two other criteria to compare the performance between the proposed algorithm and baseline methods.
% Instead, we use the following two indicators to compare the advantages and disadvantages of each method. When sample $i$ is clustered into the class $k(i)$, The first one is the $L_1$ error between the sample $i$ and the cluster $k(i)$ is calculated as below.
For the first one, we define
\begin{equation}\label{criteria:1}
    \text{L1}_n = \int_{0}^T \left| \hat{\lambda}_n(t)-\hat{\lambda}_{k(n)}^*(t) \right|dt/\sqrt{\int_{0}^T \hat{\lambda}_{k(n)}^*(t)dt},
\end{equation}
where $k(n)$ is the estimated label of sample $n$, the $\hat{\lambda}_n(t)$ is the estimated intensity function of sample $i$ via cubic spline approximation, and $\hat{\lambda}_{k(n)}^*(t)$ is the estimated intensity function of class $k(n)$. In \eqref{criteria:1}, the normalizer 
$\sqrt{\int_{0}^T \hat{\lambda}_{k(n)}^*(t)dt}$ is the estimated standard deviation of the number events for class $k(n)$. 
This helps to eliminate the influence of intensity magnitudes of different classes. 
Then the L1 error criteria is given by 
\begin{equation}\label{criteria:1:all}
    \text{L1-error} = \frac{1}{N_{in}} \sum_{n \notin \hat{\mathcal S}_{outlier}} \text{L1}_n,
\end{equation}
where $\hat{\mathcal S}_{outlier}$ is the index set of outlier returned by the proposed method (i.e. the sample with weights $w_{nk}$'s smaller than 0.1 is treated as the outlier) and $N_{in} = N - |\hat{\mathcal S}_{outlier}|$.

For the second one, we define the MLE index of $n$-th event stream as
$\text{MLE}_n(alg) := \log \operatorname{NHP}(S_n | \boldsymbol{B}_{k(n)^{alg}}^{alg})$,   
where the superscript ``$alg$" indicates one of the three algorithms. We can compute the MLE comparison ratio as
\begin{eqnarray}\label{criteria:2:out}
    \text{MLE}_{out}(alg_1, alg_2) = \frac{1}{N_{in}} \sum_{n \notin \hat{\mathcal S}_{outlier}} \mathbf 1\{ \text{MLE}_n(alg_1) > \text{MLE}_n(alg_2)\}
\end{eqnarray}
and 
\begin{eqnarray}\label{criteria:2:all}
    \text{MLE}_{all}(alg_1, alg_2) = \frac{1}{N} \sum_{n \in [N]} \mathbf 1\{ \text{MLE}_n(alg_1) > \text{MLE}_n(alg_2)\}.
\end{eqnarray}
If the index $\text{MLE}_{out}(alg_1, alg_2)$ (or $\text{MLE}_{all}(alg_1, alg_2)$) is larger than 0.5, then it indicates that``$alg_1$" performs better than ``$alg_2$".

From Table \ref{tab:real-1-1}, we can see that the proposed algorithm achieves the smallest L1-error among all the three algorithms under any choice of $K \in \{3,...,8\}$. This suggests the clusters returned by our method are more compact. From Table \ref{tab:real-1-2}, we also see that the MLE comparison ratios of the proposed method against others are uniformly greater than 0.5. This indicates that the inclusion of influence function $\phi$ and $K$-means++ type initialization indeed makes an improvement on majority of the samples. 

% The second indicator is the MLE's comparison between two methods. We know that one method work well if and only if the MLE functions of most samples under the result parameters obtained by this method are large. So we calculated and compared the ratio of the MLE function under one method to the MLE function under another method in inlier.

% After the final clustering is completed, for each cluster, we select the sample in which the next derivative weight of the robust function is less than $0.1$ as outliers.

% \begin{table}[]
%     \centering
% \begin{tabular}{c|ccc}

% \toprule  %添加表格头部粗线
% L1-error & Ours & Robust & Standard \\
% \midrule  %添加表格中横线
% K=3 & \textbf{2.678} & 2.682 & 2.989\\
% K=4 & \textbf{2.505} & 2.633 & 2.689\\
% K=5 & \textbf{2.439} & 2.506 & 2.570\\
% K=6 & \textbf{2.389} & 2.462 & 2.557\\
% K=7 & \textbf{2.386} & 2.415 & 2.503\\
% K=8 & \textbf{2.232} & 2.364 & 2.454\\
% \bottomrule %添加表格底部粗线
% \end{tabular}
%     \caption{Criterion 1 -- L1-error indices given by all three methods for IPTV data.}
%     \label{tab:real-1-1}
% \end{table}

\begin{table}[]
    \centering
    \begin{tabular}{c|cccccc}
\toprule  %添加表格头部粗线
L1-error & $K=3$ & $K=4$ & $K=5$ & $K=6$ & $K=7$ & $K=8$ \\
\midrule  %添加表格中横线
Ours & \textbf{2.678} & \textbf{2.505}  & \textbf{2.439} & \textbf{2.389} & \textbf{2.386} & \textbf{2.232} \\
Robsut & 2.682 & 2.633 & 2.506 & 2.462 & 2.415 & 2.364 \\
Standard & 2.989 & 2.689 & 2.570 & 2.557 & 2.503 & 2.454\\
\bottomrule %添加表格底部粗线
\end{tabular}
    \caption{{\small Criterion 1 -- L1-error indices given by all three methods for IPTV data.}}
    \label{tab:real-1-1}
\end{table}

\begin{table}[]
    \centering
\scalebox{0.8}{
\begin{tabular}{c|c|cc|cc|cc}
\toprule  %添加表格头部粗线
\multicolumn{1}{c|}{\multirow{2}*{MLE comparison ratio}}&\multicolumn{1}{c|}{\multirow{2}*{Outliers}}& \multicolumn{2}{|c|}{Ours vs. Standard}& \multicolumn{2}{|c|}{Ours vs. Robust}&\multicolumn{2}{|c}{Robust vs. Standard}\\
\cline{3-8}
\multicolumn{1}{c|}{} &\multicolumn{1}{c|}{} & Out & All & Out & All & Out & All \\
\midrule  %添加表格中横线
$K=3$ & 40 & 67.70 & 61.61 & 55.25 & 56.57 & 66.15 & 60.61\\
$K=4$ & 40 & 66.54 & 59.25 & 55.25 & 51.85 & 60.70 & 56.90\\
$K=5$ & 38 & 64.09 & 58.25 & 57.92 & 55.56 & 62.93 & 58.26\\
$K=6$ & 34 & 64.63 & 59.60 & 58.17 & 55.89 & 61.60 & 56.90\\
$K=7$ & 25 & 64.71 & 60.27 & 53.31 & 53.20 & 66.54 & 61.95\\
$K=8$ & 29 & 67.16 & 61.61 & 56.34 & 55.55 & 61.94 & 57.91\\
\bottomrule %添加表格底部粗线
\end{tabular}
}
    \caption{{\small Criterion 2 -- MLE comparison ratios given by all three methods for IPTV data.}}
    \label{tab:real-1-2}
\end{table}

% \begin{figure}[htbp]
% \centering
% \includegraphics[width=0.4\textwidth]{images/outlier_new1.png}
% \caption{IPtV data and outlier }
% \end{figure}

\noindent \textbf{Last.FM 1K User Dataset}~
Last.fm 1K is a public data set released by lastfm \citep{oscar_celma_2010_6090214}. It collects all listening history records (about 20 million records) of 992 users of different countries from July 2005 to May 2009. The data contains two tables. The record table includes information such as userID, event timestamp, artistID, artist\_name, songID, and song\_name, while the user feature table includes information such as gender, age, country, registration time, etc. On average, each user has about 40 events per day.

% This data set does not directly provide the characteristics corresponding to the song, but it does The artist\_name and song\_name are both real records.
% Although Last.fm 1K does not explicitly provide some user interaction behaviors, it does provide information about the timestamp of the listening event. 
% Many papers delineate sessions based on the length of the playback timestamp interval between two adjacent songs, and Based on the playback rate (the length of time between the playback timestamps of two adjacent songs/the duration of the song), it is divided into whether the user has skipped a certain song, so it can be used to do some tasks based on session or streaming recommendation.

Similar to IPTV data, we also do the preprocessing on the Last.fm data. 
From Figure \ref{fig:data2}, we again see the evidence that users' song track playing frequency is periodic with $T = 24$ hours. The size of raw data is huge so that we down-sample the data and construct the event sequence for each user with $L = 10$.
That is, we extract event streams from 10 randomly selected days for each user. 
After discarding those users with insufficient data, we have 966 users left.
In other words, we construct $N=966$ clean event sequences with $T = 24$ and $L = 10$.
Since users may come from different countries, we consider the time shift in this data set.
Again, the choice of $\rho$ and $\epsilon$ is the same as before.

From Table \ref{tab:real-2-1} and Table \ref{tab:real-2-2}, we can also see that the proposed algorithm performs the best among all the three methods in terms of both L1-error and MLE comparison ratio. 
This confirms the generality of the proposed method. Both influence function and initialization prodecure contribute to the performance improvement.

% For each sample, we use one day as the data within a period. Since the amount of data per day is sufficient, we do not need to merge it as before to increase the amount of data in a period. Instead, we just sample $10$ days of data as the experimental
% subjects. For sample data less than $10$ days, we directly select all the data. In the end, we retained $966$ samples, where more than 900 samples have the number of periods equal to 10. 

% \begin{table}[]
%     \centering
%     \begin{tabular}{c|ccc}
% \toprule  %添加表格头部粗线
% L1-error & Ours & Robust & Normal \\
% \midrule  %添加表格中横线
% K=3 & \textbf{2.246} & 2.482 & 2.585\\
% K=4 & \textbf{2.146} & 2.373 & 2.409\\
% K=5 & \textbf{2.123} & 2.308 & 2.338\\
% K=6 & \textbf{1.984} & 2.127 & 2.180\\
% K=7 & \textbf{1.934} & 2.119 & 2.158\\
% K=8 & \textbf{1.925} & 2.103 & 2.175\\
% \bottomrule %添加表格底部粗线
% \end{tabular}
%     \caption{Criterion 1 -- L1-error indices given by all three methods for Last.FM data}
%     \label{tab:real-2-1}
% \end{table}

\begin{table}[]
    \centering
    \begin{tabular}{c|cccccc}
\toprule  %添加表格头部粗线
L1-error & $K=3$ & $K=4$ & $K=5$ & $K=6$ & $K=7$ & $K=8$ \\
\midrule  %添加表格中横线
Ours & \textbf{2.246} & \textbf{2.146}  & \textbf{2.123} & \textbf{1.984} & \textbf{1.934} & \textbf{1.925} \\
Robsut & 2.482 & 2.373 & 2.308 & 2.127 & 2.119 & 2.103 \\
Standard & 2.585 & 2.409 & 2.338 & 2.180 & 2.158 & 2.175\\
\bottomrule %添加表格底部粗线
\end{tabular}
    \caption{{\small Criterion 1 -- L1-error indices given by all three methods for Last.FM data}}
    \label{tab:real-2-1}
\end{table}

\begin{table}[]
    \centering
    \scalebox{0.8}{
\begin{tabular}{c|c|cc|cc|cc}
\toprule  %添加表格头部粗线
\multicolumn{1}{c|}{\multirow{2}*{MLE comparison ratio}}&\multicolumn{1}{c|}{\multirow{2}*{Outliers}}& \multicolumn{2}{|c|}{Ours vs. Standard}& \multicolumn{2}{|c|}{Ours vs. Robust}&\multicolumn{2}{|c}{Robust vs. Standard}\\
\cline{3-8}
\multicolumn{1}{c|}{} &\multicolumn{1}{c|}{} & Out & All & Out & All & Out & All \\
\midrule  %添加表格中横线
$K=3$ & 43 & 64.57 & 62.11 & 58.94 & 59.32 & 59.26 & 56.83\\
$K=4$ & 41 & 69.41 & 66.56 & 66.81 & 64.29 & 54.59 & 52.28\\
$K=5$ & 49 & 65.10 & 62.63 & 62.70 & 62.11 & 57.14 & 54.55\\
$K=6$ & 36 & 62.15 & 60.25 & 57.74 & 56.63 & 53.23 & 51.55\\
$K=7$ & 41 & 61.19 & 59.42 & 59.57 & 57.97 & 55.78 & 53.73\\
$K=8$ & 44 & 61.17 & 59.32 & 54.34 & 52.59 & 57.38 & 56.11\\
\bottomrule %添加表格底部粗线
\end{tabular}
}
    \caption{{\small Criterion 2 -- MLE comparison ratios given by all three methods for Last.FM data.}}
    \label{tab:real-2-2}
\end{table}

% \begin{figure}[htbp]
% \centering
% \includegraphics[width=0.4\textwidth]{images/outlier_new2.png}
% \caption{Last.FM 1K User Dataset and outlier}
% \end{figure}

\section{Conclusion}\label{sec:7}

%\vspace{-4mm}

In the current literature, there is no work studying the clustering of event stream data under the outlier setting. In this work, we make an effort to solve this task and propose a robust TPP clustering framework. 
Our algorithm can be viewed as a non-parametric method which builds on the cubic spline regression.
There are two key ingredients in the new algorithm. One is the construction of a TPP-specific distance function which can be efficiently implemented. 
The other is the incorporation of Catoni's influence function which allows us to have the robust parameter training.
Under mild assumptions, the proposed method is shown to have decent performances.
Theories on convergence property, (non) asymptotic error bound, and outlier detection have been established.
Three different types of outliers are considered in the simulations and the results validate the effectiveness of the proposed method.
Two real data applications are provided. Our algorithm achieves the superior performance over other two baseline methods.

Lastly, we discuss a few potential extensions in the future work. 
(i) In the current work, we introduce a new distance function based on cubic spline regression. 
It is possible to design other types of metric which can also be computed efficiently.
(ii) In the "fine-tuning" step, we construct the pseudo likelihood function based on NHP models. 
NHP can be replaced by other types of TPP models, e.g., self-exciting processes, self-correcting processes, etc.
(iii) The current definition of outliers is individual/user-level. However, in practice, it could happen that a user behaves normally for almost all time but except for a very short period. Therefore, it may be improper to treat the whole event sequence as the outlier. Instead, we should consider the problem on the event-level.      
(iv) Although the proposed method empirically works well under any choice of $K$, it is still desired to design a guideline of choosing the best number of clusters for practitioners.

\clearpage

\bibliography{robust}

\clearpage

{
\begin{center}
\Large
    \textbf{Supplementary of ``On Robust Clustering of Temporal Point Processes"}
\end{center}
}

\section{Supporting Information of Catoni's Influence Function}

\begin{figure}[htbp]
\centering
\includegraphics[width=0.48\textwidth]{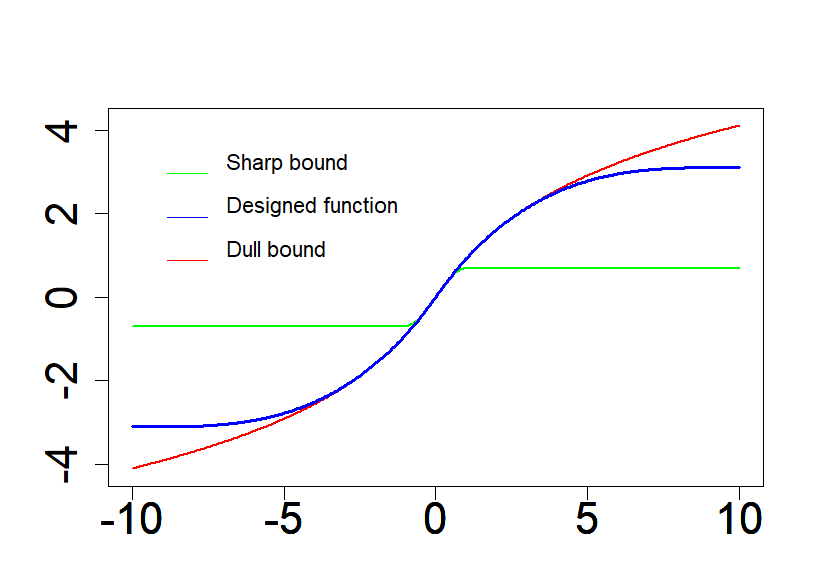}
\includegraphics[width=0.48\textwidth]{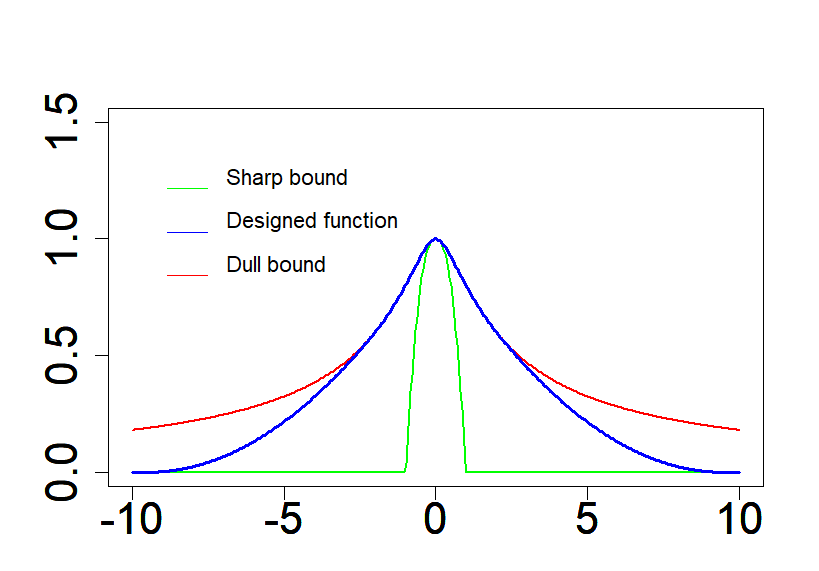}
\caption{Left figure: Catoni influence function $\phi$ and the widest influence function $\phi_{dull}$ and the narrowest influence function $\phi_{sharp}$. Right figure: First-order derivatives of $\phi$, $\phi_{dull}$ and $\phi_{sharp}$. For the definitions of $\phi_{dull}$ and $\phi_{sharp}$, please refer to \eqref{eq:example1} and \eqref{eq:example2}.}
\label{fig:cat}
\end{figure}

We provide the graphical illustrations of Catoni's influence function $\phi(x)$ and its derivative $\phi'(x)$ in Figure \ref{fig:cat}.  
% \begin{figure}[htbp]
% \centering
% \begin{minipage}[t]{0.48\textwidth}
% \centering
% \includegraphics[width=1.35\textwidth]{images/robust0.png}
% \caption{Robust function}
% \end{minipage}
% \begin{minipage}[t]{0.48\textwidth}
% \centering
% \includegraphics[width=1.35\textwidth]{images/robust1.png}
% \caption{First-order derivative function}
% \end{minipage}
% \end{figure}

The first-order derivative and second-order derivative of the function can be derived as
$$ \phi'(x)=\left\{
\begin{aligned}
&\frac{1+x}{1+x+0.5 x^2} & & x\leq 2; \\
& 0.032/3\cdot (x-9.5)^2 & & 2<x\leq 9.5; \\
& 0 & & x\geq 9.5
\end{aligned}
\right.
$$
and
$$ \phi''(x)=\left\{
\begin{aligned}
&-\frac{x+ 0.5 x^2}{(1+x+0.5 x^2)^2} & & x\leq 2; \\
& 0.064/3\cdot (x-9.5) & & 2<x\leq 9.5; \\
& 0 & & x\geq 9.5.
\end{aligned}
\right.
$$

The formula of $\phi_{dull}$ and $\phi_{sharp}$ plotted in Figure \ref{fig:cat} are given as follows.
\begin{eqnarray}\label{eq:example1}
	\phi_{dull}(x) = 
	\begin{cases}
	\log(1 + x + \frac{1}{2}|x|^2) &~ x \geq 0\\
	-\log(1 - x + \frac{1}{2}|x|^2) &~ x < 0,\\
	\end{cases}
\end{eqnarray}
and
\begin{eqnarray}\label{eq:example2}
	\phi_{sharp}(x) = 
	\begin{cases}
	- \log 2 & \text{if}~ x \leq - 1 \\
	-\log(1 - x + \frac{1}{2}|x|^2) &  \text{if} ~ - 1 \leq x \leq 0, \\
	\log(1 + x + \frac{1}{2}|x|^2) &  \text{if} ~ 0 < x \leq 1, \\
	\log 2 & \text{if}~ x \geq 1.
	\end{cases}
\end{eqnarray}

\section{Construction of Spline Basis}

Let $U=(u_0, u_1, \ldots, u_H)$ be a set of $H + 1$ non-decreasing numbers satisfying $0 = u_0 < u_1 \cdots < u_H = T$. (We may treat $T=1$ for the ease of presentation). Points $u_i$'s are called knots and the set $U$ is known as the knot vector, and the half-open interval $[u_i, u_{i+1})$ the $i$-th knot span. For practical use, the knots are usually equally spaced, i.e., $u_{i+1} - u_i$ is a constant equal to $\Delta u:=T/H$ for $0 \leq i \leq H - 1$. 
% The knots can be considered as division points that subdivide the interval $[u_0, u_m]$ into knot spans. All B-spline basis functions are supposed to have their domain on $[u_0, u_m]$. In this note, we use $u_0 = 0$ and $u_m = 1$ frequently so that the domain is the closed interval $[0,1]$.
To construct the cubic spline basis functions, we follow the classical procedure by defining $N_{i,p}(u)$ as the $i$-th B-spline basis function of degree $p$. Then its formula can be recursively written as
$$
\begin{aligned}
& N_{i, 0}(u)= \begin{cases}1 & \text { if } u_i \leq u<u_{i+1} \\
0 & \text { otherwise }\end{cases},  \\
& N_{i, p}(u)=\frac{u-u_i}{u_{i+p}-u_i} N_{i, p-1}(u)+\frac{u_{i+p+1}-u}{u_{i+p+1}-u_{i+1}} N_{i+1, p-1}(u).
\end{aligned}
$$
The above is usually referred to as the Cox-deBoor recursion formula \citep{de1972calculating}. 
% This definition looks complicated; but, it is not difficult to understand. If the degree is zero ($p = 0$), these basis functions are all step functions and this is what the first expression says. That is, basis function $N_{i,0}(u)$ is $1$ if $u$ is in the $i$-th knot span $[u_i, u_{i+1})$. For example, if we have four knots $u_0 = 0, u_1 = h, u_2 = 2h$ and $u_3 = 3h$, knot spans $0, 1$ and $2$ are $[0,1), [1,2), [2,3)$ and the basis functions of degree $0$ are $N_{0,0}(u) = 1$ on $[0,h)$ and $0$ elsewhere.
Applying the Cox-deBoor recursion formula, the first cubic spline basis function $\kappa_1(\cdot)$ can be found as follows.
{\small
$$
\begin{aligned}
& \kappa_1(u)=\begin{cases}\displaystyle\frac{1}{6\Delta u^3}u^3, & u\in[0,\Delta u], \\
\displaystyle\frac{1}{6\Delta u^3}\left((2\Delta u-u)u^2+(u-\Delta u)(4\Delta u-u)(3\Delta u-u)+(4\Delta u-u)(u-\Delta u)^2\right), & u\in[\Delta u,2\Delta u], \\
\displaystyle\frac{1}{6\Delta u^3}\left((u-4\Delta u)^2(u-2\Delta u)+(u-\Delta u)(4\Delta u-u)(3\Delta u-u)+(u-3\Delta u)^2u\right), & u\in[2\Delta u,3\Delta u], \\
\displaystyle\frac{1}{6\Delta u^3}(4\Delta u-u)^3, & u\in[3\Delta u,4\Delta u].\end{cases}
\end{aligned}
$$
}
For $h \in \{2, ..., H\}$, we can define $h$-th basis $\kappa_h(u) := \kappa_1(u - h \Delta u)$.
(When $u < h \Delta u$, $\kappa_h(u) = \kappa_1(u - h \Delta u + T)$.)
% which is called the cubic spline basic function. When we set $h=T/H$, define $\kappa_i(\cdot)$ as the basic function from $u_{i-1}$ to $u_{i+3}$, $\forall i \in\{1,2,\cdots,H\}$ as our basic function.

\section{Literature on Intensity-based Distance}

For the ease of discussion, throughout this section, we suppose all events are observed within time interval $[0,T]$, where $T$ is a fixed real number.
Most existing distances for TPPs are based on the random time change theorem \citep{brown2002time}. That is, an event stream $S = (t_1, \ldots, t_N)$ is distributed according to a TPP with intensity $\lambda^*(t)$ on the time interval $[0, T]$ if and only if the transformed sequence $Z := (v_1, \ldots, v_N) =(\Lambda^*\left(t_1\right), \ldots, \Lambda^*\left(t_N\right))$ is distributed according to a standard Poisson process on $[0, \Lambda^*(T)]$, where $\Lambda^*(t) := \int_0^t \lambda^*(u) d u$ is the cumulative intensity function.

% Intuitively, it can be viewed as a TPP analogue of how the CDF of an arbitrary random variable over $\mathbb{R}$ transforms its realizations into samples from Uniform $([0,1])$. Similarly, the compensator $\Lambda^*$ converts a random event sequence $X$ into a realization $Z$ of the standard Poisson process (SPP). In other words, we can define a statistic for a TPP with compensator $\Lambda^*$ by applying the compensator to $X$ to obtain $Z$.

% For brevity, we denote the transformed arrival times as $Z=\left(v_1, \ldots, v_N\right)=\left(\Lambda^*\left(t_1\right), \ldots, \Lambda^*\left(t_N\right)\right)$ and the length of the transformed interval as $V=\Lambda^*(T)$. One way to describe the generative process of an SPP is as follows
% $$
% N\left|V \sim \operatorname{Poisson}(V) \quad u_i\right| N, V \sim \operatorname{Uniform}([0, V]) \quad \text { for } i=1, \ldots, N .
% $$
% An SPP realization $Z=\left(v_1, \ldots, v_N\right)$ is obtained by sorting the $u_i$ 's in increasing order. This is equivalent to defining the arrival time $v_i$ as the $i$-th order statistic $u_{(i)}$. We can also represent $Z$ by the inter-event times $\left(w_1, \ldots, w_{N+1}\right)$ where $w_i=v_i-v_{i-1}$, assuming $v_0=0$ and $v_{N+1}=V$.

\cite{barnard1953time} proposed a Kolmogorov-Smirnov (KS) statistic-based metric, which quantifies the distance between observed event stream $S$ and the theoretical intensity $\lambda^{\ast}(t)$. The idea is to check whether the transformed arrival times $v_1, \ldots, v_N$ are uniformly distributed within interval $[0, T]$. To do so, it compares $\hat{F}_{\text {arr }}$, the empirical cumulative distribution function (CDF) of the arrival times, with $F_{\text {arr }}(u)=u / \Lambda^*(T)$, the CDF of the uniform random variable. 
% {\color{red}{where $V=...$ ($V$ is not defined?)}}. 
Specifically, the distance is defined as
$$
\kappa_{\text {arr }}(S, \lambda^{\ast}(\cdot)) := \sqrt{N} \cdot \sup _{u \in[0, V]}\left|\hat{F}_{\text {arr }}(u)-F_{\text {arr }}(u)\right|,$$
where $\hat{F}_{\text {arr }}(u)=\frac{1}{N} \sum_{i=1}^N \mathbf{1}\left(v_i \leq u\right)$.

Another possible metric relies on the fact that the inter-event time $w_i := v_{i+1} - v_i$ follows the standard exponential distribution (\cite{cox1966statistical}). It then compares $\hat{F}_{\text {int }}$, the empirical CDF of $w_i$'s, and $F_{\text {int }}(u) := 1-\exp (-u)$. This leads to 
$$
\kappa_{\text {int }}(S, \lambda^{\ast}(\cdot)) :=\sqrt{N} \cdot \sup _{u \in[0, \infty)}\left|\hat{F}_{\text {int }}(u)-F_{\text {int }}(u)\right|,$$ 
where $\hat{F}_{\text {int }}(u)=\frac{1}{N+1} \sum_{i=1}^{N+1} \mathbf{1}\left(w_i \leq u\right)$.

Although metrics $\kappa_{\text {arr}}$ and $\kappa_{\text {int}}$ are popular in testing the goodness-of-fit of various Poisson processes \cite{daley2003introduction, gerhard2011applying, alizadeh2013scalable, kim2014call, li2018learning, tao2018common}, they still have many limitations. 
They suffer severe non-identifiability issues. Two very different event streams can be very close under such metrics.  
% (1) They suffer severe non-identifiability issues.
% Consider the following two different intensity functions 
% \begin{eqnarray}
%  \lambda_1^{\ast}(t) = 
%  \begin{cases}
%      1 &  0 \leq t < \frac{T}{2} ;\\
%      2 &  \frac{T}{2} \leq t \leq T ,
%  \end{cases}
% ~ \text{and} ~ 
% \lambda_2^{\ast}(t) = 
%  \begin{cases}
%      2 &  0 \leq t < \frac{T}{2} ;\\
%      1 &  \frac{T}{2} \leq t \leq T.
%  \end{cases}
% \end{eqnarray}
% Unfortunately, $\kappa_{\text {arr}}$ and $\kappa_{\text {int}}$ cannot differentiate between time sequences generated from 
% $\lambda_1^{\ast}(t)$ and $\lambda_2^{\ast}(t)$.
% (2) They are insensitive {(\color{red}confusing)} to the event count-removing all events $\frac{V}{2}<v_i \leq V$ from an SPP realization $Z$ will only result in just a single atypically large inter-event time $w_i$, which changes the value of $\kappa_{\text {int }}(Z)$ at most by $\frac{1}{N+1}$.
% \begin{eqnarray}
%  \lambda_1^{\ast}(t) = 
%  \begin{cases}
%      1 &  0 \leq t < \frac{T}{2} ;\\
%      0 &  \frac{T}{2} \leq t \leq T ,
%  \end{cases}
% ~ \text{and} ~ 
% \lambda_2^{\ast}(t) = 1 &  0 \leq t < {T} ;
% \end{eqnarray}
More failure modes of $\kappa_{\text {arr}}$ and $\kappa_{\text {int}}$ can be found in \cite{pillow2009time}.
% For example, KS arrival only checks if the arrival times $v_i$ are distributed uniformly, conditioned on the event count $N$. But what if the observed $N$ is itself extremely unlikely under the SPP? KS inter-event can be similarly insensitive to the event count-removing all events $\frac{V}{2}<v_i \leq V$ from an SPP realization $Z$ will only result in just a single atypically large inter-event time $w_i$, which changes the value of $\kappa_{\text {int }}(Z)$ at most by $\frac{1}{N+1}$. Other failure modes can be found in \cite{pillow2009time}. 

Taking into account the above problems, \cite{shchur2021detecting} proposed a sum-of-squared-spacings metric, 
$$
\kappa_{sss}(S, \lambda^{\ast}(\cdot)) := \frac{1}{\Lambda^*(T)} \sum_{i=1}^{N+1} w_i^2=\frac{1}{\Lambda^*(T)} \sum_{i=1}^{N+1}\left(v_i-v_{i-1}\right)^2,
$$
which extends the idea in \cite{greenwood1946statistical}. 
%sum-of-squared-spacings statistic proposed as a test of uniformity for fixed length samples by
As we can see, the above method can measure the closeness between the sample and the specific distribution well. However, they fail to meet the data requirements in our scenarios. To be more specific, we can only observe the sample data and has no information of model specification, which means that $\lambda^{\ast}(\cdot)$ or $\Lambda^*(\cdot)$ is unknown. For any two samples $S_1$ and $S_2$, of course, we can consider to estimate $\Lambda_1^*(\cdot)$ ($\Lambda_2^*(\cdot)$) based on sample $S_1$ ($S_2$) first, and then calculate the above KS-type distance between sample $S_2$ ($S_1$) and the estimated $\Lambda_1^*(\cdot)$ ($\Lambda_2^*(\cdot)$). Unfortunately, this procedure makes it not symmetric about $S_1$ and $S_2$ and also fails to satisfy the triangle inequality. As a result, it is not a proper metric distance.

\section{Additional Figures in Numerical Studies}

\begin{figure}[htbp]
\centering
\includegraphics[width=0.48\textwidth]{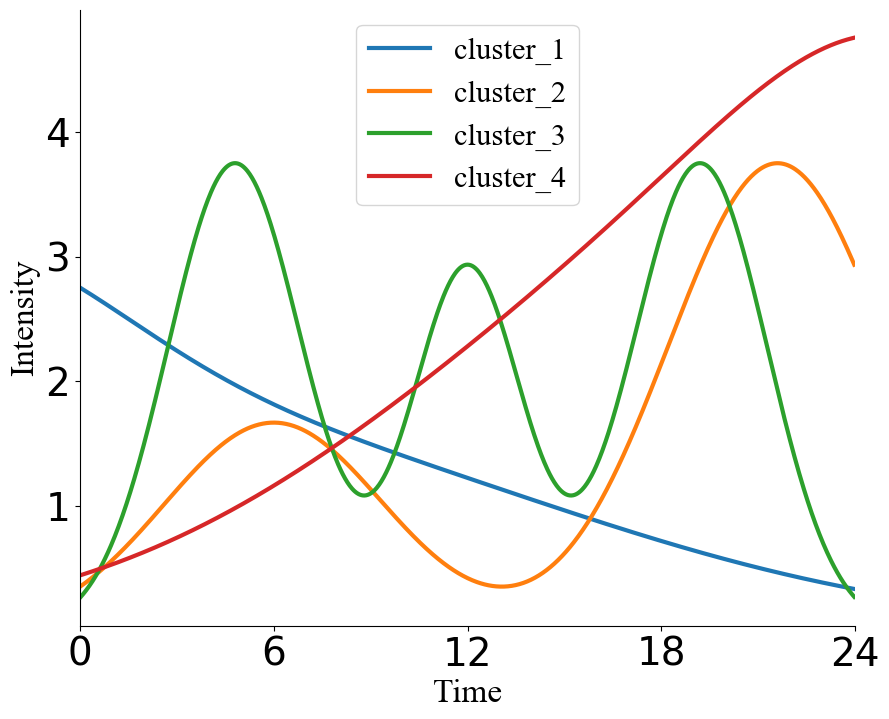}
\includegraphics[width=0.48\textwidth]{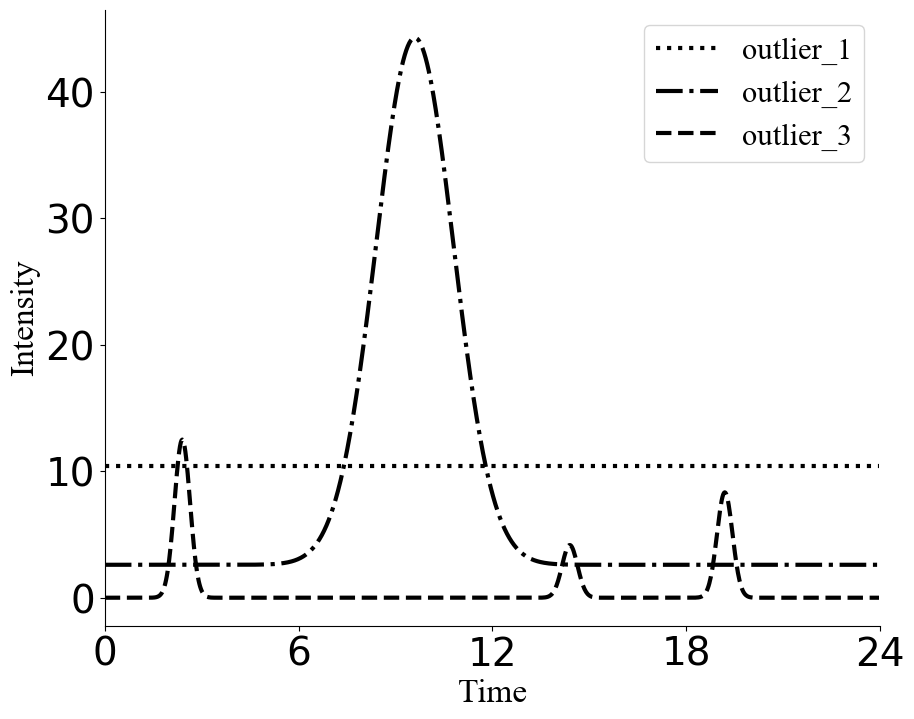}
\caption{Left: Intensity functions of inlier event streams from 4 classes. Right: Intensity functions of outlier event streams of three types. Due to the randomness of $\lambda_{out1}$ - $\lambda_{out3}$, curves are shown with one random realization of $u$.}\label{fig:intensity}
\end{figure}

To help readers to gain more intuitions, the curves of intensity function considered in simulation studies are shown in Figure \ref{fig:intensity}.

\begin{figure}[htbp]
\centering
\includegraphics[width=0.7\textwidth]{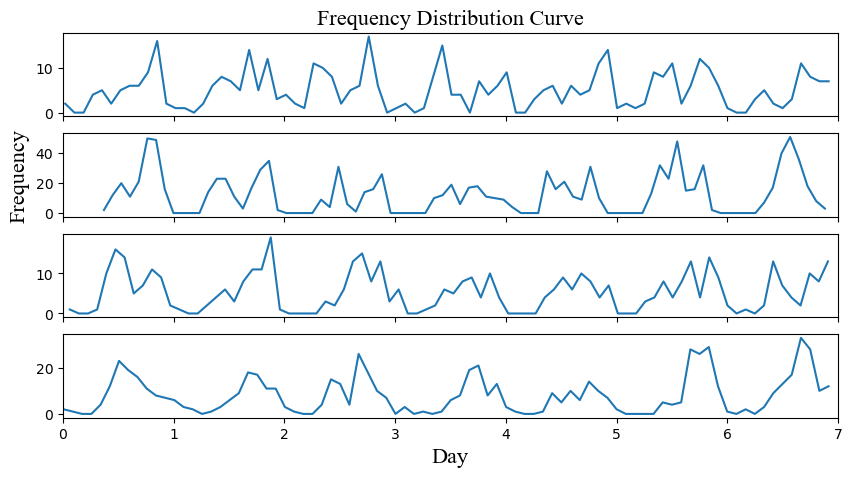}
\caption{IPTV data: the frequency plot of four randomly selected households.}\label{fig:data1}
\end{figure}

\begin{figure}[htbp]
\centering
\includegraphics[width=0.7\textwidth]{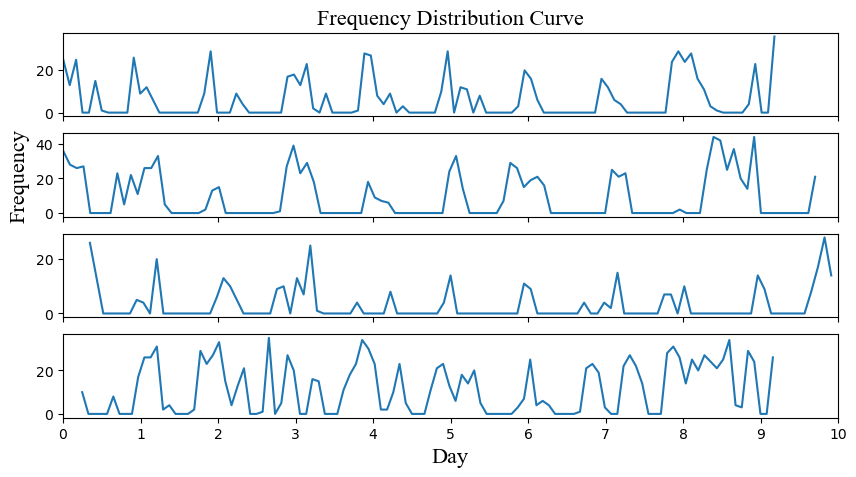}
\caption{Last.FM 1K User Dataset: the frequency plot of four randomly selected users.}\label{fig:data2}
\end{figure}

The frequency plots of two real data sets are given in Figure \ref{fig:data1} and Figure \ref{fig:data2}. It empirically indicates the existence of daily effect in user behaviors, i.e., the period of event sequences can be viewed as 24 hours.

\clearpage

\section{Proof of Propositions}

\noindent \textbf{Proof of Proposition \ref{pro:poi}}
First, we consider the case where $f$ is a constant value function, such as $f$ being always equal to $1$. If $X$ follows a Poisson distribution with parameter $\lambda$, we prove that the variance of $\sqrt{X}$ is approximately $1/4+O(1/\lambda)$. In general, for a smooth $g(X)$, we can do a Taylor expansion around the mean $\lambda=\mathbb{E}(X)$, so we have
$$
g(X)=g(\lambda)+g^{\prime}(\lambda)(X-\lambda)+\frac{g^{\prime \prime}(\lambda)}{2 !}(X-\lambda)^2+\frac{g^{\prime \prime \prime}(\lambda)}{3 !}(X-\lambda)^3+\cdots.
$$
Therefore,
$$
\mathbb{E}[g(X)]=g(\lambda)+\frac{g^{\prime \prime}(\lambda)}{2 !} m_2+\frac{g^{\prime \prime \prime}(\lambda)}{3 !} m_3+\cdots,
$$
where $m_i$ is the $i$-th centered moment. In our case $m_2=m_3=\lambda$, thus
$$
\mathbb{E}[\sqrt{X}]=\sqrt{\lambda}-\frac{\lambda^{-1 / 2}}{8}+\frac{\lambda^{-3 / 2}}{16}+\cdots,
$$
which indicates that the expected value is approximately $\sqrt{\lambda}$. 
Taking square of it, it gives
$$
\left(\mathbb{E}[\sqrt{X}]\right)^2 \approx \lambda-\frac{1}{4}+\frac{9}{64 \lambda}+\ldots.
$$
Then
$$
\operatorname{Var}(\sqrt{X}) \approx \frac{1}{4}-\frac{9}{64 \lambda}+\ldots,
$$
which is approximately $1 / 4$ for large $\lambda$.

Next, we divide the interval $[0,T]$ into $n$ segments, each of which is $0=a_0<a_1<\cdots<a_{n-1}<a_{n}=T$. Write $X_i:=\frac{1}{\sqrt{N(T)}}\sum_{t_j\in(a_{i-1},a_i)}f(t_j)$, then $\operatorname{var}(X_i)\approx \frac{\int_{a_{i-1}}^{a_i}f^2(t)dt}{T}\cdot( \frac{1}{4}-\frac{9}{64 \lambda}+\ldots)$. So the variance of $\frac{1}{\sqrt{N(T)}}\sum_{t_j}f(t_j)$ is $\sum_i \operatorname{var}(X_i)=\frac{\int_0^T f^2(t)dt}{T}\cdot( \frac{1}{4}-\frac{9}{64 \lambda}+\ldots)$.
This completes the proof.

\medskip

\noindent \textbf{Proof of Proposition \ref{pro:mstep}}:
By the definition of $\hat{\mu}_{\phi}^{(t)}(\boldsymbol{B}_k)$, we know that
$$
\begin{aligned}
    \frac{\partial }{\partial \boldsymbol{B}_k} & \left\{\sum_{n=1}^N  r_{n k}^{(t)}  \cdot L(S_n) \cdot  \phi_{\rho}\left(\log \operatorname{NHP}\left(S_n \mid \boldsymbol{B}_k\right) /  L(S_n) - \hat{\mu}_{\phi}^{(t)}(\boldsymbol{B}_k)\right)\right\}=0,
\end{aligned}
$$
which implies
$$
\begin{aligned}
   & \frac{\partial\hat{\mu}_{\phi}^{(t)}(\boldsymbol{B}_k)}{\partial\boldsymbol{B}_k} \\
    & =\sum_{n=1}^N\frac{r_{n k}^{(t)} \phi'_{\rho}\left(\log \operatorname{NHP}\left(S_n \mid \boldsymbol{B}_k\right) / L(S_n) -\hat{\mu}_{\phi}(\boldsymbol{B}^{(t-1)}_k)\right)}{\sum_{n=1}^N r_{n k}^{(t)}  \phi'_{\rho}\left(\log \operatorname{NHP}\left(S_n \mid \boldsymbol{B}_k\right) / L(S_n)-\hat{\mu}_{\phi}(\boldsymbol{B}^{(t-1)}_k)\right) L(S_n) }\cdot\frac{\partial\log \operatorname{NHP}\left(S_n \mid \boldsymbol{B}_k\right)}{\partial\boldsymbol{B}_k}.
\end{aligned}$$
Plugging $B_k = B_k^{(t-1)}$ into the above formula, we get $=\varrho_k^{(t)}$. This completes the proof.

\section{Proof of Theorem \ref{thm:2} and Theorem \ref{thm:initia}}

We first provide a lemma showing that the ``outlier screening" procedure can eliminate all outliers with high probability.

\begin{lemma}\label{lem1}
    % Under Assumptions 1 and 2, for each sequence in $S$, we randomly select $M$ other sequences to calculate the distance to the sequence. Then calculate its $\alpha$-quantile, select the smallest $\beta$ part and add it to the inlier set. Repeat it.
    Under Assumption \ref{asm:1} and \ref{asm:2}, steps 3-5 in Algorithm \ref{alg:kmean+} eliminate all outliers with high probability.
\end{lemma}
\textbf{Proof of Lemma \ref{lem1}} Without loss of generality, we consider Cluster $1$. Assume that Cluster $1$ accounts for $\alpha_1$ proportion of the set $\mathcal{S}$. Select $M$ samples from an $N$-element set. It is easy to know that the Cluster $1$ part and others obey the binomial distribution $ B(M,\alpha_1)$. Then the probability of $\alpha$-quantile being smaller than $r_{max}$ is $p:=\sum_k \mathbb{P}(X\in\mathcal{C}_{k})\cdot\mathbb{P}(X_{dis}\geq M\cdot\alpha)=\sum_k\left(\alpha_k\sum_{i\geq \alpha\cdot M} \tbinom{M}{i}\alpha_k^i\cdot (1-\alpha_k)^{M-i}\right)$. We choose a suitable $\alpha$ such that $p=\sum_k \alpha_k \cdot(1-\delta_1)$, where $\delta_1$ is a small enough positive number. Then choose $\beta$ such that $\sum_{i\geq\beta\cdot N'}\tbinom{N'}{i}p^i(1-p)^{N'-i} >1-\delta_2$. Repeat it until we choose enough samples, and at the same time, we avoid selecting outliers with a high probability. This completes the proof.

\bigskip

Next we show that the proposed ``inlier weighting" procedure can produce a set of good initial centers.
In the following proof, we consider an arbitrary pseudo-metric $d$ which has quasi-triangular properties, that is, $ d(x,z)\leq M(d(x,y)+d(y,z))$ for all $x,y,z \in \mathcal{S}$. 
For our proposed distance function, it holds $M \equiv 1$.

\textbf{Overview of Proof of Theorem \ref{thm:2}}. In order to find the upper bound of the $\Upsilon$, we use mathematical induction to prove that the upper bound of the objective function $\Upsilon$ can be controlled after adding several centers. Lemma \ref{lem:phi} proves the case of one-step addition and Lemma \ref{lem:4} generalizes to the general case.
As defined previously, we know that under the optimal center set $\mathcal{C}_{\mathrm{OPT}}$, each sequence will be classified into the same class of an element in $\mathcal{C}_{\mathrm{OPT}}$, so we can divide $\mathcal{S}_{in}$ into $K$ sub-sets. Let $A$ be an arbitrary sub-set.

\begin{lemma}
Let $S$ be a set of sequences with center $c(S)$, and let $z$ be an arbitrary sequence. Then
$\sum_{x\in S}d(x,z)^2-M\sum_{x\in S}d(x,c(S))^2\leq 2M^2|S|\cdot d(c(S),z)^2.$
\end{lemma}

\begin{lemma}\label{lem:phi}
     Let $\mathcal{C}$ be an arbitrary set of centers. Define $\Upsilon(A):=\sum_{a\in A} \min_{c\in \mathcal{C}}d\left(a,c\right)^2$, $\Upsilon_{\mathrm{OPT}}(A):=\sum_{a\in A} \min_{c\in \mathcal{C}_{\mathrm{OPT}}}d\left(a,c\right)^2$. If we add a random center to $\mathcal{C}$ from $A$, chosen with $D^2$ weighting, then $\mathbb{E}[\Upsilon(A)] \leq 16M^4 \Upsilon_{\mathrm{OPT}}(A)$.
\end{lemma}

\noindent \textbf{Proof of Lemma \ref{lem:phi}} The probability that we choose some fixed $a_0$ as our center is precisely $\frac{D\left(a_0\right)^2}{\sum_{a \in A} D(a)^2}$. Furthermore, after choosing the center $a_0$, a sequence $a$ will contribute precisely $\min \left(D(a),d(a,a_0)\right)^2$ to the potential. Therefore,
$$
\mathbb{E}[\Upsilon(A)]=\sum_{a_0 \in A} \frac{D\left(a_0\right)^2}{\sum_{a \in A} D(a)^2} \sum_{a \in A} \min \left(D(a),d(a,a_0)\right)^2 .
$$
Note by the triangle inequality that $D\left(a_0\right) \leq$ $M(D(a)+d(a,a_0))$ for all $a, a_0$. From this, the powermean inequality implies that $D\left(a_0\right)^2 \leq 2M^2(D(a)^2+d(a,a_0)^2)$. Summing over all $a$, we then have that $D\left(a_0\right)^2 \leq \frac{2M^2}{|A|} \sum_{a \in A} D(a)^2+\frac{2M^2}{|A|} \sum_{a \in A}d(a,a_0)^2$. Then $\mathbb{E}[\Upsilon(A)]$ is at most
$$
\begin{array}{r}
\frac{2M^2}{|A|} \cdot \sum_{a_0 \in A} \frac{\sum_{a \in A} D(a)^2}{\sum_{a \in A} D(a)^2} \cdot \sum_{a \in A} \min \left(D(a),d(a,a_0)\right)^2 \\
+\frac{2M^2}{|A|} \cdot \sum_{a_0 \in A} \frac{\sum_{a \in A}d(a,a_0)^2}{\sum_{a \in A} D(a)^2} \cdot \sum_{a \in A} \min \left(D(a),d(a,a_0)\right)^2 .
\end{array}
$$
In the first expression, we substitute $\min \left(D(a),d(a,a_0)\right)^2 \leq d(a,a_0)^2$, and in the second expression, we substitute $\min \left(D(a),d(a,a_0)\right)^2 \leq D(a)^2$. Simplifying, we then have,
$$
\begin{aligned}
\mathbb{E}[\Upsilon(A)] & \leq \frac{4M^2}{|A|} \cdot \sum_{a_0 \in A} \sum_{a \in A}d(a,a_0)^2  =16M^4 \Upsilon_{\mathrm{OPT}}(A) .
\end{aligned}
$$
This completes the proof.

\begin{lemma}\label{lem:4}
  Let $\mathcal{C}$ be the current center set, and write $\Upsilon:=\Upsilon(\mathcal{S})$. Choose $u>0$ "uncovered" class, and let $\mathcal{S}_u$ denote the set of sequences in these class. Also let $\mathcal{S}_c=\mathcal{S}-\mathcal{S}_u$. Now suppose we add $t \leq u$ random centers to $\mathcal{C}$, chosen with $D^2$ weighting. Let $\mathcal{C}^{\prime}$ denote the new center set, and let $\Upsilon^{\prime}:=\Upsilon^{\prime}(\mathcal{S})$ denote the corresponding potential. Then, $\mathbb{E}\left[\Upsilon^{\prime}\right]$ is at most,
$$
\left(\Upsilon\left(\mathcal{S}_c\right)+16 M^4\Upsilon_{\mathrm{OPT}}\left(\mathcal{S}_u\right)\right) \cdot\left(1+H_t\right)+\frac{u-t}{u} \cdot \Upsilon\left(\mathcal{S}_u\right)
$$
Here, $H_t$ denotes the harmonic sum, $1+\frac{1}{2}+\cdots+\frac{1}{t}$.
\end{lemma}

\noindent \textbf{Proof of Lemma \ref{lem:4}} 
 We prove the conclusion by induction, showing that if the result holds for $(t-1, u)$ and $(t-1, u-1)$, then it also holds for $(t, u)$. Therefore, it suffices to check $t=0, u>0$ and $t=u=1$ as our base cases.

If $t=0$ and $u>0$, the result follows from the fact that $1+H_t=\frac{u-t}{u}=1$. Next, suppose $t=u=1$. We choose our one new center from one uncovered class with probability exactly $\frac{\Upsilon\left(\mathcal{S}_u\right)}{\Upsilon}$. In this case, Lemma \ref{lem:phi} guarantees that $\mathbb{E}\left[\Upsilon^{\prime}\right] \leq \Upsilon\left(\mathcal{S}_c\right)+16M^4\Upsilon_{\mathrm{OPT}}\left(\mathcal{S}_u\right)$. Since $\Upsilon^{\prime} \leq \Upsilon$, even if we choose a center from a covered class, we have
$$
\begin{aligned}
\mathbb{E}\left[\Upsilon^{\prime}\right] & \leq \frac{\Upsilon\left(\mathcal{S}_u\right)}{\Upsilon} \cdot\left(\Upsilon\left(\mathcal{S}_c\right)+16 M^4\Upsilon_{\mathrm{OPT}}\left(\mathcal{S}_u\right)\right)+\frac{\Upsilon\left(\mathcal{S}_c\right)}{\Upsilon} \cdot \Upsilon\\
& \leq 2 \Upsilon\left(\mathcal{S}_c\right)+16 M^4\Upsilon_{\mathrm{OPT}}\left(\mathcal{S}_u\right)
\end{aligned}
$$
Since $1+H_t=2$ here, we have shown the result holds for both base cases.

We now proceed to prove the inductive step. It is convenient here to consider two cases. First, suppose we choose our first center from a covered class. As above, this happens with probability exactly $\frac{\Upsilon\left(\mathcal{S}_c\right)}{\Upsilon}$. Note that this new center can only decrease $\Upsilon$. We apply the inductive hypothesis with the same choice of covered class, but with $t$ decreased by $1$. It follows that our contribution to $\mathbb{E}\left[\Upsilon^{\prime}\right]$ in this case is at most,
$$
\begin{gathered}
\frac{\Upsilon\left(\mathcal{S}_c\right)}{\Upsilon} \cdot\left(\left(\Upsilon\left(\mathcal{S}_c\right)+16M^4\Upsilon_{\mathrm{OPT}}\left(\mathcal{S}_u\right)\right) \cdot\left(1+H_{t-1}\right)+\frac{u-t+1}{u} \cdot \Upsilon\left(\mathcal{S}_u\right)\right).
\end{gathered}
$$
On the other hand, suppose we choose our first center from some uncovered class $A$. This happens with probability $\frac{\Upsilon(A)}{\Upsilon}$. Let $p_a$ denote the probability that we choose $a \in A$ as our center, given the center is somewhere in $A$, and let $\Upsilon_a$ denote $\Upsilon(A)$ after we choose $a$ as our center. Once again we apply our inductive hypothesis, as well as decrease both $t$ and $u$ by 1.
It follows that our contribution to $\mathbb{E}\left[\Upsilon_{\mathrm{OPT}}\right]$ in this case is at most,
$$
\begin{aligned}
& \frac{\Upsilon(A)}{\Upsilon} \cdot \sum_{a \in A} p_a\left(\left(\Upsilon\left(\mathcal{S}_c\right)+\Upsilon_a+16 M^4\Upsilon_{\mathrm{OPT}}\left(\mathcal{S}_u\right)-16 M^4\Upsilon_{\mathrm{OPT}}(A)\right)\right. \\
& \left.\cdot\left(1+H_{t-1}\right)+\frac{u-t}{u-1} \cdot\left(\Upsilon\left(\mathcal{S}_u\right)-\Upsilon(A)\right)\right) \\
& \leq \frac{\Upsilon(A)}{\Upsilon} \cdot\left(\left(\Upsilon\left(\mathcal{S}_c\right)+16M^4\Upsilon_{\mathrm{OPT}}\left(\mathcal{S}_u\right)\right) \cdot\left(1+H_{t-1}\right)+\frac{u-t}{u-1} \cdot\left(\Upsilon\left(\mathcal{S}_u\right)-\Upsilon(A)\right)\right).
\end{aligned}
$$
The last step here follows from the fact that $\sum_{a \in A} p_a \Upsilon_a \leq 16M^4 \Upsilon_{\mathrm{OPT}}(A)$, which is implied by Lemma \ref{lem:phi}.

Now, the power-mean inequality implies that $\sum_{A \subset \mathcal{S}_u} \Upsilon(A)^2 \geq \frac{1}{u} \cdot \Upsilon\left(\mathcal{S}_u\right)^2$. Therefore, if we sum over all uncovered class $A$, we obtain a contribution at most,
$$
\begin{aligned}
& \frac{\Upsilon\left(\mathcal{S}_u\right)}{\Upsilon} \cdot\left(\Upsilon\left(\mathcal{S}_c\right)+16M^4\Upsilon_{\mathrm{OPT}}\left(\mathcal{S}_u\right)\right) \cdot\left(1+H_{t-1}\right) + \frac{1}{\Upsilon} \cdot \frac{u-t}{u-1} \cdot\left(\Upsilon\left(\mathcal{S}_u\right)^2-\frac{1}{u} \cdot \Upsilon\left(\mathcal{S}_u\right)^2\right) \\
&=\frac{\Upsilon\left(\mathcal{S}_u\right)}{\Upsilon} \cdot\left(\left(\Upsilon\left(\mathcal{S}_c\right)+16M^4\Upsilon_{\mathrm{OPT}}\left(\mathcal{S}_u\right)\right) \cdot\left(1+H_{t-1}\right)+\frac{u-t}{u} \cdot \Upsilon\left(\mathcal{S}_u\right)\right) .
\end{aligned}
$$
Combining the potential contribution to $\mathbb{E}\left[\Upsilon^{\prime}\right]$ from both cases, we now obtain the desired bound:
$$
\begin{aligned}
\mathbb{E}\left[\Upsilon^{\prime}\right] \leq & \left(\Upsilon\left(\mathcal{S}_c\right)+16M^4\Upsilon_{\mathrm{OPT}}\left(\mathcal{S}_u\right)\right) \cdot\left(1+H_{t-1}\right) +\frac{u-t}{u} \cdot \Upsilon\left(\mathcal{S}_u\right)+\frac{\Upsilon\left(\mathcal{S}_c\right)}{\Upsilon} \cdot \frac{\Upsilon\left(\mathcal{S}_u\right)}{u} \\
\leq & \left(\Upsilon\left(\mathcal{S}_c\right)+16M^4\Upsilon_{\mathrm{OPT}}\left(\mathcal{S}_u\right)\right) \cdot\left(1+H_{t-1}+\frac{1}{u}\right) +\frac{u-t}{u} \cdot \Upsilon\left(\mathcal{S}_u\right).
\end{aligned}
$$
The inductive step now follows from the fact that $\frac{1}{n} \leq \frac{1}{t}$.

\noindent \textbf{Proof of Theorem \ref{thm:2}}
Consider the clustering $\mathcal{C}$ after we have completed Step 1. Let $A$ denote the $\mathcal{C}_{\mathrm{OPT}}$ cluster in which we chose the first center. Applying Lemma \ref{lem:4} with $t=u=k-1$ and with $A$ being the only covered class, we have,
$$
\mathbb{E}\left[\Upsilon_{\mathrm{OPT}}\right] \leq\left(\Upsilon(A)+16M^4 \Upsilon_{\mathrm{OPT}}-16M^4 \Upsilon_{\mathrm{OPT}}(A)\right) \cdot\left(1+H_{k-1}\right) .
$$
The result now follows from Lemma \ref{lem:phi}, and from the fact that $H_{k-1} \leq 1+\ln k$.

\noindent \textbf{Proof of Theorem \ref{thm:initia}} By Assumption 2, we know that there are at least $\alpha$ proportion of samples here that are not classified into the correct class. Denote the correctly classified set as $\mathcal{S}_{right}$, and the incorrectly classified set as $\mathcal{S}_{wrong}$. Then 

\begin{align}
\Upsilon_{lack}=\sum_{x\in\mathcal{S}_{wrong}} \min_{c\in \mathcal{C}_{lack}}d\left(x,c\right)^2+\sum_{x\in\mathcal{S}_{right}} \min_{c\in \mathcal{C}_{lack}}d\left(x,c\right)^2.
\end{align}

We consider the part $\mathcal{S}_{right}$ first, we know that for each sample, there is an estimated function of cubic spline approximation, which is $\hat{\lambda}(t)=\sum_{h=1}^H b_h\kappa_h(t)$. 
When sequences $x$ and $c$ are generated from the same class, the distance between them is $d(x,c)=\int_0^T\left|\hat{\lambda_x}(t)/\sqrt{M_x}-\hat{\lambda_c}(t)/\sqrt{M_c}\right|dt\leq \sum_{h=1}^H \left|b_h^x/\sqrt{M_x}-b_h^c/\sqrt{M_c}\right|\int_0^T\kappa_h(t)dt $. 
% Under assumption that $x$ and $c$ all sampled with parameter $\lambda(t)=\sum_{h=1}^H b_h\kappa_h(t)$, 
%By definition, it is easy to know that
%When $x$ samples from the distribution with the intensity function $\lambda(t)$, we know that $\left\|\nabla\log p(x,b)\right\|_2\sim O(L^{-1/2})$. At the same time,the eigenvalues of the second derivative $\Delta\log p(x,b)$ is greater than a constant that is independent of $L(S)$. By the concavity, we know that there is only one optimal value point of $\log p(x,b)$. So $\left\|b^x-b\right\|\sim O(L^{-1/2})$. In the same way we get that $\left\|b^c-b\right\|\sim O(L^{-1/2})$.
Thus we know $d(x,c)\sim O(L^{-1/2})$. 
As $L(S)\rightarrow\infty$, we get that $\Upsilon_{\mathrm{OPT}}/\Upsilon_{lack} \sim O(L^{-1/2})$.

\section{Proof of Theorem \ref{thm:local} and Theorem \ref{thm:robust}}

We first provide several supporting results regarding the properties of Poisson random variables and Poisson processes.

Let $h:[-1, \infty) \rightarrow \mathbb{R}$ be the function defined by $h(u)   := 2 \frac{(1+u) \ln (1+u)-u}{u^2}$.
\begin{lemma}\label{lem:poi}
Let $X \sim \operatorname{Poisson}(\lambda)$ with $\lambda>0$. Then, for any $x>0$, we have
$$
\mathbb{P}\left(X \geq  \lambda+x\right) \leq \exp\left(-\frac{x^2}{2  \lambda} h\left(\frac{x}{\lambda}\right)\right)
$$
and, for any $0<x< \lambda$,
$$
\mathbb{P}\left(X \leq  \lambda-x\right) \leq \exp\left(-\frac{x^2}{2 \lambda} h\left(-\frac{x}{\lambda}\right)\right) .
$$
In particular, this implies that $\mathbb{P}\left(X \geq  \lambda+x\right), \mathbb{P}\left(X \leq \lambda-x\right) \leq \exp\left(-\frac{x^2}{2(\lambda+x)}\right)$, for $x>0$; from which
$$
\mathbb{P}\left(|X-\lambda| \geq x\right) \leq 2 \exp\left(-\frac{x^2}{2(\lambda+x)}\right), \quad x>0.
$$
\end{lemma}

\noindent \textbf{Proof of Lemma \ref{lem:poi}} Recall that if $\left(Y^{(n)}\right)_{n \geq 1}$ is a sequence of independent random variables such that $Y^{(n)}$ follows a {$\operatorname{Binomial}\left(n, \frac{\lambda}{n}\right)$} distribution, then $\left(Y^{(n)}\right)_{n \geq 1}$ converges in law to $X$, a random variable with Poisson $(\lambda)$ distribution. In particular, since convergence in law corresponds to pointwise convergence of distribution functions, this implies that, for any $t \in \mathbb{R}$,
$$
\mathbb{P}\left(Y^{(n)} \geq t\right) \underset{n \rightarrow \infty}{\longrightarrow} \mathbb{P}\left(X \geq t\right).
$$
For any fixed $n \geq 1$,  by the definition, we can write $Y^{(n)}$ as $Y^{(n)}=\sum_{k=1}^n Y_k^{(n)}$, where $Y_1^{(n)}, \ldots, Y_n^{(n)}$ are i.i.d. random variables with $\operatorname{Bernoulli}\left(\frac{\lambda}{n}\right)$ distribution. Note that $\mathbb{E}\left[Y^{(n)}\right]=\lambda$ and $\operatorname{Var}\left[Y^{(n)}\right]=\lambda\left(1-\frac{\lambda}{n}\right) \leq \lambda$. As $\mathbb{E}\left[Y_k^{(n)}\right]=\frac{\lambda}{n}$ and $\left|Y_k^{(n)}\right| \leq 1$ for all $1 \leq k \leq n$, we can apply Bennett's inequality \citep{10.1093/acprof:oso/9780199535255.001.0001}, to obtain, for any $t \geq 0$,
$$
\mathbb{P}\left(Y^{(n)} \geq \lambda+x\right)=\mathbb{P}\left(Y^{(n)} \geq \mathbb{E}\left[Y^{(n)}\right]+x\right) \leq \exp\left(-\frac{x^2}{2 \lambda} h\left(\frac{x}{\lambda}\right)\right).
$$
Taking the limit as $n$ goes to $\infty$, we obtain that $\mathbb{P}\left(X \geq \lambda+x\right) \leq \exp\left(-\frac{x^2}{2 \lambda} h\left(\frac{x}{\lambda}\right)\right)$.

% For sequence $S=\{s_1,s_2,\dots,s_n\}$, assume its cycle time is $T(s)$, and length is $L(s)$.

\begin{lemma}[Bernstein's inequality \citep{vershynin2018high}]\label{lemma:Bern}
Let $X_1, \ldots, X_N$ be independent, mean zero, sub-exponential random variables, and $a=\left(a_1, \ldots, a_N\right) \in \mathbb{R}^N$. Then, for every $t \geq 0$, we have
$$
\mathbb{P}\left(\left|\sum_{i=1}^N a_i X_i\right| \geq t\right) \leq 2 \exp \left[-c \min \left(\frac{t^2}{K^2\|a\|_2^2}, \frac{t}{K\|a\|_{\infty}}\right)\right],
$$
where $K=\max _i\left\|X_i\right\|_{\psi_1}$ and $\|X\|_{\psi_1}:=\inf \{t>0: \mathbb{E} \exp (|X| / t) \leq 2\}$.
\end{lemma}

If $S$ is sampled from an NHP with intensity $\lambda_{\ast}(t)$, according to lemma \ref{lem:poi}, we can utilize Lemma \ref{lemma:Bern} to bound the number of events $m(S)$. That is,
\begin{align*}
    &\mathbb{P}\left(\left|m(S)-\int_0^T\lambda_{\ast}(t)dt\right|>t\right)\\
    &\leq 2\exp\left( -\frac{t^2}{2\int_0^T\lambda_{\ast}(t)dt}h\left(\frac{t}{\int_0^T\lambda_{\ast}(t)dt}\right)\right)\\
    &\leq 2\exp\left( -\frac{t^2}{2(t+\int_0^T\lambda_{\ast}(t)dt)}\right)\\
    & \leq 2\exp\left( -\frac{\log(2)+\sqrt{\log(2)(\log(2)+2\int_0^T\lambda_{\ast}(t)dt)}}{2\log(2)+\sqrt{\log(2)(\log(2)+2\int_0^T\lambda_{\ast}(t)dt)+2\int_0^T\lambda_{\ast}(t)dt}}t\right)\\
    & := 2\exp\left( -K_0t\right) .
\end{align*}
The last inequality comes from the property that probability is always less than $1$. Moreover, we can use Lemma \ref{lemma:Bern} to prove that the log-likelihood is sub-exponential, i.e., its tail probability decays exponentially fast.

\begin{lemma}\label{lem:log-fun}
  When event sequence $S$ is sampled from the NHP process with parameter $\lambda_*$, its log-likelihood function $\log \operatorname{NHP}(S\mid \boldsymbol{B}_i)$ follows a sub-exponential distribution.
\end{lemma}

\noindent \textbf{Proof of Lemma \ref{lem:log-fun}}
Divide the interval $[0,T]$ into $\mathcal{M}$ small intervals $[a_0,a_1],\cdots,[a_{\mathcal{M}-1},a_\mathcal{M}]$, where $0=a_0<a_1<\cdots<a_\mathcal{M}=T$. Within the small interval $[a_i,a_{i+1}]$, there is approximately a homogeneous Poisson process with intensity $\lambda(a_i+\eta)$, where $\eta<a_{i+1}-a_{i}$. At this point we can divide the log-likelihood function into $\mathcal{M}$ parts $F_1,\cdots,F_{\mathcal{M}}$, where $F_{\ell}:=\sum_{t_i\in [a_{\ell-1},a_{\ell}]} \log(t_i)$. 
At this time $F_{\ell}/log(a_i+\eta)$ approximately obeys the homogeneous Poisson process with the parameter $\lambda(a_i+\eta)\cdot(a_{i+1}-a_{i})$, so its variance is $\lambda(a_i+\eta)(a_{i+1}-a_{i})\cdot\log(\lambda(a_i+\eta))^2$. 
According to Lemma \ref{lem:poi}, each of $F_\ell$ follows a sub-exponential distribution. Using Lemma \ref{lemma:Bern}, we know that 
\begin{align*}
    \mathbb{P}\left(\left|\log \operatorname{NHP}(S\mid \boldsymbol{B}_i)/L(S)-\mu_{avg}\right| \geq t\right) \leq 2 \exp \left[-c \min \left(\frac {L(S)^2 t^2}{C^2\max\log(\lambda_{*})^2}, \frac{L(S)t}{C\max\log(\lambda_{*})}\right)\right],
\end{align*}
where $C$ is a finite constant depend on $\boldsymbol{B}_i$ and $\mu_{avg}:=\mathbb{E}_{S\sim \lambda_*} \log \operatorname{NHP}(S\mid \boldsymbol{B}_i)/L(S)$.

Similar to the derivative function of $\log \operatorname{NHP}(S\mid \boldsymbol{B}_i)$, there is
\begin{align*}
    \mathbb{P}\left(\left|\frac{\partial\log \operatorname{NHP}(S\mid \boldsymbol{B}_i)}{\partial \boldsymbol{B}_i}/L(S)-\mu_{avg}\right| \geq t\right) \leq 2 \exp \left[-c \min \left(\frac {L(S)^2 t^2}{C^2(\max\frac{\kappa_{\max}}{\lambda_{*}(t)})^2}, \frac{L(S)t}{C\max\frac{\kappa_{\max}}{\lambda_{*}(t)}}\right)\right].
\end{align*}

\begin{corollary}\label{col:sub}

According to proposition 2.7.1 from \citep{vershynin2018high}, $m(S)$ follow a sub-exponential distribution. From Lemma \ref{lemma:Bern}, we know that for $m(S)$ with $L$ periods, it follows a sub-exponential distribution as well, and $\mathbb{P}\left(\left|m(S)/L-\int_0^T\lambda_{\ast}(t)dt\right|>t\right)\leq 2\exp(-K_0 L t)$. 
Take a small enough $\delta>0$, we have $\mathbb{P}\left(m(S)/L>m_c\right)<\delta$ when $m_c\geq \int_0^T\lambda_{\ast}(t)dt+\log(2/\delta)/(L\cdot K_0)$. Define $C_0:=m_c\cdot L$, which can be viewed as the high probability bound of number of events in event sequence $S$.
\end{corollary}

\noindent \textbf{Overview of Proof Theorem \ref{thm:local}}.~
In order to prove the local convergence property of the proposed algorithm, we need to check the following three key important aspects.
(i) What is the difference
$|\mu(\boldsymbol{B}_k|\boldsymbol{B}_k') - \mu(\boldsymbol{B}_k|\boldsymbol{B}_k'')|$ when $\boldsymbol{B}_k'$ and $\boldsymbol{B}_k''$ are close; see Theorem \ref{thm:gamma}.
(ii) What is the difference between sample gradient $\varrho_k^{(t)}$ and population gradient $\nabla \mu(\boldsymbol{B}_k|\boldsymbol{B}_k^{(t-1)}) $ (``$\nabla$" stands for the derivative with respect to parameter $\boldsymbol{B}_k$); see Lemma \ref{lem:grad:concen}.
(iii) The local concavity of $\mu(\boldsymbol{B}_k|\boldsymbol{B}_k^{(t)})$ holds around $\boldsymbol{B}_k = \boldsymbol{B}_k^{\ast}$; see Lemma \ref{lem:convex}.

Define the weight $w_k(S;\mathbf{B})={\pi_k \operatorname{NHP}(S\mid \boldsymbol{B}_k)}/{\sum_j\pi_j \operatorname{NHP}(S\mid \boldsymbol{B}_j)}$ for $k \in [K]$.

\begin{lemma}\label{lem:weight}
    If $\left\|\boldsymbol{B}_k-\boldsymbol{B}_k^*\right\|_1 <a/(T\cdot\kappa_{\max})$ for $\forall k\in [K]$, there exists a constant $G>0$ such that
\begin{align*}
    \mathbb{E}_S\left[ w_k(S ; \mathbf{B})\left(1-w_k(S ; \mathbf{B})\right)\left\|\frac{\partial \log \operatorname{NHP}({S} \mid \boldsymbol{B}_k)}{\partial \boldsymbol{B}_k}\right\|^p\right]\sim O(L(S)^p\exp(-G\cdot L(S)))
\end{align*}
for $p = 1,2$.
    
\end{lemma}
\noindent \textbf{Proof of Lemma \ref{lem:weight}}
Without loss of generality, we prove the claim for $k=1$. Taking the expectation of $S$, we get
$$
\begin{aligned}
& \mathbb{E}_S \left[w_1(S ; \mathbf{B})\left(1-w_1(S ; \mathbf{B})\right)\left\|\frac{\partial \log \operatorname{NHP}({S} \mid \boldsymbol{B}_1)}{\partial \boldsymbol{B}_1}\right\|^p\right] \\
= & \sum_{i \in[K]} \pi_i \mathbb{E}_{s \sim \mathcal{POI}\left(\boldsymbol{B}_i^*\right)} \left[w_1(S ; \mathbf{B})\left(1-w_1(S ; \mathbf{B})\right)\left\|\frac{\partial \log \operatorname{NHP}({S} \mid \boldsymbol{B}_1)}{\partial \boldsymbol{B}_1}\right\|^p \right] \\
\leq & \pi_1  \mathbb{E}_{s \sim \mathcal{POI}\left(\boldsymbol{B}_1^*\right)}\left[ w_1(S ; \mathbf{B})\left(1-w_1(S ; \mathbf{B})\right)\left\|\frac{\partial \log \operatorname{NHP}({S} \mid \boldsymbol{B}_1)}{\partial \boldsymbol{B}_1}\right\|^p\right]\\
&+\sum_{i \neq 1} \pi_i  \mathbb{E}_{s \sim \mathcal{POI}\left(\boldsymbol{B}_i^*\right)} \left[w_1(S ; \mathbf{B})\left(1-w_1(S ; \mathbf{B})\right)\left\|\frac{\partial \log \operatorname{NHP}({S} \mid \boldsymbol{B}_1)}{\partial \boldsymbol{B}_1}\right\|^p\right] .
\end{aligned}
$$
For the the first term, we define event
$\mathcal{E}_r^{(1)}=\left\{S: S \sim \mathcal{POI}\left(\boldsymbol{B}_1^*\right) ;\left\|\frac{\partial \log \operatorname{NHP}({S} \mid \boldsymbol{B}_1^*)}{\partial \boldsymbol{B}_1}\right\|  \leq r\cdot L(S)\right\}$ for some $r>0$. According to the assumption that $\left\|\boldsymbol{B}_1-\boldsymbol{B}_1^*\right\|_1\leq a/(T*\kappa_{\max})$, we know that $\max\left|\lambda_{\boldsymbol{B}_1}(s)-\lambda_{\boldsymbol{B}_1^*}(s)\right|\leq a/T$. Then for $S \in \mathcal{E}_r^{(1)}$, using triangle inequality, we have
$$
\begin{aligned}
& \left|\sum_t^{m(S)}\frac{\kappa_h(s_t)}{\lambda_{\boldsymbol{B}_1}(s_t)}-\int_0^{T}\kappa_h(x)dx \right|\\
\leq & \left|\sum_t^{m(S)}\frac{\kappa_h(s_t)}{\lambda_{\boldsymbol{B}_1^*}(s_t)}-\int_0^{T}\kappa_h(x)dx \right|+\left|\sum_t^{m(S)}\kappa_h(s_t)(\frac{1}{\lambda_{\boldsymbol{B}_1}(s_t)}-\frac{1}{\lambda_{\boldsymbol{B}_1^*}(s_t)})\right|\\
\leq & L(S)\cdot r +\frac{m(S)a}{T\tau^2}, ~ \forall h\in \{1,\cdots,H\}.\\
\end{aligned}
$$
Because $\left|\lambda_{\boldsymbol{B}_i}(t)-\lambda_{\boldsymbol{B}_i^*}(t)\right|< a/T$ for $i=1,2,\dots,K$, then we have $\log  \operatorname{NHP}(S\mid \boldsymbol{B}_1)=\sum_i\log\lambda_{\boldsymbol{B}_1}(t_i)-\int\lambda_{\boldsymbol{B}_1}(s)ds\geq\log \operatorname{NHP}(S\mid \boldsymbol{B}_1^*)-m(S)\log\left(\frac{\tau+a/T}{\tau}\right)-a\cdot L(S) .$

For $j\neq1$, $\log \operatorname{NHP}(S\mid \boldsymbol{B}_j)-\log \operatorname{NHP}(S\mid \boldsymbol{B}_j^*)=\sum_i\log\frac{\lambda_{\boldsymbol{B}_j}(t_i)}{\lambda_{\boldsymbol{B}_j^*}(t_i)}-\int(\lambda_{\boldsymbol{B}_j}(s)-\lambda_{\boldsymbol{B}_j^*}(s))ds\leq a\cdot L(S)+m(S)\log\left(\frac{\tau+a/T}{\tau}\right)$.
By Assumption \ref{asm:5}, we know that $\log  \operatorname{NHP}(S\mid \boldsymbol{B}_j)\leq \log \operatorname{NHP}(S\mid \boldsymbol{B}_1^*)-C\cdot L(S)+a\cdot L(S)+m(S)\log\left(\frac{\tau+a/T(S)}{\tau}\right)$.
Then we get that 
\begin{align*}
&~\mathbb{E}_S\left[1-w_1(S; \mathbf{B})| \mathcal{E}_r^{(1)}\right] \\
&\leq \frac{1-\pi_1}{\pi_1}\frac{ \operatorname{NHP}(S\mid \boldsymbol{B}_i)}{ \operatorname{NHP}(S\mid \boldsymbol{B}_1)}\\
&\leq \frac{1-\pi_1}{\pi_1}\exp{\left(2a\cdot L(S)+2m(S)\log(\frac{\tau+a/T(S)}{\tau})-C\cdot L(S)\right)}
*\left(r\cdot L(S)+\frac{a}{\tau^2}\frac{m(S)}{T(S)}\right).
\end{align*}

For $ \mathcal{E}_r^{c}$ part, we now have $\left\|\frac{\partial \log \operatorname{NHP}({S} \mid \boldsymbol{B}_1)}{\partial \boldsymbol{B}_1}\right\| >r\cdot L(S)$.
We define
\begin{align*}
    M_h&:=\int_0^{L(S)*T}\frac{\kappa_h(t)}{\lambda_{\boldsymbol{B}_1}(t)}dN(t)-\int_0^{L(S)*T}\kappa_h(x)dx\\
    &=\sum_{l=1}^{L(S)} \int_{(l-1)*T}^{l*T}\frac{\kappa_h(t)}{\lambda_{\boldsymbol{B}_1}(t)}dN(t)-\int_{(l-1)*T}^{l*T(S)}\kappa_h(x)dx\\
    &=\sum_{l=1}^{L(S)} X_l,
\end{align*}
where $X_l$'s are independent. According to Lemma \ref{lem:log-fun}, there exists $c_0>0$ such that
\begin{align*}
    \mathbb{P}\left(|M_h/L(S)|\geq t\right)\leq 2\exp{\left(-\frac{t L(S)}{c_0}\right)}.
\end{align*}

Obviously we have $w_1(S ; \mathbf{B})\left(1-w_1(S ; \mathbf{B})\right)\leq 1/4$. Then
\begin{align*}
&\mathbb{E}_S\left[ w_1(S ; \mathbf{B})\left(1-w_1(S ; \mathbf{B})\right)\left\|\frac{\partial \log \operatorname{NHP}({S} \mid \boldsymbol{B}_1)}{\partial \boldsymbol{B}_1}\right\|^p \mid \mathcal{E}_r^{c}\right]\\
&\leq \frac{1}{4}\int_{r}^\infty t^p d\mathbb{P}\left(\left\|\frac{\partial \log \operatorname{NHP}({S} \mid \boldsymbol{B}_1)}{\partial \boldsymbol{B}_1}\right\|\geq t\cdot L(S)\right)\\
&= \frac{1}{4}\big(r^p\cdot L(S)\mathbb{P}\left(\left\|\frac{\partial \log \operatorname{NHP}({S} \mid \boldsymbol{B}_1)}{\partial \boldsymbol{B}_1}\right\|\geq r\cdot L(S)\right) \\
& +\int_{r}^\infty pt^{p-1} \mathbb{P}\left(\left\|\frac{\partial \log \operatorname{NHP}({S} \mid \boldsymbol{B}_1)}{\partial \boldsymbol{B}_1}\right\|\geq t\cdot L(S)\right)dt \big)\\
&\leq \frac{1}{2}\left(r^p L(S)\exp{\left(-\frac{r L(S)}{c_0}\right)}+\int_{r}^\infty pt^{p-1}\exp{\left(-\frac{t L(S)}{c_0}\right)}dt\right).
\end{align*}

For fixed $r\geq 0$, when $L(S)\rightarrow\infty$, it is easy to know that 
\[\frac{1}{2}\left(r^p L(S)\exp{\left(-\frac{r L(S)}{c_0}\right)}+\int_{r}^\infty pt^{p-1}\exp{\left(-\frac{t L(S)}{c_0}\right)}dt\right)\rightarrow 0 .\] 

Next we consider the remainder of the gradient. For $i\neq 1$,
\begin{align*}
 &\pi_i  \mathbb{E}_{s \sim \mathcal{POI}\left(\boldsymbol{B}_i^*\right)}\left[ w_1(S ; \mathbf{B})\left\|\frac{\partial \log \operatorname{NHP}({S} \mid \boldsymbol{B}_1)}{\partial \boldsymbol{B}_1}\right\|^p\right]  \\
 &=\int_{\left\|\frac{\partial \log \operatorname{NHP}(\mathbf{S} \mid \boldsymbol{B}_i^*)}{\partial \boldsymbol{B}_i}\right\|<r\cdot L(S)}\frac{\pi_1 \operatorname{NHP}(S\mid \boldsymbol{B}_1)\pi_i \operatorname{NHP}(S\mid \boldsymbol{B}_i^*)}{\sum_j\pi_j \operatorname{NHP}(S\mid \boldsymbol{B}_j)}\left\|\frac{\partial \log \operatorname{NHP}({S} \mid \boldsymbol{B}_1)}{\partial \boldsymbol{B}_1}\right\|^p dS \\
 &+\int_{\left\|\frac{\partial \log \operatorname{NHP}(\mathbf{S} \mid \boldsymbol{B}_i^*)}{\partial \boldsymbol{B}_i}\right\|>r\cdot L(S)}\frac{\pi_1 \operatorname{NHP}(S\mid \boldsymbol{B}_1)\pi_i \operatorname{NHP}(S\mid \boldsymbol{B}_i^*)}{\sum_j\pi_j \operatorname{NHP}(S\mid \boldsymbol{B}_j)}\left\|\frac{\partial \log \operatorname{NHP}({S} \mid \boldsymbol{B}_1)}{\partial \boldsymbol{B}_1}\right\|^p dS.
\end{align*}
When $\left\|\frac{\partial \log \operatorname{NHP}({S} \mid \boldsymbol{B}_i^*)}{\partial \boldsymbol{B}_i}\right\|<r\cdot L(S)$, we have $\frac{ \operatorname{NHP}(S\mid \boldsymbol{B}_i)}{ \operatorname{NHP}(S\mid \boldsymbol{B}_i^*)}\leq \exp{\left(a\cdot L(S)+m(S)\log(\frac{\tau+a/T}{\tau})\right)}$ and 
$\frac{ \operatorname{NHP}(S\mid \boldsymbol{B}_i^*)}{ \operatorname{NHP}(S\mid \boldsymbol{B}_i)}\leq \exp{\left(a\cdot L(S)+m(S)\log(\frac{\tau+a/T}{\tau})\right)}$. Then it holds
\begin{align*}
    I_1&\leq \frac{\pi_i \operatorname{NHP}(S\mid \boldsymbol{B}_i^*)}{\pi_i \operatorname{NHP}(S\mid \boldsymbol{B}_i)}\cdot\int_{\left\|\frac{\partial \log \operatorname{NHP}(\mathbf{S} \mid \boldsymbol{B}_i^*)}{\partial \boldsymbol{B}_i}\right\|<r\cdot L(S)}\pi_1 \operatorname{NHP}(S\mid \boldsymbol{B}_1)\left\|\frac{\partial \log \operatorname{NHP}({S} \mid \boldsymbol{B}_1)}{\partial \boldsymbol{B}_1}\right\|^p dS\\
    &\leq\pi_1\exp{\left(aL(S)+m(S)\log(\frac{\tau+a/T}{\tau})\right)}\int_{\left\|\frac{\partial \log \operatorname{NHP}(\mathbf{S} \mid \boldsymbol{B}_i^*)}{\partial \boldsymbol{B}_i}\right\|<r\cdot L(S)} \operatorname{NHP}(S\mid \boldsymbol{B}_1)\left\|\frac{\partial \log \operatorname{NHP}({S} \mid \boldsymbol{B}_1)}{\partial \boldsymbol{B}_1}\right\|^p dS\\
    &\leq\pi_1\exp{\left(aL(S)+m(S)\log(\frac{\tau+a/T}{\tau})\right)}\\
    &\cdot \int_{\left\|\frac{\partial \log \operatorname{NHP}(\mathbf{S} \mid \boldsymbol{B}_i^*)}{\partial \boldsymbol{B}_i}\right\|<r\cdot L(S)} \operatorname{NHP}(S\mid \boldsymbol{B}_i^*)\cdot \exp{\left(-CL(S)+2aL(S)+2m(S)\log(\frac{\tau+a/T}{\tau})\right)}(C_0 L(S))^pdS\\
    &\leq \pi_1\exp{\left(-CL(S)+2aL(S)+2m(S)\log(\frac{\tau+a/T}{\tau}))\right)}*(C_0 L(S))^p ,
\end{align*}
where $C_0$ is the upper bound of $\left\|\frac{\partial \log \operatorname{NHP}({S} \mid \boldsymbol{B}_i)}{\partial \boldsymbol{B}_i}\right\|$, $\forall i=1,\cdots,K$ with probability of $1-\delta$.

When $\left\|\frac{\partial \log \operatorname{NHP}(\mathbf{S} \mid \boldsymbol{B}_i^*)}{\partial \boldsymbol{B}_i}\right\|>r\cdot L(S)$ and $L(S)\rightarrow \infty$, it holds
\begin{align*}
    I_2 &= \frac{\pi_1 \operatorname{NHP}(S\mid \boldsymbol{B}_1)}{\sum_j\pi_j \operatorname{NHP}(S\mid \boldsymbol{B}_j)}\cdot\int_{\left\|\frac{\partial \log \operatorname{NHP}(\mathbf{S} \mid \boldsymbol{B}_i^*)}{\partial \boldsymbol{B}_i}\right\|>r\cdot L(S)}\pi_i \operatorname{NHP}(S\mid \boldsymbol{B}_i^*)\left\|\frac{\partial \log \operatorname{NHP}({S} \mid \boldsymbol{B}_1)}{\partial \boldsymbol{B}_1}\right\|^p dS\\
    &\leq \int_{\left\|\frac{\partial \log \operatorname{NHP}(\mathbf{S} \mid \boldsymbol{B}_i^*)}{\partial \boldsymbol{B}_i}\right\|>r\cdot L(S)}\pi_i \operatorname{NHP}(S\mid \boldsymbol{B}_i^*)\left\|\frac{\partial \log \operatorname{NHP}({S} \mid \boldsymbol{B}_1)}{\partial \boldsymbol{B}_1}\right\|^p dS\\
    &\leq \pi_i(C_0 L(S))^p\int_{\left\|\frac{\partial \log \operatorname{NHP}(\mathbf{S} \mid \boldsymbol{B}_i^*)}{\partial \boldsymbol{B}_i}\right\|>r\cdot L(S)} \operatorname{NHP}(S\mid \boldsymbol{B}_i^*)dS\\
    &\leq 2\pi_i(C_0 L(S))^p\exp{\left(-\frac{t L(S)}{c_0}\right)}dS,
\end{align*}
where we use the same conclusion obtained above that $\mathbb{P}\left(\left\|\frac{\partial \log \operatorname{NHP}({S} \mid \boldsymbol{B}_i^*)}{\partial \boldsymbol{B}_i}\right\|/L(S)\geq t\right)\leq 2\exp{\left(-\frac{t L(S)}{c_0}\right)}$. 
We take $G=\min\{C_{gap}-2a-2m_c\log(\frac{\tau+a/T(S)}{\tau}),t/c_0\}$, where $\mathbb{P}\left(|M(S)/L(S)|\geq m_c\right)<\delta$ for small enough $\delta>0$. Thus we get the result.

\begin{lemma}\label{lem:9}
    If $\left\|\boldsymbol{B}_k - \boldsymbol{B}_k^*\right\|_1 <a/(T\cdot\kappa_{\max})$ for $\forall k \in [K]$, 
    %that is $\left|\lambda_k(t)-\lambda_k^*(t)\right|< a/T$ for $k=1,2,\dots,K$, 
    then it holds 
    $$\left\|\nabla w_k(S,\mathbf{B})\right\|\sim O(L(S)\exp(-G\cdot L(S))).$$
\end{lemma}

\noindent\textbf{Proof of Lemma \ref{lem:9}}~
Without loss of generality, we prove the lemma for $k=1$. Recall the definition of $w_1(S ; \mathbf{B})$ , for any given $S$, consider the function $\mathbf{B}\rightarrow w_1(S ; \mathbf{B})$, it is easy to know that 
\begin{align*}
\nabla w_1(S;\mathbf{B})=\left(\begin{array}{c}
-w_1(S ; \mathbf{B})\left(1-w_1(S ; \mathbf{B})\right)\displaystyle\frac{\partial \log \operatorname{NHP}({S} \mid \boldsymbol{B}_1)}{\partial \boldsymbol{B}_1}\\
w_1(S ; \mathbf{B})w_2(S ; \mathbf{B})\displaystyle\frac{\partial \log \operatorname{NHP}({S} \mid \boldsymbol{B}_2)}{\partial \boldsymbol{B}_2}\\
\vdots\\ 
w_1(S ; \mathbf{B})w_K(S ; \mathbf{B})\displaystyle\frac{\partial \log \operatorname{NHP}({S} \mid \boldsymbol{B}_K)}{\partial \boldsymbol{B}_k}\\ 
\end{array}\right),
\end{align*}
where 
\begin{align*}
\frac{\partial \log \operatorname{NHP}({S} \mid \boldsymbol{B}_i)}{\partial \boldsymbol{B}_i}=\left(\begin{array}{c}
\displaystyle\sum_t^{m(S)}\frac{\kappa_1(s_t)}{\lambda_{\boldsymbol{B}_i}(s_t)}-\int_0^{T}\kappa_1(x)dx\\
\vdots\\
\displaystyle\sum_t^{m(S)}\frac{\kappa_H(s_t)}{\lambda_{\boldsymbol{B}_i}(s_t)}-\int_0^{T}\kappa_H(x)dx\\
\end{array}\right)^\top.
\end{align*}

To calculate the upper bound of $\left\|\nabla w_i(S,\mathbf{B})\right\|$, we start by considering the first line. 
By Lemma \ref{lem:weight}, it is easy to know that the first line is of order $O(L(S)\exp(-G\cdot L(S)))$. Then we turn to other lines. Note that 
\begin{align*}
\mathbb{E}_S\left[ w_1(S ; \mathbf{B})w_i(S ; \mathbf{B})\left\|\frac{\partial \log \operatorname{NHP}({S} \mid \boldsymbol{B}_i)}{\partial \boldsymbol{B}_i}\right\|\right]\leq \mathbb{E}_S\left[ w_i(S ; \mathbf{B})\left(1-w_i(S ; \mathbf{B})\right)\left\|\frac{\partial \log \operatorname{NHP}({S} \mid \boldsymbol{B}_i)}{\partial \boldsymbol{B}_i}\right\|\right]
\end{align*}
for $\forall i \neq 1$. Therefore the upper bound of line $i$ has the same order as that of line $1$.

\begin{lemma}\label{lem:weight2}
    If $\left\|\boldsymbol{B}_i-\boldsymbol{B}_i^*\right\|_1 <a/(T\cdot\kappa_{\max})$, then $\forall i,j\in [K]$, we have
\begin{align*}
    \mathbb{E}_S\left[ w_i(S ; \mathbf{B})w_j(S ; \mathbf{B})\left\|\frac{\partial \log \operatorname{NHP}({S} \mid \boldsymbol{B}_i)}{\partial \boldsymbol{B}_i}\right\|\cdot\left\|\frac{\partial \log \operatorname{NHP}({S} \mid \boldsymbol{B}_j)}{\partial \boldsymbol{B}_j}\right\|\right]\sim O(L(S)^2\exp(-G\cdot L(S))).
\end{align*}
    
\end{lemma}

\noindent\textbf{Proof of Lemma \ref{lem:weight2}}~
Taking the expectation with respect to $S$, we get
$$
\begin{aligned}
& \mathbb{E}_S\left[ w_i(S ; \mathbf{B})w_j(S ; \mathbf{B})\left\|\frac{\partial \log \operatorname{NHP}({S} \mid \boldsymbol{B}_i)}{\partial \boldsymbol{B}_i}\right\|\cdot\left\|\frac{\partial \log \operatorname{NHP}({S} \mid \boldsymbol{B}_j)}{\partial \boldsymbol{B}_j}\right\|\right] \\
\leq & \mathbb{E}_S\left[ w_i(S ; \mathbf{B})w_j(S ; \mathbf{B})\left\|\frac{\partial \log \operatorname{NHP}({S} \mid \boldsymbol{B}_i)}{\partial \boldsymbol{B}_i}\right\|\cdot\left\|\frac{\partial \log \operatorname{NHP}({S} \mid \boldsymbol{B}_j)}{\partial \boldsymbol{B}_j}\right\| \mid\mathcal{E}_0\right]\mathbb{P}(\mathcal{E}_0)\\
&+\sum_{k} \pi_k  \mathbb{E}_{s \sim \mathcal{POI}\left(\boldsymbol{B}_k^*\right)} \left[ w_i(S ; \mathbf{B})w_j(S ; \boldsymbol{B})\left\|\frac{\partial \log \operatorname{NHP}({S} \mid \boldsymbol{B}_i)}{\partial \boldsymbol{B}_i}\right\|\cdot\left\|\frac{\partial \log \operatorname{NHP}({S} \mid \boldsymbol{B}_j)}{\partial \boldsymbol{B}_j}\right\|\mid \right.\\
& \left.\left\|\frac{\partial \log \operatorname{NHP}({S} \mid \boldsymbol{B}_k)}{\partial \boldsymbol{B}_k}\right\|\leq r\right] .
\end{aligned}
$$

Next we consider the remainder of the gradient. When $\left\|\frac{\partial \log \operatorname{NHP}({S} \mid \boldsymbol{B}_k^*)}{\partial \boldsymbol{B}_k}\right\|<r\cdot L(S)$, we have $\frac{ \operatorname{NHP}(S\mid \boldsymbol{B}_k^*)}{ \operatorname{NHP}(S\mid \boldsymbol{B}_k)}\leq \exp{\left(a\cdot L(S)+m(S)\log(\frac{\tau+a/T}{\tau})\right)}$. Then for $I_k$,
\begin{align*}
I_k&=\int_{S}\frac{\pi_i \operatorname{NHP}(S\mid \boldsymbol{B}_i)\pi_j \operatorname{NHP}(S\mid \boldsymbol{B}_j)\pi_k \operatorname{NHP}(S\mid \boldsymbol{B}_k^*)}{(\sum_j\pi_j \operatorname{NHP}(S\mid \boldsymbol{B}_j))^2}\left\|\frac{\partial \log \operatorname{NHP}({S} \mid \boldsymbol{B}_i)}{\partial \boldsymbol{B}_i}\right\|\cdot\left\|\frac{\partial \log \operatorname{NHP}({S} \mid \boldsymbol{B}_j)}{\partial \boldsymbol{B}_j}\right\|dS \\
&\leq\int_{S}\frac{\pi_i \operatorname{NHP}(S\mid \boldsymbol{B}_i)\pi_j \operatorname{NHP}(S\mid \boldsymbol{B}_j)\pi_k \operatorname{NHP}(S\mid \boldsymbol{B}_k)\exp{\left(aL(S)+m(S)\log(\frac{\tau+a/T}{\tau})\right)}}{(\sum_j\pi_j \operatorname{NHP}(S\mid \boldsymbol{B}_j))^2}\\
&\cdot\left\|\frac{\partial \log \operatorname{NHP}({S} \mid \boldsymbol{B}_i)}{\partial \boldsymbol{B}_i}\right\| \left\|\frac{\partial \log \operatorname{NHP}({S} \mid \boldsymbol{B}_j)}{\partial \boldsymbol{B}_j}\right\|dS.
\end{align*}
Because $i\neq j$, it is easy to know that at least one of ${i,j}$ is not equal to $k$. Without loss of generality, assume that $i\neq k$, we have
\begin{align*}
    I_k&=\pi_i\frac{\pi_j \operatorname{NHP}(S\mid \boldsymbol{B}_j)\pi_k \operatorname{NHP}(S\mid \boldsymbol{B}_k)\exp{\left(aL(S)+m(S)\log(\frac{\tau+a/T}{\tau})\right)}}{(\sum_j\pi_j \operatorname{NHP}(S\mid \boldsymbol{B}_j))^2}\\
    &\cdot\int_{S} \operatorname{NHP}(S\mid \boldsymbol{B}_i)\left\|\frac{\partial \log \operatorname{NHP}({S} \mid \boldsymbol{B}_i)}{\partial \boldsymbol{B}_i}\right\|\cdot\left\|\frac{\partial \log \operatorname{NHP}({S} \mid \boldsymbol{B}_j)}{\partial \boldsymbol{B}_j}\right\|dS\\
    &\leq\pi_i\exp{\left(aL(S)+m(S)\log(\frac{\tau+a/T}{\tau})\right)}\int_{S} \operatorname{NHP}(S\mid \boldsymbol{B}_i)\left\|\frac{\partial \log \operatorname{NHP}({S} \mid \boldsymbol{B}_i)}{\partial \boldsymbol{B}_i}\right\|\\
    & \cdot\left\|\frac{\partial \log \operatorname{NHP}({S} \mid \boldsymbol{B}_j)}{\partial \boldsymbol{B}_j}\right\|dS\\
    &\leq\pi_i\exp{\left(aL(S)+m(S)\log(\frac{\tau+a/T}{\tau})\right)}\\
    &\cdot \int_{S} \operatorname{NHP}(S\mid \boldsymbol{B}_k^*)*\exp{\left(-CL(S)+aL(S)+m(S)\log(\frac{\tau+a/T}{\tau})\right)}(C_0 L(S))^2dS\\
    &\leq \pi_1\exp{\left(-CL(S)+2aL(S)+2m(S)\log(\frac{\tau+a/T}{\tau}))\right)}*(C_0 L(S))^2,
\end{align*}
where $C_0$ is the upper bound of $\left\|\frac{\partial \log \operatorname{NHP}({S} \mid \boldsymbol{B}_i^*)}{\partial \boldsymbol{B}_i}\right\|$, $\forall i=1,\cdots,K$ with probability of $1-\delta$.

When $\left\|\frac{\partial \log \operatorname{NHP}({S} \mid \boldsymbol{B}_i^*)}{\partial \boldsymbol{B}_i}\right\|>r\cdot L(S)$, if $L(S)\rightarrow \infty$, 
\begin{align*}
    I_0 &= \frac{\pi_1 \operatorname{NHP}(S\mid \boldsymbol{B}_1)}{\sum_j\pi_j \operatorname{NHP}(S\mid \boldsymbol{B}_j)}\cdot\int_{\left\|\frac{\partial \log \operatorname{NHP}(\mathbf{S} \mid \boldsymbol{B}_i^*)}{\partial \boldsymbol{B}_i}\right\|>r\cdot L(S)}\pi_i \operatorname{NHP}(S\mid \boldsymbol{B}_i^*)\left\|\frac{\partial \log \operatorname{NHP}({S} \mid \boldsymbol{B}_1)}{\partial \boldsymbol{B}_1}\right\|dS\\
    &\leq \int_{\left\|\frac{\partial \log \operatorname{NHP}(\mathbf{S} \mid \boldsymbol{B}_i^*)}{\partial \boldsymbol{B}_i}\right\|>r\cdot L(S)}\pi_i \operatorname{NHP}(S\mid \boldsymbol{B}_i^*)\left\|\frac{\partial \log \operatorname{NHP}({S} \mid \boldsymbol{B}_1)}{\partial \boldsymbol{B}_1}\right\|dS\\
    &\leq \pi_iC_0 L(S)\int_{\left\|\frac{\partial \log \operatorname{NHP}(\mathbf{S} \mid \boldsymbol{B}_i^*)}{\partial \boldsymbol{B}_i}\right\|>r\cdot L(S)} \operatorname{NHP}(S\mid \boldsymbol{B}_i^*)dS\\
    &\leq 2\pi_iC_0 L(S)\exp{\left(-\frac{t L(S)}{c_0}\right)}dS,
\end{align*}
where we use the same conclusion obtained above that $\mathbb{P}\left(\left\|\frac{\partial \log \operatorname{NHP}({S} \mid \boldsymbol{B}_i^*)}{\partial \boldsymbol{B}_i}\right\|/L(S)\geq t\right)\leq 2\exp{\left(-\frac{t L(S)}{c_0}\right)}$. 
We still take $G=\min\{C_{gap}-2a-2m_c\log(\frac{\tau+a/T(S)}{\tau}),t/c_0\}$, where \\ $\mathbb{P}\left(|M(S)/L(S)|\geq m_c\right)<\delta$ for small enough $\delta>0$. Thus we get the result.

\begin{lemma} \label{lem:convex} 
% $\boldsymbol{B}_i^+=\arg\max_{\boldsymbol{B}_i}\mu(\boldsymbol{B}_i\mid \boldsymbol{B}_i^{(t)}) $ 
Function $\mu(\boldsymbol{B}_k \mid \boldsymbol{B}_k^{(t)})$ 
is a locally concave function with high probability for $k =1,2,\dots,K$.
\end{lemma}

\noindent\textbf{Proof of Lemma \ref{lem:convex}}~
Without loss of generality, we let $k = 1$. 
We abuse the notation by treating $\alpha = \rho$ in the following proof.
By taking the first derivative of the estimating equation, we have
\begin{align*}
0&=\nabla_{\boldsymbol{B}_1}\left(\sum_{i = 1}^N w_1(S_i;\mathbf{B}^{(t)})\phi_\alpha\left(\log \operatorname{NHP}(S_i\mid\boldsymbol{B}_1)/L(S_i)-\mu(\boldsymbol{B}_1\mid \boldsymbol{B}_1^{(t)})\right)\right)\\
&=\sum_{i}^N w_1(S_i;\mathbf{B}^{(t)})\phi_\alpha^{'}\left(\log \operatorname{NHP}(S_i\mid\boldsymbol{B}_1)/L(S_i)-\mu(\boldsymbol{B}_1\mid \boldsymbol{B}_1^{(t)})\right) \\
& \cdot \left(\nabla\log \operatorname{NHP}(S_i\mid\boldsymbol{B}_1)/L(S_i)-\nabla\mu(\boldsymbol{B}_1\mid \boldsymbol{B}_1^{(t)}) \right).
\end{align*}
By taking the second derivative, we have
\begin{align*}
0&=\sum_{i}^n w_1(S_i;\mathbf{B}^{(t)})\nabla_{\boldsymbol{B}_1}^2\phi_\alpha\left(\log \operatorname{NHP}(S_i\mid\boldsymbol{B}_1)/L(S_i)-\mu(\boldsymbol{B}_1\mid \boldsymbol{B}_1^{(t)})\right)\\
&=\sum_{i}^n w_1(S_i;\mathbf{B}^{(t)})\phi_\alpha^{'}\left(\log \operatorname{NHP}(S_i\mid\boldsymbol{B}_1)/L(S_i)-\mu(\boldsymbol{B}_1\mid \boldsymbol{B}_1^{(t)})\right) \\
& \cdot \left(\nabla^2\log \operatorname{NHP}(S_i\mid\boldsymbol{B}_1)/L(S_i)-\nabla^2\mu(\boldsymbol{B}_1\mid \boldsymbol{B}_1^{(t)}) \right)\\
&+\sum_{i}^n w_1(S_i;\mathbf{B}^{(t)})\phi_\alpha^{''}\left(\log \operatorname{NHP}(S_i\mid\boldsymbol{B}_1)/L(S_i)-\mu(\boldsymbol{B}_1\mid \boldsymbol{B}_1^{(t)})\right) \\
& \cdot \alpha\left(\nabla\log \operatorname{NHP}(S_i\mid\boldsymbol{B}_1)/L(S_i)-\nabla\mu(\boldsymbol{B}_1\mid \boldsymbol{B}_1^{(t)}) \right)^2.
\end{align*}
With a high probability, there exists $c_{\phi}$ such that $c_{\phi}|\phi^{'}(\eta)|>|\phi^{''}(\eta)|$, where $\eta\in(-9.5+2/c_{\phi},9.5-2/c_{\phi})$.
By Matrix Chernoff inequalities (Lemma \ref{lem:mat}), as $L(S)\rightarrow\infty$, we claim that $\lambda_{min} \left(\nabla^2\log \operatorname{NHP}(S_i\mid\boldsymbol{B}_1)/L(S_i)\right)- c_{\phi}\alpha\lambda_{max}\left(\nabla(\log \operatorname{NHP}(S_i\mid\boldsymbol{B}_1)/L(S_i))^2\right)\succeq 0$.
Next we explain the reasons. Write $S_i$ as $\{S_{i,1},S_{i,2},\cdots,S_{i,m(S)}\}$, then

\begin{eqnarray*} \label{part0_9}
    \begin{bmatrix}
        \displaystyle\nabla\frac{\log \operatorname{NHP}(S_i\mid\boldsymbol{B}_1)}{L(S_i)}
    \end{bmatrix}^2
     &=&
    \begin{bmatrix}
        \displaystyle\sum_{t=1}^{m(S)}\frac{\kappa_1(S_{i,t})}{\lambda_{\boldsymbol{B}_1}(S_{i,t})\cdot L(S_i)}-\int_0^{T}\kappa_1(x)dx\\
        \vdots \\
       \displaystyle\sum_{t=1}^{m(S)}\frac{\kappa_H(S_{i,t})}{\lambda_{\boldsymbol{B}_1}(S_{i,t})\cdot L(S_i)}-\int_0^{T}\kappa_H(x)dx
    \end{bmatrix}\\
    & &\times
    \begin{bmatrix}
        \displaystyle\sum_{t=1}^{m(S)}\frac{\kappa_1(S_{i,t})}{\lambda_{\boldsymbol{B}_1}(S_{i,t})\cdot L(S_i)}-\int_0^{T}\kappa_1(x)dx\\
        \vdots \\
       \displaystyle\sum_{t=1}^{m(S)}\frac{\kappa_H(S_{i,t})}{\lambda_{\boldsymbol{B}_1}(S_{i,t})\cdot L(S_i)}-\int_0^{T}\kappa_H(x)dx
    \end{bmatrix}^\top\\
    & =: & G\cdot G^\top .
\end{eqnarray*}
Therefore the largest eigenvalue of $\nabla\log \operatorname{NHP}(S_i\mid\boldsymbol{B}_1)/L(S_i)$ is the 2-norm of vector $G$. For each component of $G$, we know that $\mathbb{E}\left[\sum_{t=1}^{m(S)}\frac{\kappa_h(S_{i,t})}{\lambda_{\boldsymbol{B}_1}(S_{i,t})}/L(S_i)-\int_0^{T}\kappa_h(x)dx\right]=\mathbb{E}\left[\int_0^T \frac{\kappa_h(S_{i,t})}{\lambda_{\boldsymbol{B}_1}(S_{i,t})}dN(t)\right]/L(S_i)-\int_0^{T}\kappa_h(x)dx=\int_0^T \frac{\kappa_h(t)}{\lambda_{\boldsymbol{B}_1}(t)}\cdot \lambda_{\boldsymbol{B}_1}(t)dt/L(S_i)-\int_0^{T}\kappa_h(x)dx=0, \forall h=1,\cdots,H$. When $S_i$ is generated from the Poisson process with the intensity function $\lambda_{\boldsymbol{B}_1}(\cdot)$, we know that $\left\|G\right\|_2 \sim O(L^{-1/2})$ with high probability. Thus, we get the result that $\alpha c_{\phi}\lambda_{max}\left(\nabla(\log \operatorname{NHP}(S_i\mid\boldsymbol{B}_1^*)/L(S_i))^2\right)\sim O(\alpha L^{-1/2})\rightarrow 0$ as $L\rightarrow 0$. For fixed $\boldsymbol{B}_1^{(t)}$, we also know that $\left\|G\right\|\sim O(L^{-1/2})$, while $\lambda_{min} \left(\nabla^2\log \operatorname{NHP}(S_i\mid\boldsymbol{B}_1)/L(S_i)\right) \sim O(1)$.
Because of the continuity of $\phi'$ and $\phi''$, it is easy to confirm the continuity of $\nabla^2\mu(\boldsymbol{B}_1\mid \boldsymbol{B}_1^{(t)})$.

\begin{lemma}[Matrix Chernoff I \citep{tropp2012user}]\label{lem:mat}
Consider a finite sequence of independent, random, self-adjoint matrices  $\{\mathbf{X}_k\}$ with dimension $d$.
Assume that each random matrix satisfies: 
$\mathbf{X}_k \succeq \mathbf{0} \quad \text{and} \quad \lambda_{\text{max}}(\mathbf{X}_k) \leq R$ almost surely.
Define
$$
\mu_{\text{min}} = \lambda_{\text{min}}\left( \sum_k \mathbb{E}\, \mathbf{X}_k \right) \quad \text{and} \quad
\mu_{\text{max}} = \lambda_{\text{max}}\left( \sum_k \mathbb{E}\, \mathbf{X}_k \right).
$$
Then we have 
{\small
\begin{align*}
\mathbb{P} \left( \lambda_{\text{min}}\left( \sum_k \mathbf{X}_k \right) \leq (1-\delta)\mu_{\text{min}}  \right) \leq d \cdot \left[ \frac{e^{-\delta}}{(1-\delta)^{1-\delta}} \right]^{\mu_{\text{min}}/R} \quad \text{for } \delta\in [0,1)\\
\mathbb{P} \left( \lambda_{\text{max}}\left( \sum_k \mathbf{X}_k \right) \geq (1+\delta)\mu_{\text{max}}  \right) \leq d \cdot \left[ \frac{e^{\delta}}{(1+\delta)^{1+\delta}} \right]^{\mu_{\text{max}}/R} \quad \text{for } \delta \geq 0.
\end{align*}
}
\end{lemma}

\begin{theorem}\label{thm:gamma}
    For $k=\{1,2,\cdots,K\}$, 
    $\left\|\boldsymbol{B}_k-\boldsymbol{B}_k^*\right\|_1 <a/(T\cdot\kappa_{\max})$ and $\left\|\nabla \mu(\boldsymbol{B}_k^t\mid \boldsymbol{B}_k^t)-\nabla \mu(\boldsymbol{B}_k^t\mid \boldsymbol{B}_k^*)\right\|\leq \gamma\left\|\boldsymbol{B}_k^t-\boldsymbol{B}_k^*\right\| $.
    We take the tuning parameter $\alpha$ sufficiently small. Then $\gamma-\frac{\lambda_{min}}{4}\leq O(L^{-1/2})$ as $L(S)\rightarrow\infty$, $\gamma\rightarrow\lambda_{min}/4$.
\end{theorem}

\noindent\textbf{Proof of Theorem \ref{thm:gamma}}
% Consider the difference of the gradient corresponding to $\boldsymbol{B}_i$, 
Without loss of generality, we only consider $k=1$.
{\small
\begin{align*}
&\nabla \mu(\boldsymbol{B}_1^t\mid \boldsymbol{B}_1^t)-\nabla \mu(\boldsymbol{B}_1^t\mid \boldsymbol{B}_1^*)=\\
&\mathbb{E}_S 
\Big( w_1\left(S ; \boldsymbol{B}^t\right)\phi'_\alpha\left(\log \operatorname{NHP}(S\mid \boldsymbol{B}^t_1 ))/L(S)-\mu(\boldsymbol{B}_1^t\mid \boldsymbol{B}_1^t)\right ) \\
& -w_1\left(S ; \boldsymbol{B}^*\right)\phi'_\alpha\left(\log \operatorname{NHP}(S\mid \boldsymbol{B}^t_1 ))/L(S)-\mu(\boldsymbol{B}_1^t\mid \boldsymbol{B}_1^*)\right)\Big)\\
&\cdot\alpha\nabla\log \operatorname{NHP}(S\mid \boldsymbol{B}_1^t ))/L(S).
\end{align*}
}
\vspace{-5mm}
For any given $S$, 
% consider the function $\boldsymbol{B} \rightarrow w_1(S ; \boldsymbol{B})$, 
we have
{\small 
\begin{align*}
&\nabla\left( w_1(S ; \mathbf{B})\phi'_\alpha\left(\log \operatorname{NHP}(S\mid \mathbf{B} ))/L(S)-\mu(\boldsymbol{B}_1\mid \boldsymbol{B}_1^t)\right)\right)\\
&=\nabla w_1(S ; \mathbf{B})\cdot\phi'_\alpha\left(\log \operatorname{NHP}(S\mid \mathbf{B} ))/L(S)-\mu(\boldsymbol{B}_1\mid \boldsymbol{B}_1^t)\right)\\
& +w_1(S ; \mathbf{B})\cdot\nabla \phi'_\alpha\left(\log \operatorname{NHP}(S\mid \mathbf{B} ))/L(S)-\mu(\boldsymbol{B}_1\mid \boldsymbol{B}_1^t)\right)
\\
&=\left[\begin{array}{c}
-w_1(S ; \mathbf{B})\left(1-w_1(S ; \mathbf{B})\right)\displaystyle\frac{\partial \log \operatorname{NHP}({S} \mid \boldsymbol{B}_1)}{\partial \boldsymbol{B}_1}\\
w_1(S ; \mathbf{B})w_2(S ; \mathbf{B})\displaystyle\frac{\partial \log \operatorname{NHP}({S} \mid \boldsymbol{B}_2)}{\partial \boldsymbol{B}_2}\\
\vdots\\ 
w_1(S ; \mathbf{B})w_K(S ; \mathbf{B})\displaystyle\frac{\partial \log \operatorname{NHP}({S} \mid \boldsymbol{B}_K)}{\partial \boldsymbol{B}_k}\\ 
\end{array}\right]\cdot \phi'_\alpha\left(\log \operatorname{NHP}(S\mid \boldsymbol{B}_1 ))/L(S)-\mu(\boldsymbol{B}_1\mid \boldsymbol{B}_1^t)\right)\\
&+w_1(S ; \mathbf{B})\cdot \phi''_\alpha\left(\log \operatorname{NHP}(S\mid \boldsymbol{B}_1 ))/L(S)-\mu(\boldsymbol{B}_1\mid \boldsymbol{B}_1^t)\right) \\
&\cdot (1-w_1\left(S ; \mathbf{B}\right)\phi'_\alpha\left(\log \operatorname{NHP}(S\mid \mathbf{B} ))-\mu(\boldsymbol{B}_1\mid \boldsymbol{B}_1^t)\right))\alpha\nabla\log \operatorname{NHP}(S\mid \mathbf{B} ))/L(S).
\end{align*}
}

Let $\mathbf{B}^u=\mathbf{B}^*+u\left(\mathbf{B}^t-\mathbf{B}^*\right), \forall u \in[0,1]$. By Taylor's expansion, we have
{\small
\begin{align*}
& \left\|\mathbb{E}_S 
\left( w_1\left(S ; \mathbf{B}^t\right)\phi'_\alpha\left(\log \operatorname{NHP}(S\mid \mathbf{B} ))-\mu(\boldsymbol{B}_1\mid \boldsymbol{B}_1^t)\right)-w_1\left(S ; \mathbf{B}^*\right)\phi'_\alpha\left(\log \operatorname{NHP}(S\mid \mathbf{B} ))-\mu(\mathbf{B}\mid \mathbf{B}^*)\right)\right) \right.\quad\\ 
& \left.\quad\cdot\alpha\nabla\log \operatorname{NHP}(S\mid \boldsymbol{B}^t_1 ))/L(S)\right\|\\
= & \left\|\mathbb{E}\left[\int_{u=0}^1 \nabla w_1(S ; \mathbf{B^u})\phi'_\alpha\left(\log \operatorname{NHP}(S\mid \boldsymbol{B}_1 ))-\mu(\boldsymbol{B}_1\mid \boldsymbol{B}_1^u)\right)du\cdot\alpha\nabla\log \operatorname{NHP}(S\mid \boldsymbol{B}_1^t ))/L(S)\right]\right\| \\
\leq &\left \|\mathbb{E} \int_{u=0}^1  w_1(S ; \mathbf{B}^u)\left(1-w_1(S ; \mathbf{B}^u)\right)\frac{\partial \log \operatorname{NHP} (\mathbf{S} \mid \boldsymbol{B}_1^u)}{\partial \boldsymbol{B}_1}^\top\left(\boldsymbol{B}_1^t-\boldsymbol{B}_1^*\right)\cdot\alpha\frac{\partial \log \operatorname{NHP}(\mathbf{S} \mid \boldsymbol{B}_1^t)}{\partial \boldsymbol{B}_1}/L(S) d u \right.\quad\\
& \left. \quad-\sum_{i\neq 1}\mathbb{E} \int_{u=0}^1 w_1(S ; \mathbf{B}^u)w_i(S ; \mathbf{B}^u)\frac{\partial \log \operatorname{NHP}(\mathbf{S} \mid \boldsymbol{B}_i^u)}{\partial \boldsymbol{B}_i}^\top\left(\boldsymbol{B}_i^t-\boldsymbol{B}_i^*\right)\cdot\alpha\frac{\partial \log \operatorname{NHP}(\mathbf{S} \mid \boldsymbol{B}_1^t)}{\partial \boldsymbol{B}_1}/L(S) d u \right\|\cdot\phi'_{max} \\
& +{\left\|\mathbb{E} \int_{u=0}^1 w_1(S ; \mathbf{B})(1-w_1(S ; \mathbf{B})\phi'(\cdot))\cdot \alpha\frac{\partial \log \operatorname{NHP} (\mathbf{S} \mid \boldsymbol{B}_1^u)}{\partial \boldsymbol{B}_1}^\top\left(\boldsymbol{B}_1^t-\boldsymbol{B}_1^*\right)\cdot\alpha\frac{\partial \log \operatorname{NHP}(\mathbf{S} \mid \boldsymbol{B}_1^t)}{L(S)^2 \partial \boldsymbol{B}_1} d u \right\|\cdot\phi''_{max}}\\
\leq & U_1\left\|\boldsymbol{B}_1^t-\boldsymbol{B}_1^*\right\|_2+\sum_{i \neq 1} U_i\left\|\boldsymbol{B}_i^t-\boldsymbol{B}_i^*\right\|_2\\
&\quad+\underbrace{\sup _{u \in[0,1]}\left\|\mathbb{E}  w_1(S ; \mathbf{B})(1-w_1(S ; \mathbf{B})\phi'(\cdot))\cdot \alpha^2/L(S)^2\frac{\partial \log \operatorname{NHP} (\mathbf{S} \mid \boldsymbol{B}_1^u)}{\partial \boldsymbol{B}_1}\frac{\partial \log \operatorname{NHP}(\mathbf{S} \mid \boldsymbol{B}_1^t)}{\partial \boldsymbol{B}_1}^\top d u \right\|_2}_{I_0}\\
&\cdot\phi'_{max}\phi''_{max}\cdot\left\|\boldsymbol{B}_1^t-\boldsymbol{B}_1^*\right\|_2,
\end{align*}
}
where
$$
\begin{aligned}
& U_1=\sup _{u \in[0,1]}\left \|\mathbb{E}  w_1(S ; \mathbf{B}^u)\left(1-w_1(S ; \mathbf{B}^u)\right)\alpha/L(S)\frac{\partial \log \operatorname{NHP}(\mathbf{S} \mid \boldsymbol{B}_1^t)}{\partial \boldsymbol{B}_1}\frac{\partial \log \operatorname{NHP} (\mathbf{S} \mid \boldsymbol{B}_1^u)}{\partial \boldsymbol{B}_1}^T\right\|_{2} \\
& U_i=\sup _{u \in[0,1]}\left \|\mathbb{E}  w_1(S ; \mathbf{B}^u)w_i(S ; \mathbf{B}^u)\alpha/L(S)\frac{\partial \log \operatorname{NHP}(\mathbf{S} \mid \boldsymbol{B}_1^t)}{\partial \boldsymbol{B}_1}\frac{\partial \log \operatorname{NHP}(\mathbf{S} \mid \boldsymbol{B}_i^u)}{\partial \boldsymbol{B}_i}^T\right\|_{2}.
\end{aligned}
$$

For $U_1$, by triangle inequality, we have
\begin{align*}
U_1&\leq \sup _{u \in[0,1]}\left \|\mathbb{E}  w_1(S ; \mathbf{B}^u)\left(1-w_1(S ; \mathbf{B}^u)\right)\alpha/L(S)\frac{\partial \log \operatorname{NHP} (\mathbf{S} \mid \boldsymbol{B}_1^u)}{\partial \boldsymbol{B}_1}\frac{\partial \log \operatorname{NHP} (\mathbf{S} \mid \boldsymbol{B}_1^u)}{\partial \boldsymbol{B}_1}^T\right\|_{2}\\
&+\sup _{u \in[0,1]}\left \|\mathbb{E}  w_1(S ; \mathbf{B}^u)\left(1-w_1(S ; \mathbf{B}^u)\right)\alpha/L(S)\frac{\partial \log \operatorname{NHP}(\mathbf{S} \mid \boldsymbol{B}_1^u)^2}{\partial \boldsymbol{B}_1^2}(\boldsymbol{B}_1^u-\boldsymbol{B}_1^t)\frac{\partial \log \operatorname{NHP} (\mathbf{S} \mid \boldsymbol{B}_1^u)}{\partial \boldsymbol{B}_1}^T\right\|_{2}\\
&\leq \sup _{u \in[0,1]}\left \|\mathbb{E}  w_1(S ; \mathbf{B}^u)\left(1-w_1(S ; \mathbf{B}^u)\right)\alpha/L(S)\frac{\partial \log \operatorname{NHP} (\mathbf{S} \mid \boldsymbol{B}_1^u)}{\partial \boldsymbol{B}_1}\frac{\partial \log \operatorname{NHP} (\mathbf{S} \mid \boldsymbol{B}_1^u)}{\partial \boldsymbol{B}_1}^T\right\|_{2}\\
&+a\sup _{u \in[0,1]}\left \|\mathbb{E}  w_1(S ; \mathbf{B}^u)\left(1-w_1(S ; \mathbf{B}^u)\right)\frac{\partial \log \operatorname{NHP} (\mathbf{S} \mid \boldsymbol{B}_1^u)}{\partial \boldsymbol{B}_1}\right\|\cdot\left\|\frac{\partial \log \operatorname{NHP}(\mathbf{S} \mid \boldsymbol{B}_1^u)^2}{\partial \boldsymbol{B}_1^2}/L(S)\right\|.
\end{align*}
According to Lemma \ref{lem:weight} , we know that $U_1\sim O(\exp(-G\cdot L)\cdot L)$ . When $L\rightarrow\infty$, $U_1\rightarrow0$.
Similarly, for $U_i, i \neq 1$,
\begin{align*}
U_i&\leq \sup _{u \in[0,1]}\left \|\mathbb{E}  w_1(S ; \mathbf{B}^u)w_i(S ; \mathbf{B}^u)\alpha/L(S)\frac{\partial \log \operatorname{NHP} (\mathbf{S} \mid \boldsymbol{B}_1^u)}{\partial \boldsymbol{B}_1}\frac{\partial \log \operatorname{NHP}(\mathbf{S} \mid \boldsymbol{B}_i^u)}{\partial \boldsymbol{B}_i}^\top\right\|_{2}\\
&+a\sup _{u \in[0,1]}\left \|\mathbb{E}  w_1(S ; \mathbf{B}^u)w_i(S ; \mathbf{B}^u)\frac{\partial \log \operatorname{NHP}(\mathbf{S} \mid \boldsymbol{B}_i^u)}{\partial \boldsymbol{B}_i}\right\|\cdot\left\|\frac{\partial \log \operatorname{NHP}(\mathbf{S} \mid \boldsymbol{B}_1^u)^2}{\partial \boldsymbol{B}_1^2}/L(S)\right\|.
\end{align*}
Refer to Lemma \ref{lem:weight2}, we can get that $U_i\rightarrow0$.

When $S$ is sampled from class $i\neq 1$, $w_1(S;\mathbf{B})\sim\exp(-GL)$ and it can be checked that $I_0\rightarrow 0$ at this time (like Lemma \ref{lem:weight}). So we only consider the situation when $S$ is sampled from class $1$. For $I_0$ by triangle inequality we have,
\begin{align*}
I_0&\leq \left\|\mathbb{E}  w_1(S ; \mathbf{B})(1-w_1(S ; \mathbf{B})\phi'(\cdot))\cdot \alpha^2/L(S)^2\frac{\partial \log \operatorname{NHP}(\mathbf{S} \mid \boldsymbol{B}_1^*)}{\partial \boldsymbol{B}_1}\frac{\partial \log \operatorname{NHP}(\mathbf{S} \mid \boldsymbol{B}_1^*)}{\partial \boldsymbol{B}_1}^\top d u \right\|_2\\
&+2a\sup _{u \in[0,1]}\left \|\mathbb{E}  w_1(S ; \mathbf{B})(1-w_1(S ; \mathbf{B})\phi'(\cdot))\cdot \alpha^2/L(S)^2\frac{\partial \log \operatorname{NHP}(\mathbf{S} \mid \boldsymbol{B}_1^u)^2}{\partial \boldsymbol{B}_1^2}\frac{\partial \log \operatorname{NHP} (\mathbf{S} \mid \boldsymbol{B}_1^*)}{\partial \boldsymbol{B}_1}^\top\right\|_{2}\\
&+a^2\sup _{u \in[0,1]}\left \|\mathbb{E}  w_1(S ; \mathbf{B})(1-w_1(S ; \mathbf{B})\phi'(\cdot))\cdot \alpha^2\right\|\cdot\left\|\frac{\partial \log \operatorname{NHP}(\mathbf{S} \mid \boldsymbol{B}_1^u)^2}{\partial \boldsymbol{B}_1^2}/L(S)\right\|^2.
\end{align*}

There exists an upper bound of $\left\|\frac{\partial \log \operatorname{NHP}(\mathbf{S} \mid \boldsymbol{B}_1^u)^2}{\partial \boldsymbol{B}_1^2}/L(S)\right\|$ with a high probability. Taking $a\leq \frac{\lambda_{min}}{4}/\alpha\left\|\frac{\partial \log \operatorname{NHP}(\mathbf{S} \mid \boldsymbol{B}_1^u)^2}{\partial \boldsymbol{B}_1^2}/L(S)\right\|$, we have $I_0\rightarrow \frac{\lambda_{min}}{4}$ and $\gamma\rightarrow \frac{\lambda_{min}}{4}$ when $L(S) \rightarrow \infty$.

\begin{lemma}\label{lem:grad:concen}
    For cluster $i$, we write 
    {\small
    \begin{align*}
        &\nabla \mu(\boldsymbol{B}_i\mid \boldsymbol{B}_i^{(t)})_{\mathcal{S}} ~~ (\equiv \varrho_i^{(t)} ) \\
        &:=\frac{\frac{1}{N}\sum_{n\in\mathcal{S}} w_1(S_n;\mathbf{B})\phi_\rho^{'}\left(\log \operatorname{NHP}(S_n\mid\boldsymbol{B}_i)/L(S_n)-\mu(\boldsymbol{B}_i\mid \boldsymbol{B}_i^{(t)})_{\mathcal{S}}\right)\cdot\nabla\log \operatorname{NHP}(S_n\mid\boldsymbol{B}_i)/L(S_n)}{\frac{1}{N}\sum_{n}^N w_1(S_n;\mathbf{B})\phi_\rho^{'}\left(\log \operatorname{NHP}(S_n\mid\boldsymbol{B}_i)/L(S_n)-\mu(\boldsymbol{B}_i\mid \boldsymbol{B}_i^{(t)})_{\mathcal{S}}\right)},\\
        & \nabla \mu(\boldsymbol{B}_i\mid \boldsymbol{B}_i^{(t)})\\
        &:=\frac{E w_1(S_n;\mathbf{B})\phi_\rho^{'}\left(\log \operatorname{NHP}(S_n\mid\boldsymbol{B}_i)/L(S_n)-\mu(\boldsymbol{B}_i\mid \boldsymbol{B}_i^{(t)})\right)\cdot\nabla\log \operatorname{NHP}(S_n\mid\boldsymbol{B}_i)/L(S_n)}{E w_1(S_n;\mathbf{B})\phi_\rho^{'}\left(\log \operatorname{NHP}(S_n\mid\boldsymbol{B}_i)/L(S_n)-\mu(\boldsymbol{B}_i\mid \boldsymbol{B}_i^{(t)})\right)}.
    \end{align*}
    }
    Then we have $\left\|\nabla \mu(\boldsymbol{B}_i\mid \boldsymbol{B}_i^{(t)})_{\mathcal{S}}-\nabla \mu(\boldsymbol{B}_i\mid \boldsymbol{B}_i^{(t)})\right\|\leq  O(L\exp(-GL)/\sqrt{N}+(\rho+1)(1/\sqrt{NL}+\frac{\rho v}{L} + \frac{\log N}{\rho N} + \frac{\eta}{\rho}))$.
    
\end{lemma}

\noindent\textbf{Proof of Lemma \ref{lem:grad:concen}}~
Recall that $\mathcal{S}=\mathcal{S}_{\operatorname{inlier}}\cup\mathcal{S}_{\operatorname{outlier}}$ with $\mathcal{S}_{\operatorname{inlier}} = \mathcal{S}_1 \cup ... \cup \mathcal{S}_K$. We define 
{\small
    \begin{align*}
        &\widetilde{\nabla \mu(\boldsymbol{B}_i\mid \boldsymbol{B}_i^{(t)})}_{\mathcal{S}_{\operatorname{inlier}}} \nonumber \\
        &:=\frac{\frac{1}{N}\sum_{n\in\mathcal{S}_{\operatorname{inlier}}}w_1(S_n;\mathbf{B})\phi_\rho^{'}\left(\log \operatorname{NHP}(S_n\mid\boldsymbol{B}_i)/L(S_n)-\mu(\boldsymbol{B}_i\mid \boldsymbol{B}_i^{(t)})_{\mathcal{S}_{\operatorname{inlier}}}\right)\cdot\nabla\log \operatorname{NHP}(S_n\mid\boldsymbol{B}_i)/L(S_n)}{\frac{1}{N}\sum_{n\in \mathcal{S}_{\operatorname{inlier}}} w_1(S_n;\mathbf{B})\phi_\rho^{'}\left(\log \operatorname{NHP}(S_n\mid\boldsymbol{B}_i)/L(S_n)-\mu(\boldsymbol{B}_i\mid \boldsymbol{B}_i^{(t)})_{\mathcal{S}_{\operatorname{inlier}}}\right)}\\
        &:=\frac{A}{B},
    \end{align*}
}
which is the gradient based on the inlier samples only.
By triangle inequality, we have 
\begin{align*}
    &\left\|\nabla \mu(\boldsymbol{B}_i\mid \boldsymbol{B}_i^{(t)})_{\mathcal{S}}-\nabla \mu(\boldsymbol{B}_i\mid \boldsymbol{B}_i^{(t)})\right\|\\
    &\leq \underbrace{\left\|\nabla \mu(\boldsymbol{B}_i\mid \boldsymbol{B}_i^{(t)})_{\mathcal{S}}-\widetilde{\nabla \mu(\boldsymbol{B}_i\mid \boldsymbol{B}_i^{(t)})}_{\mathcal{S}_{\operatorname{inlier}}}\right\|}_{I_1}+\underbrace{\left\|\widetilde{\nabla \mu(\boldsymbol{B}_i\mid \boldsymbol{B}_i^{(t)})}_{\mathcal{S}_{\operatorname{inlier}}}-\nabla \mu(\boldsymbol{B}_i\mid \boldsymbol{B}_i^{(t)})\right\|}_{I_2}.
\end{align*}

We consider the part $I_2$ first. According to Lemma \ref{lemma:error:bound} and Lemma \ref{lemma:mu:bound}, the deviation of $\mu(\boldsymbol{B}_i\mid \boldsymbol{B}_i^{(t)})_{\mathcal{S}_{\operatorname{inlier}}}$ from $\mathbb E[\log \operatorname{NHP}(S_n\mid\boldsymbol{B}_i)/L(S_n)]$ is $  O((\rho v)/L + \log N/(\rho N) + \eta/\rho + L^2 \exp\{-GL\} + \rho^2/\sqrt{L})$, so $\left|\log \operatorname{NHP}(S_n\mid\boldsymbol{B}_i)/L(S_n)-\mu(\boldsymbol{B}_i\mid \boldsymbol{B}_i^{(t)})_{\mathcal{S}_{\operatorname{inlier}}}\right|\sim O(1/\sqrt{L}+(\rho v)/L + \log N/(\rho N) + \eta/\rho +  L^2 \exp\{-GL\})$. 
The standard deviation of\\
$\phi_\rho^{'}\left(\log \operatorname{NHP}(S_n\mid\boldsymbol{B}_i)/L(S_n)-\mu(\boldsymbol{B}_i\mid \boldsymbol{B}_i^{(t)})_{\mathcal{S}_{\operatorname{inlier}}}\right)$ is $O(\rho/\sqrt{L}+(\rho^2 v)/L + \log N/N + {\eta} + \rho L^2 \exp\{-L\})$, so the standard deviation of $B$ is $O(\rho/\sqrt{NL}+(\rho^2 v)/L + \log N/N + {\eta} + \rho L^2 \exp\{-L\})$.
The standard deviation of part $A$ is similar to part $B$.
Similarly, the standard deviation of $\left\|\sum_N\nabla\log \operatorname{NHP}(S_n\mid\boldsymbol{B}_i)/NL\right\|$ is $O(1/\sqrt{NL})$, then $I_2\sim O(\rho/\sqrt{NL}+(\rho^2 v)/L + \log N/N + {\eta} + \rho L^2 \exp\{-L\})$.

Next we consider the part $I_1$. Again by  Lemma \ref{lemma:error:bound}, $ \left|\mu(\boldsymbol{B}_i\mid \boldsymbol{B}_i^{(t)})_{\mathcal{S}}- \mu(\boldsymbol{B}_i\mid \boldsymbol{B}_i^{(t)})_{\mathcal{S}_{\operatorname{inlier}}}\right|\sim O((\rho v)/L + \log N/(\rho N) + \eta/\rho + L^2 \exp\{-GL\})$. Note that
\begin{align*}
    &\frac{1}{N}\sum_{n\in\mathcal{S}} w_1(S_n;\mathbf{B})\phi_\rho^{'}\left(\log \operatorname{NHP}(S_n\mid\boldsymbol{B}_i)/L(S_n)-\mu(\boldsymbol{B}_i\mid \boldsymbol{B}_i^{(t)})_{\mathcal{S}}\right)\cdot\nabla\log \operatorname{NHP}(S_n\mid\boldsymbol{B}_i)/L(S_n)\\
    &=\underbrace{\frac{1}{N}\sum_{n\in\mathcal{S}_1} w_1(S_n;\mathbf{B})\phi_\rho^{'}\left(\log \operatorname{NHP}(S_n\mid\boldsymbol{B}_i)/L(S_n)-\mu(\boldsymbol{B}_i\mid \boldsymbol{B}_i^{(t)})_{\mathcal{S}}\right)\cdot\nabla\log \operatorname{NHP}(S_n\mid\boldsymbol{B}_i)/L(S_n)}_{W_1}\\
    &+\underbrace{\frac{1}{N}\sum_{n\in\mathcal{S}_{\operatorname{inlier}}\backslash\mathcal{S}_1} w_1(S_n;\mathbf{B})\phi_\rho^{'}\left(\log \operatorname{NHP}(S_n\mid\boldsymbol{B}_i)/L(S_n)-\mu(\boldsymbol{B}_i\mid \boldsymbol{B}_i^{(t)})_{\mathcal{S}}\right)\cdot\nabla\log \operatorname{NHP}(S_n\mid\boldsymbol{B}_i)/L(S_n)}_{W_2}\\
    &+\underbrace{\frac{1}{N}\sum_{n\in\mathcal{S}_{\operatorname{outlier}}} w_1(S_n;\mathbf{B})\phi_\rho^{'}\left(\log \operatorname{NHP}(S_n\mid\boldsymbol{B}_i)/L(S_n)-\mu(\boldsymbol{B}_i\mid \boldsymbol{B}_i^{(t)})_{\mathcal{S}}\right)\cdot\nabla\log \operatorname{NHP}(S_n\mid\boldsymbol{B}_i)/L(S_n)}_{W_3}.
\end{align*}
According to Lemma \ref{lem:weight}, $\left\|W_2\right\|\leq \left\|N^{-1}\sum_{n\in\mathcal{S}_{\operatorname{inlier}}\backslash\mathcal{S}_1} w_1(S_n;\mathbf{B})\cdot\nabla\log \operatorname{NHP}(S_n\mid\boldsymbol{B}_i)/L(S_n)\right\|\sim O(L\exp(-GL))$, so $\left\|W_2-EW_2\right\|\sim O(L\exp(-GL)/\sqrt{N})$. Similarly, $\left\|W_1-EW_1\right\|\sim O(L\exp(-GL)/\sqrt{N})$.
When $\left|\log \operatorname{NHP}(S_n\mid\boldsymbol{B}_i)/L(S_n)-\mu(\boldsymbol{B}_i\mid \boldsymbol{B}_i^{(t)})_{\mathcal{S}}\right|<9.5$,  
the gradient of outlier are less than a constant $c_{out}$ with a high probability, then $\left\|W_3\right\|\leq O(\eta/\rho)$. 
Then $\left\|W_1+W_2+W_3-A\right\|\leq\left\|W_1-A\right\|+\left\|W_2\right\|+\left\|W_3\right\|\sim O(L\exp(-GL)/\sqrt{N}+\eta/\rho)$. 
The standard deviation of part $A$ is similar to part $B$.
Hence $\left\|I_1\right\|\leq O(L\exp(-GL)/\sqrt{N}+\frac{\rho v}{L} + \frac{\log N}{\rho N} + \frac{\eta}{\rho})$.

In summary, $\left\|\nabla \mu(\boldsymbol{B}_i\mid \boldsymbol{B}_i^{(t)})_{\mathcal{S}}-\nabla \mu(\boldsymbol{B}_i\mid \boldsymbol{B}_i^{(t)})\right\|\leq I_1+I_2 \leq  O(L\exp(-GL)/\sqrt{N}+(\rho+1)(1/\sqrt{NL}+(\rho v)/L + \frac{\log N}{\rho N} + \frac{\eta}{\rho}))$.

\medskip

\noindent \textbf{Proof of Theorem \ref{thm:local}}~ 
 % $\nabla_{\boldsymbol{B}_i} \mu(\boldsymbol{B}_i\mid \boldsymbol{B}_i^*)=\mathbb{E}_S w_i\left(S ; \boldsymbol{B}^*\right)\phi'_\alpha\left(\log \operatorname{NHP}(S\mid \boldsymbol{B}_i ))-\mu(\boldsymbol{B}_i\mid\boldsymbol{B}_i^*)\right)\cdot\alpha\nabla\log \operatorname{NHP}(S\mid \boldsymbol{B}_i ))$. Without loss of generality, we only show the claim for $i=1$. That is equivalent of saying, if $S \sim \operatorname{Poi}\left(\pi, \boldsymbol{B}_1^*\right)$, we have 
 % \[\mathbb{E}_S w_i\left(S ; \boldsymbol{B}^*\right)\phi_\alpha\left(\log \operatorname{NHP}(S\mid \boldsymbol{B}_1^* ))-\mu(\boldsymbol{B}_1^*)\right)\alpha\nabla\log \operatorname{NHP}(S\mid \boldsymbol{B}_1^* ))=0.\]
Recall the update rule and definition of $\nabla \mu(\boldsymbol{B}_1 | \boldsymbol{B}_1^{(t)})$, we know that 
\[\boldsymbol{B}_1^{(t+1)} = 
\boldsymbol{B}_1^{(t)} - \text{lr} \cdot \varrho_1^{(t)} = \boldsymbol{B}_1^{(t)} - \text{lr} \cdot \nabla \mu(\boldsymbol{B}_1 | \boldsymbol{B}_1^{(t)})_{\mathcal S}.\]
By triangle inequality and  Theorem \ref{thm:gamma}, we have
$$
\begin{aligned}
&\left\|\boldsymbol{B}_1^{(t+1)}-\boldsymbol{B}_1^*\right\|  =\left\|\boldsymbol{B}_1^{(t)}-\boldsymbol{B}_1^*+ \text{lr} \cdot \nabla \mu(\boldsymbol{B}_1\mid \boldsymbol{B}_1^{(t)})_{\mathcal{S}}\right\| \\
& \leq\left\|\boldsymbol{B}_1^{(t)}-\boldsymbol{B}_1^*+ \text{lr} \cdot \nabla \mu(\boldsymbol{B}_1\mid \boldsymbol{B}_1^*)\right\|+  \text{lr} \cdot \left\|\nabla \mu(\boldsymbol{B}_1\mid \boldsymbol{B}_1^{(t)})-\nabla \mu(\boldsymbol{B}_1\mid \boldsymbol{B}_1^*)\right\|\\
& +  \text{lr} \cdot \left\|\nabla \mu(\boldsymbol{B}_1\mid \boldsymbol{B}_1^{(t)})-\nabla \mu(\boldsymbol{B}_1\mid \boldsymbol{B}_1^{(t)})_{\mathcal{S}}\right\|  \\
& \leq \frac{\lambda_{\max }-\lambda_{\min }}{\lambda_{\max }+\lambda_{\min }}\left\|\boldsymbol{B}_1^{(t)}-\boldsymbol{B}_1^*\right\|+\frac{2}{\lambda_{\max }+\lambda_{\min }} \gamma\left\|\boldsymbol{B}_1^{(t)}-\boldsymbol{B}_1^*\right\|+\epsilon^{unif}\\
& \leq \frac{\lambda_{\max }-\lambda_{\min }+2 \gamma}{\lambda_{\max }+\lambda_{\min }}\left\|\boldsymbol{B}_1^t-\boldsymbol{B}_1^*\right\|+\epsilon^{unif}.
\end{aligned}
$$
To see why the second inequality holds, note that, for any $\boldsymbol{B}_1'$ with $\|\boldsymbol{B}_1' - \boldsymbol{B}^{\ast}\| \leq a$, $\Delta \mu(\boldsymbol{B}_1\mid \boldsymbol{B}_1')$  has the largest eigenvalue $-\lambda_{\min }$ and smallest eigenvalue $-\lambda_{\max }$. Applying the classical result for gradient descent with step size $\text{lr} = 2/(\lambda_{\max }+\lambda_{\min })$, it guarantees (see \cite{nesterov2003introductory})
$$
\left\|\boldsymbol{B}_1^t-\boldsymbol{B}_1^*+ \text{lr} \cdot \nabla \mu(\boldsymbol{B}_1\mid \boldsymbol{B}_1^*)\right\| \leq \frac{\lambda_{\max }-\lambda_{\min }}{\lambda_{\max }+\lambda_{\min }}\left\|\boldsymbol{B}_1^t-\boldsymbol{B}_1^*\right\|.
$$
This completes the proof.

\begin{lemma}\label{lem:robust}
    Assume $S=\{t_0,t_1,\cdots\}$ sample from the non-homogeneous poisson process 
    with the parameter $\mathcal{B}$, then the variance of log-likelihood function is $\int_0^T \lambda_{\mathcal{B}}(t)\cdot\log\left( \lambda_{\mathcal{B}}(t)\right)^2 dt$.
\end{lemma}
\noindent \textbf{Proof of Lemma \ref{lem:robust}}~
 See \cite{kalbfleisch2011statistical}.

\begin{lemma}\label{thm:var}
    Assume $S=\{t_0,t_1,\cdots\}$ sample from the non-homogeneous Poisson process 
    with the parameter $\mathcal{B}$, and its period is $T$ and its number of periods is $L(S)$. Then the variance of its log-likelihood function is $O(L(S)^{-1})$.
\end{lemma}

\noindent \textbf{Proof of Lemma \ref{thm:var}}
For the sequence $S$ of length $L(S)$, we write the log-likelihood function as $Y:=\sum_{h=1}^{L(S)}X_h$, where $X_h:=\sum_{t_j\in \left((h-1)\cdot T,h\cdot T\right]}\log \lambda_{\mathcal{B}}(t_j)-\int_{0}^T \lambda_{\mathcal{B}}(t)dt$. According to Lemma \ref{lem:robust}, it is known that the variance of each $X_h$ is $\sigma_X^2=\int_0^T \lambda_{\mathcal{B}}(t)\cdot\log\left( \lambda_{\mathcal{B}}(t)\right)^2 dt$. Assume that the mean of $X_h$ is $\mu_X$.
Using the Chebyshev's inequality, we know that 
$$
\mathbb{P}\left(|X_h - \mu_X| \geq k\sigma_X\right) = \mathbb{P}\left((X_h - \mu_X)^2 \geq k^2\sigma_X^2\right) \leq \frac{1}{k^2}, \forall k>1.
$$
Since each $X_h$ is independent and identically distributed, it is easy to know that the variance of $Y/L(S)$ is $\sigma_Y^2=\sigma_X^2/L(S)$. Then take $k=4.5$, we have 
$$
\mathbb{P}\left(|Y/L(S) - \mu_Y/L(S)| \geq k\sigma_Y \right)=\mathbb{P}\left(|Y/L(S) - \mu_X| \geq k\sigma_X/\sqrt{L(S)} \right)\leq \frac{1}{k^2} < 0.05.
$$

\begin{lemma}\label{lem:robust+}
    Assume $\mathcal{S}=\{s_1,s_2,\cdots\}$ sample from the non-homogeneous Poisson process 
    with the parameter $\mathbf{B}$, and its period is $T$ and its number of periods is $L$. For each sample $s_n\in\mathcal{S}$, when we select robust parameter $\alpha\sim O(L^{\beta}), 0<\beta<1/2$. Then as $L\rightarrow\infty$, the weight function $\phi'_{\alpha}\left(\log \operatorname{NHP}\left(s_n \mid \mathbf{B}\right)/L(s_n)-\hat{\mu}_{\phi}(\mathbf{B})\right)$ tends to $1$ with a high probability. If $s_o$ is an outlier sample, as $L\rightarrow\infty$, the weight function $\phi'_{\alpha}\left(\log \operatorname{NHP}\left(s_o \mid \mathbf{B}\right)/L(s_n)-\hat{\mu}_{\phi}(\mathbf{B})\right)$ tends to $0$ with a high probability.
\end{lemma}

\noindent \textbf{Proof of Lemma \ref{lem:robust+}} 
By Lemma \ref{thm:var}, we know that the standard deviation of the log-likelihood functions for each sample is $O(L^{-1/2})$. From Lemma \ref{lemma:error:bound}, we know that $\hat \mu_{\phi}(\mathbf{B}) - \mu^{\ast}(\mathbf{B}) = O_p((\rho v)/L + log N/(\rho N) + \eta/\rho + L^2 \exp\{-GL\})$. So we have 
\begin{align*}
& \log \operatorname{HP}\left(s_n \mid \mathbf{B}\right)/L(s_n)-\hat{\mu}_{\phi}(\mathbf{B})\sim O\left(L^{-1/2} + \frac{\rho v}{L} + \frac{\log N}{\rho N} + \frac{\eta}{\rho} + L^2 \exp\{-GL\}\right)\\
\Rightarrow & \alpha\left(\log \operatorname{HP}\left(s_n \mid \mathbf{B}\right)/L(s_n)-\hat{\mu}_{\phi}(\mathbf{B})\right)\sim O(L^{\beta-1/2} + L^{2 + \beta} \exp\{-GL\})\rightarrow0
\end{align*}
for any $\alpha = O(L^{\beta})$ with $0 < \beta < 1/2$,
when $L\rightarrow\infty$. Looking back at the definition of robust function \eqref{eq:rob}, we can easily know that $\lim_{x\rightarrow0}\phi(x)=1$. At this time there is $\phi'_{\alpha}\left(\log \operatorname{HP}\left(s_n \mid \mathbf{B}\right)/L(s_n)-\hat{\mu}_{\phi}(\mathbf{B})\right)\rightarrow1$.
For outlier $s_o$ we have
\begin{align*}
& \log \operatorname{HP}\left(s_o \mid \mathbf{B}\right)/L(s_n)-\hat{\mu}_{\phi}(\mathbf{B})\sim O(1),
\end{align*}
which implies
\begin{align*}
\Rightarrow & \alpha\left(\log \operatorname{HP}\left(s_o \mid \mathbf{B}\right)/L(s_n)-\hat{\mu}_{\phi}(\mathbf{B})\right)\sim O(L^{\beta})\rightarrow\infty
\end{align*}
when $L\rightarrow\infty$. Because of $\lim_{x\rightarrow\infty}\phi(x)=0$, so we have $\phi'_{\alpha}\left(\log \operatorname{HP}\left(s_o \mid \mathbf{B}\right)/L(s_n)-\hat{\mu}_{\phi}(\mathbf{B})\right)\rightarrow0$.

\noindent \textbf{Proof of Theorem \ref{thm:robust}} 
According to Lemma \ref{lem:robust+},  we know that  the weight function will tend to $0$ for all outliers as $L\rightarrow\infty$. Therefore we can distinguish almost all outliers with a high probability by setting the cutoff as 0.1.

\begin{remark}
In all the above proofs, we do not take into account the shift parameter.
The local convergence result could be still applied, if the algorithm starts with the true shift parameter and $\left\|\boldsymbol{B}_k^{(0)}-\boldsymbol{B}_k^*\right\|$ is small enough for $k \in \{1,2,\cdots,K\}$.
\end{remark}

\section{Proof of Theorem \ref{thm:dis}}

Here we would like to point out that we say the event sequence $S$ is different from $S'$ if their induced intensity $\hat \lambda_S / \sqrt{M}$'s are different.
Otherwise, we treat them as the same event sequence.

\noindent \textbf{Proof of Theorem \ref{thm:dis}}
It is easy to know that the distance between an object and itself is always zero and the distance between distinct objects is always positive. Moreover, the distance from $S_A$ to $S_B$ is always the same as the distance from $S_B$ to $S_A$. We only need to prove that $d(S_A,S_B)$ satisfies the triangle inequality.

By definition we know that $d(S_A,S_B)=\int_{0}^T \left|\hat{\lambda}_A\left(t\right)/\sqrt{M_A}-\hat{\lambda}_B\left(t+\delta_B\right)/\sqrt{M_B}\right|dt$, where $\delta_B=\arg min_{\delta_B}\int_{0}^T \left|\hat{\lambda}_A\left(t\right)/\sqrt{M_A}-\hat{\lambda}_B\left(t+\delta_B\right)/\sqrt{M_B}\right|dt $.
In the same way we define $\delta_C$. Then
\begin{align*}
d(S_B,S_C)&\leq \int_{0}^T \left|\hat{\lambda}_C\left(t+\delta_C\right)/\sqrt{M_C}-\hat{\lambda}_B\left(t+\delta_B\right)/\sqrt{M_B}\right|dt\\
&\leq \int_{0}^T \left|\hat{\lambda}_C\left(t+\delta_C\right)/\sqrt{M_C}-\hat{\lambda}_A\left(t\right)/\sqrt{M_A}\right|dt+\int_{0}^T \left|\hat{\lambda}_A\left(t\right)/\sqrt{M_A}-\hat{\lambda}_B\left(t+\delta_B\right)/\sqrt{M_B}\right|dt\\
&=d(S_A,S_B)+d(S_A,S_C).
\end{align*}
This completes the proof.

\section{Supporting Results of $\hat \mu_{\phi}^{(t)}(\boldsymbol{B}_k)$ and $\mu(\boldsymbol{B}_k|\boldsymbol{B}_k^{\ast})$}

In this section, we provide two supporting lemmas to characterize the difference between $\hat \mu_{\phi}^{(t)}(\boldsymbol{B}_k)$ and $\mu(\boldsymbol{B}_k|\boldsymbol{B}_k^{\ast})$.

\begin{lemma}\label{lemma:error:bound}
     When $\|\hat{\boldsymbol{B}}_k^{(t)} - \boldsymbol{B}_k^{\ast}\| \leq a$ and $\eta := |\mathcal S_{outlier}| / N < \frac{1}{4 \cdot (\log 5 + 1.5)}$, it holds 
    \begin{eqnarray}\label{eq:error:mu:key}
    |\hat \mu_{\phi}^{(t)}(\boldsymbol{B}_k) -  \mu^{\ast}(\boldsymbol{B}_k)| =O_p\left(\frac{\rho v}{L} + \frac{\log N}{\rho N} + \frac{\eta}{\rho} + L^2 \exp\{-GL\}\right),
    \end{eqnarray}
    where $\mu^{\ast}(\boldsymbol{B}_k) = \mathbb E_{S \sim \lambda_k^{\ast}}[\log \operatorname{NHP}(S|\boldsymbol{B}_k)]$ and $v := \sup_{\boldsymbol{B}_k} \mathbb E[(\log \operatorname{NHP}(S|\boldsymbol{B}_k))^2]$ ($S$ is an event sequence on $[0,T]$ generated according to $\lambda_k^{\ast}(t)$).
\end{lemma}

\noindent \textbf{Proof of Lemma \ref{lemma:error:bound}}
First, we define $\bar \mu_{\phi}^{(t)}(\boldsymbol{B}_k)$ to be the solution to 
\begin{eqnarray}\label{eq:mu:bar}
\sum_{n=1}^N  1  / L(S_n) \cdot \phi_{\rho} \left(\log \operatorname{NHP}\left(S_n \mid \boldsymbol{B}_k\right) / L(S_n)- \mu \right)=0
\end{eqnarray}
with respect to $\mu$. We can show that 
\begin{eqnarray}\label{eq:1:diff}
   \bar \mu_{\phi}^{(t)}(\boldsymbol{B}_k) - \hat \mu_{\phi}^{(t)}(\boldsymbol{B}_k)| = O_p(L^2 \exp\{-GL\}). 
\end{eqnarray}
To see this, we compare the difference between
\[\frac{1}{N}\sum_{n=1}^N  1  / L(S_n) \cdot \phi_{\rho} \left(\log \operatorname{NHP}\left(S_n \mid \boldsymbol{B}_k\right) / L(S_n)- \bar \mu_{\phi}^{(t)}(\boldsymbol{B}_k) \right)\]
and 
\[\frac{1}{N} \sum_{n=1}^N  r_{n k}^{(t)}  / L(S_n) \cdot \phi_{\rho} \left(\log \operatorname{NHP}\left(S_n \mid \boldsymbol{B}_k\right) / L(S_n)- \bar \mu_{\phi}^{(t)}(\boldsymbol{B}_k) \right).\]
By the previous analysis, we have already shown that $|r_{nk}^{(t)} - 1| = O_p(L \exp\{-G L\})$.
Then such difference is bounded by $C L \exp\{-G L\} \cdot \sum_{n} L(S_n)  \phi_{\rho} \left(\log \operatorname{NHP}\left(S_n \mid \boldsymbol{B}_k\right) / L(S_n)- \bar \mu_{\phi}^{(t)}(\boldsymbol{B}_k) \right)$ which is order of $\exp\{- GL\} (\eta/\rho + \log L)$ and is less than $L \exp\{-GL\}$. (Here we use the fact that $\eta / \rho \rightarrow 0$).
By the definition of $\bar \mu_{\phi}^{(t)}(\boldsymbol{B}_k)$, we have 
\[|\frac{1}{N} \sum_{n=1}^N  r_{n k}^{(t)}  / L(S_n) \cdot \phi_{\rho} \left(\log \operatorname{NHP}\left(S_n \mid \boldsymbol{B}_k\right) / L(S_n)- \bar \mu_{\phi}^{(t)}(\boldsymbol{B}_k) \right)| \leq L \exp\{-GL\}.\]
It can be also checked that 
$ \nabla_{\mu} \big(N^{-1}\sum_{n=1}^N  r_{n k}^{(t)}  / L(S_n) \cdot \phi_{\rho} \left(\log \operatorname{NHP}\left(S_n \mid \boldsymbol{B}_k\right) / L(S_n)- \mu \right) \big) \geq 1/2L$ for all bounded $\mu$ with probability 1.
Therefore,
\begin{eqnarray}
    & & \frac{1}{2L} |\bar \mu_{\phi}^{(t)}(\boldsymbol{B}_k) - \hat \mu_{\phi}^{(t)}(\boldsymbol{B}_k)| \nonumber \\
    &\leq& 
    |\frac{1}{N} \sum_{n=1}^N  r_{n k}^{(t)}  / L(S_n) \cdot \phi_{\rho} \left(\log \operatorname{NHP}\left(S_n \mid \boldsymbol{B}_k\right) / L(S_n)- \bar \mu_{\phi}^{(t)}(\boldsymbol{B}_k) \right)| \leq L \exp\{-GL\}, \nonumber
\end{eqnarray}
which gives the desired result \eqref{eq:1:diff}.

Next, we construct 
\begin{align}\label{eq:B}
B_{+, \boldsymbol{B}_k}(\mu) &= (\mu^{\ast}(\boldsymbol{B}_k) - \mu) + \frac{\rho}{2}(\frac{v^{\ast}(\boldsymbol{B}_k)}{L} + (\mu^{\ast}(\boldsymbol{B}_k) - \mu)^2) + \frac{ 2 \log N }{\pi_k^{\ast} N \rho}, \\
B_{-, \boldsymbol{B}_k}(\mu) &= (\mu^{\ast}(\boldsymbol{B}_k) - \mu) - \frac{\rho}{2}(\frac{v^{\ast}(\boldsymbol{B}_k)}{L} + (\mu^{\ast}(\boldsymbol{B}_k) - \mu)^2) - \frac{ 2 \log N }{\pi_k^{\ast} N \rho},  \nonumber
\end{align}
where $v^{\ast}(\boldsymbol{B}_k) = \mathbb E_{S \sim \lambda_k^{\ast}}[(\log \operatorname{NHP}(S|\boldsymbol{B}_k))^2]$, to put the upper and lower bounds on $\phi_{\rho}$ in \eqref{eq:mu:robust}.
Following the proof of Theorem 3.1 in \cite{bhatt2022minimax} and the compactness of parameter space, we can have
\begin{eqnarray}\label{eq:bd:simple}
    |\bar \mu_{\phi}^{(t)}(\boldsymbol{B}_k) - \mu^{\ast}(\boldsymbol{B}_k)| =O_p\left(\frac{\rho v}{L} + \frac{\log N}{\rho N} + \frac{\eta}{\rho}\right)
\end{eqnarray}
for all $\boldsymbol{B}_k$, where $v = \max_{\boldsymbol{B}_k} v^{\ast}(\boldsymbol{B}_k)$.
Combining \eqref{eq:1:diff} and \eqref{eq:bd:simple}, we prove the lemma.

\begin{lemma}\label{lemma:mu:bound}
     It holds 
     \begin{eqnarray}\label{eq:mu:w}
     | \mu(\boldsymbol{B}_k\mid \boldsymbol{B}_k^*) - \mu^{\ast}(\boldsymbol{B}_k)| = O\left(L^2 \exp\{-GL\} + \rho^2\sqrt{\frac{1}{L}}\right) ,
     \end{eqnarray}
     where $\mu^{\ast}(\boldsymbol{B}_k)$ is defined the same as that in Lemma \ref{lemma:error:bound}.
\end{lemma}

\noindent \textbf{Proof of Lemma \ref{lemma:mu:bound}}
We first define $\bar \mu(\boldsymbol{B}_k\mid \boldsymbol{B}_k^*)$ to be the solution to 
\[\mathbb{E}_S [\phi_\rho\left(\log \operatorname{NHP}(S\mid \boldsymbol{B}_k ))/L(S)-\mu\right)]=0\]
with respect to $\mu$. By the same procedure as in the first part of proof of Lemma \ref{lemma:error:bound}, we can show that 
\begin{eqnarray}\label{eq:diff:mu:1}
    |\mu(\boldsymbol{B}_k\mid \boldsymbol{B}_k^*) - \bar \mu(\boldsymbol{B}_k\mid \boldsymbol{B}_k^*)| \leq L^2 \exp\{-GL\}.
\end{eqnarray}
Next we compute the bound of 
$|\mathbb{E}_S [\phi_\rho\left(\log \operatorname{NHP}(S\mid \boldsymbol{B}_k ))/L(S) - \mu^{\ast}(\boldsymbol{B}_k)\right)]|$.
Note that $\phi_{\rho}(x) = x - \rho^2x^3/6 + o(\rho^2 x^3)$ by Taylor expansion.
Therefore, for sufficiently small $\rho$, we have 
\begin{eqnarray}
    & &|\mathbb{E}_S [\phi_\rho\left(\log \operatorname{NHP}(S\mid \boldsymbol{B}_k ))/L(S) - \mu^{\ast}(\boldsymbol{B}_k)\right)]| \nonumber \\
    &\leq& \frac{\rho^2}{3}|\mathbb{E}_S [(\log \operatorname{NHP}(S\mid \boldsymbol{B}_k ))/L(S) - \mu^{\ast}(\boldsymbol{B}_k))^3]| \nonumber \\
    &\leq&  \frac{\rho^2}{3} \Big(\mathbb{E}_S [(\log \operatorname{NHP}(S\mid \boldsymbol{B}_k ))/L(S) - \mu^{\ast}(\boldsymbol{B}_k))^6]\Big)^{1/2} \nonumber \\
    &=& O\left(\rho^2 \sqrt{\frac{1}{L}}\right).
\end{eqnarray}
Lastly, note that $\nabla_{\mu} (\mathbb{E}_S [\phi_\rho\left(\log \operatorname{NHP}(S\mid \boldsymbol{B}_k ))/L(S)-\mu\right)]) \geq 1/2$. Therefore, we have 
\[| \bar \mu(\boldsymbol{B}_k\mid \boldsymbol{B}_k^*) - \mu^{\ast}(\boldsymbol{B}_k)| \leq 2 |\mathbb{E}_S [\phi_\rho\left(\log \operatorname{NHP}(S\mid \boldsymbol{B}_k ))/L(S) - \mu^{\ast}(\boldsymbol{B}_k)\right)]| = O\left(\rho^2\sqrt{\frac{1}{L}}\right).\]
In summary, we get the desired result
\[\mu(\boldsymbol{B}_k\mid \boldsymbol{B}_k^*) - \mu^{\ast}(\boldsymbol{B}_k) = O\left(L^2 \exp\{-GL\} + \rho^2\sqrt{\frac{1}{L}}\right). \]

\end{document}